\documentclass[pra, reprint, superscriptaddress, amsmath,amssymb, aps]{revtex4-2}

\usepackage{amsmath}
\usepackage{amsthm}
\newtheorem{theorem}{Theorem}
\newtheorem{lemma}{Lemma}
\newtheorem{remark}{Remark}
\newtheorem{definition}{Definition}

\usepackage{tcolorbox}
\usepackage{esint}
\usepackage{bbold}
\usepackage{amssymb}
\usepackage{graphicx}
\usepackage{textcomp}
\usepackage[margin=0.8in]{geometry}
\usepackage{relsize}
\usepackage{url}
\usepackage[colorlinks=true, allcolors=camblue]{hyperref}
\usepackage{blkarray}
\usepackage{geometry}
\usepackage{cancel}
\usepackage{tikz}
\usepackage{natbib}
\usepackage{stmaryrd}
\usepackage[utf8]{inputenc}
\usepackage[T1]{fontenc}
\usepackage[caption=false]{subfig}
\usetikzlibrary{quantikz2}
\usepackage{multirow}
\usepackage{mathtools}
\mathtoolsset{showonlyrefs}
\usepackage{booktabs}
\usepackage{array}
\usepackage{bbding}

\definecolor{camblue}{cmyk}{0.2443, 0.0000, 0.1250, 0.3098}

\usepackage{xcolor}
\usepackage{lipsum}

\begin{document}

\title{Classification of Kirkwood-Dirac positivity-preserving unitaries}
\author{Jędrzej Burkat}
\email{jbb55@cam.ac.uk}
\affiliation{Cavendish Laboratory, Department of Physics, University of Cambridge, CB3 0US, UK}
\author{Sergii Strelchuk}
\affiliation{Department of Computer Science, University of Oxford, OX1 3QD, UK}

\begin{abstract}
    Kirkwood--Dirac (KD) quasiprobability distributions are increasingly used across quantum information science, yet their computational significance remains unclear. We classify unitary dynamics that preserve KD positivity and connect this structure to classical simulation. In contrast to the discrete Wigner setting, we show that KD positivity preservation, stochastic evolution of quasiprobabilities, and preservation of total non-positivity \textit{do not coincide}. We identify three classes of positivity-preserving unitaries: type~\textbf{I} gates, which are exactly the KD-stochastic unitaries; type~\textbf{II} gates, which are non-stochastic yet preserve total non-positivity by permuting and conjugating KD entries; and, for Fourier-conjugate bases in dimension $d=pq$, type~\textbf{III} gates, which preserve KD positivity but can change the total non-positivity of non-real distributions. The classification is complete for Haar-random bases and for Fourier-conjugate dimensions $d=p^k$ and $d=pq$, with $p, q$ distinct primes. Adapting the sampling algorithm of Pashayan et al. [PRL 115, 070501], we simulate all of these positivity-preserving circuits efficiently on KD-positive inputs; type~\textbf{III} gates, however, create a sharp distinction between real inputs, which remain efficiently simulable, and non-real inputs, whose sampling overhead can grow exponentially. Consequently, for $d=pq$ no resource theory can both treat the KD total non-positivity as a monotone and admit every efficiently simulable positivity-preserving unitary as a free operation.

\end{abstract}

\maketitle
\section{Introduction}

Demarcation between quantum and classical computational power is a central theme of quantum information science. Given the ability to implement a fixed set of quantum gates on an input state, whether the resulting circuit is powerful is typically understood in terms of its classical simulation complexity. The most famous example of this is the stabilizer formalism, which provides an efficient simulation scheme for Clifford circuits on stabilizer states that becomes inefficient upon introduction of additional non-stabilizer resources \cite{gottesman1998}. Concurrently, phase-space formulations of quantum mechanics have sought to identify non-classical phenomena in terms of non-positivity in quasiprobability distributions \cite{Spekkens_2008, Ferraro_2012,Mari_2012}. In his seminal paper, Gross \cite{Gross_2006} showed that stabilizer states form the only pure states with positive Wigner quasiprobability distributions, and that Clifford gates exhaust the set of positivity-preserving unitary evolutions. As a result, the discrete Wigner distribution gained a new operational interpretation, and has become a popular tool for theorists seeking to understand when a quantum system achieves computational advantage \cite{Ferrie_2008, Veitch_2012, PhysRevA.71.042302, Veitch_2014, PhysRevX.5.021003, Howard_2017, PhysRevA.101.012350, RevModPhys.84.621, Mari_2012,Pashayan_2015}. As the Clifford gates rarely constitute natively implementable operations in quantum hardware, other models of computation with sharp classical-quantum boundaries have been proposed \cite{Jozsa_2008, aaronson2010, terhal2004}. A promising candidate for phase-space formulations of non-Clifford computation is the Kirkwood--Dirac (KD) distribution \cite{PhysRev.44.31, RevModPhys.17.195}, which offers great flexibility for describing experimental state preparation and measurement in arbitrary reference bases \cite{Lostaglio_2023}. The utility of the KD distribution for finite-dimensional systems has led to its revival, with successful applications to quantum metrology \cite{PhysRevA.76.012119, PhysRevLett.112.070405, Arvidsson2020, Das_2023, Lupu_Gladstein_2022, Arvidsson_Shukur_2021}, thermodynamics \cite{PhysRevA.95.012120, PhysRevE.90.032137, Miller_2017, Levy_2020, PhysRevLett.125.230603, PhysRevLett.122.040404}, and the foundations of quantum mechanics \cite{Wagner_2024,schmid2024kirkwooddirac, PhysRevLett.113.200401, Williams_2008, Hofmann2024} (see \cite{arvidssonshukur2024properties} for a comprehensive review). However, the KD distribution has received little attention in the context of quantum computation, where KD positivity lacks an operational meaning in terms of efficient classical simulability. A natural question is also whether this distribution can resolve issues faced by the Wigner distribution in even-dimensional systems \cite{Raussendorf_2017}. In this work, we address these questions by classifying unitary evolutions which preserve KD positivity, and thus yield a natural candidate for free operations in resource theories built on Kirkwood--Dirac quasiprobabilities. As an application of our findings we adapt the quasiprobabilistic Born rule probability estimation algorithm of \cite{Pashayan_2015} to KD distributions, which explicitly bounds simulation runtime with respect to the non-positivity of the initial state and the non-positivity induced by its unitary evolution. For our identified classes of positivity-preserving unitaries, we show that the algorithm proceeds efficiently on any KD-positive input. However, for a special class of positivity-preserving unitaries, the algorithm can blow up exponentially in runtime even if the input contains a very small amount of non-positivity. Our insights lead us to identify a fundamental obstruction for resource theories defined on Fourier-conjugate KD distributions in squarefree semiprime dimensions.

\begin{table*}[t]
\centering
\small
\defcitealias{Pashayan_2015}{PWB'15}
\renewcommand{\arraystretch}{1.3}
\setlength{\tabcolsep}{6pt}
\begin{tabular}{lccc}
\toprule
\textbf{Property} & \textbf{Type I} & \textbf{Type II} & \textbf{Type III} \\
\midrule
Positivity Preservation
  & \Checkmark
  & \Checkmark
  & \Checkmark \\[2pt]
Total Non-Positivity Preservation
  & \Checkmark
  & \Checkmark
  & \XSolidBrush \\[2pt]
Induces Stochastic Superoperator
  & \Checkmark
  & \XSolidBrush
  & \XSolidBrush \\[2pt]
Induces Covariant Transformation
  & \Checkmark
  & \Checkmark
  & \Checkmark \ \textit{(real inputs only)} \\ [2pt]
  Efficient Simulation in \citetalias{Pashayan_2015} 
  & \Checkmark
  & \Checkmark
  & \Checkmark \ \textit{(real inputs only)} \\ [2pt]
\multirow{2}{*}{Modification in \citetalias{Pashayan_2015}}
&\multirow{2}{*}{None required.}
& Set $I_k = \gamma(I_{k-1})$,
& Set $I_k = \gamma(I_{k-1})$ on real \\
& & conjugate phases. &  inputs. Otherwise, none.  \\ [2pt]
Existence for Haar-random $V$
& \Checkmark \ \textit{(}$\{ e^{i \theta} \mathbb{1}\}$ \textit{only)}
& \XSolidBrush
& \XSolidBrush \\ [2pt]
Existence for $V = \mathrm{DFT}_{p^k}$
& \Checkmark
& \Checkmark
& \XSolidBrush \\ [2pt]
Existence for $V = \mathrm{DFT}_{pq}$
& \Checkmark
& \Checkmark
& \Checkmark \\ [2pt]

\bottomrule
\end{tabular}
\caption{ \textit{Summary of Results.} The table summarises the properties of type \textbf{I}, \textbf{II}, and \textbf{III} unitaries identified in this work. The first four rows follow from Lemmas \ref{lemma:nonpospresunitaries}, \ref{cor:compositefourier} and Theorems \ref{thm:ustar}, \ref{theorem:genperm} in Section \ref{section:kdstochasticunitaries}. The next two rows follow from Section \ref{section:kdsimulation}, where $\gamma$ is the induced permutation of indices in $Q(U \rho U^\dagger)_{ij} = Q(\rho)_{\gamma^{-1}(ij)}$ (types \textbf{I}, \textbf{III}) and $Q(U \rho U^\dagger)_{ij} = Q(\rho)^*_{\gamma^{-1}(ij)}$ (type \textbf{II}). For type \textbf{III} unitaries, `real inputs' refer to states $\rho$ satisfying $Q(\rho)_{ij} \in \mathbb{R}$ for all $i, j$, directly prior to the action of $U$. The last three rows provide a complete classification of positivity-preserving unitaries for three types of KD distributions (see end of Section \ref{section:kdstochasticunitaries}), where the results for Haar-random $V$ hold with probability $1$.} 
\label{tab:unitary-types}
\end{table*}

This article is structured as follows: In Section \ref{section:preliminaries} we review the Kirkwood--Dirac distribution, as well as its vectorised form. The latter gives a useful representation in which transformations of KD distributions induced by quantum channels take on the form of linear matrix multiplication. Section \ref{section:kdstochasticunitaries} contains our main results. We first classify unitaries which preserve KD positivity on pure states: up to phase, such unitaries always act as permutations on $\mathcal{A} \cup \mathcal{B} \cup \mathcal{C}$, where $\mathcal{A, B}$ are the reference bases of the distribution, and $\mathcal{C}$ is the set of non-basis KD-positive states. We identify three types of positivity-preserving unitaries: type \textbf{I} unitaries act as automorphisms on the sets $\mathcal{A, B, C}$, type \textbf{II} unitaries act as bijections between $\mathcal{A}$ and $\mathcal{B}$, and type \textbf{III} unitaries act as bijections among $\mathcal{A}$, $\mathcal{B}$ and subsets of $\mathcal{C}$. We then prove structural results: whenever $\mathcal{C} = \emptyset$, only type \textbf{I} or \textbf{II} unitaries can exist. Both types are covariant transformations which permute (and for type \textbf{II} additionally conjugate) the entries of the KD distribution, thus preserving the total non-positivity of any input. Out of the three, only type \textbf{I} unitaries correspond to KD-stochastic unitaries, i.e., evolution which induces a stochastic superoperator in the vectorised picture. This stands in contrast to the discrete Wigner distribution, for which all positivity-preserving (Clifford) unitaries induce stochastic superoperators. The existence of positivity-preserving unitaries is predicated on symmetries between the reference bases of the distribution. By identifying necessary conditions for their existence, we show that for KD distributions defined on Haar-randomly sampled bases $\mathcal{A, B}$, no non-trivial positivity-preserving unitaries exist with probability 1. We then focus on distributions defined on Fourier-conjugate bases. Recent work \cite{debievre2025,Spriet2026} has shown that for $V = \mathrm{DFT}_d$ the Weyl--Heisenberg group is positivity-preserving. In our classification such unitaries fall into type \textbf{I}. Similarly, any product of type \textbf{I} unitaries with the DFT matrix itself forms a type \textbf{II} unitary. We prove that for $d = p^k$ (with $p$ prime), positivity-preserving unitaries can only be of type \textbf{I} or \textbf{II}. On the other hand, for any $d = pq$ (with $p, q$ distinct primes) we prove the additional existence of type \textbf{III} unitaries: whilst such unitaries are shown to be positivity-preserving on any KD-positive input, they can change the total non-positivity of distributions containing complex entries. Furthermore, they do not form elements of the Weyl--Heisenberg group and their action is only guaranteed to be covariant on real-valued distributions. Section \ref{section:kdsimulation} applies our findings to the quantum simulation algorithm of \cite{Pashayan_2015}: the original method relies on superoperator stochasticity (satisfied by type \textbf{I}) to be efficient; we show how to adapt it to also efficiently simulate type \textbf{II} unitaries. The existence of type \textbf{III} unitaries poses an additional problem for the algorithm, due to their general lack of covariance. We show that for circuits composed of type \textbf{III} unitaries, if an input distribution is real-valued the simulation can proceed efficiently; however, if the input's distribution contains complex entries then the simulation runtime can blow up exponentially in the worst case. Our insights lead to a no-go theorem for resource theories defined on $V = \mathrm{DFT}^{\otimes n}_{pq}$ distributions: any such theory cannot simultaneously (i) include the efficiently simulable, positivity-preserving type \textbf{III} unitaries in its free operations, and (ii) treat the total non-positivity of the distribution as a monotone under the free operations. Our findings are summarised in Table \ref{tab:unitary-types}. The appendices mostly contain additional standalone results: Appendix \ref{appendix:cycletest} shows how to apply the cycle test algorithm of \cite{Wagner_2024} for experimental measurement of KD superoperator elements. Appendix \ref{section:kdconvolutions} contains additional structural results on Fourier-conjugate KD distributions. Appendix \ref{section:kdbounds} contains new bounds on the magnitudes of quasiprobabilities for any KD distribution. Finally, Appendix \ref{app:type3proofs} presents an extended proof of Theorem \ref{thm:ustar} showing that our identified type \textbf{III} unitaries are positivity-preserving beyond pure states, including for many-qudit systems where $V = \mathrm{DFT}^{\otimes n}_{pq}$.

\section{Preliminaries}
\label{section:preliminaries}

In this work, we will consider informationally complete Kirkwood--Dirac (KD) quasiprobability distributions defined with respect to two orthonormal bases of a $d$-dimensional Hilbert space. Here, we state all the relevant properties. For a thorough review of KD distributions and their applications, we recommend \cite{arvidssonshukur2024properties}. 

\begin{definition}[Kirkwood--Dirac Distributions \cite{arvidssonshukur2024properties}]
    For a quantum state $\rho$, the Kirkwood--Dirac distribution is the set of complex numbers (quasiprobabilities) $Q(\rho)_{ij}$ given by:

    \begin{equation} \label{eq:kddef}
        Q(\rho)_{ij} = \langle b_j | a_i \rangle \langle a_i | \rho | b_j \rangle, \quad (i, j) \in \llbracket d \rrbracket^2,
    \end{equation}

\noindent where $\mathcal{A} = \{|a_i \rangle \}$ and $\mathcal{B} = \{|b_j \rangle \}$ are the reference bases of the distribution, related by a transition matrix $V$ such that $|b_i \rangle = V |a_i \rangle$. If $V$ is such that $m_{\mathcal{A,B}} = \min_{ij} |\langle a_i | b_j \rangle| > 0$, the KD distribution is said to be informationally complete. 
\end{definition}

The name `quasiprobability' arises from the fact that the quantities $Q(\rho)_{ij}$ satisfy some, but not all, of the Kolmogorov axioms. For instance, they sum to $1$ and correctly reproduce Born rule probabilities upon marginalisation:

\begin{align}
    \sum_{j} Q(\rho)_{ij} &= \langle a_i | \rho | a_i \rangle,& \sum_{i} Q(\rho)_{ij} &= \langle b_j | \rho | b_j \rangle. 
\end{align}

\noindent As the KD distribution is invariant under a global change of basis (where $|a_i \rangle \rightarrow U |a_i \rangle$, $|b_j \rangle \rightarrow U |b_j \rangle$, $\rho \rightarrow U \rho U^\dagger$), it is often convenient to take $\{|a_i \rangle \}$ to be the computational basis and $|b_j \rangle = V |a_j \rangle$ for a unitary transition matrix $V$. For mutually unbiased bases (MUBs), $|\langle a_i | b_j \rangle| = m_{\mathcal{A,B}} = 1/\sqrt{d}$ for all $(i, j)$, and the corresponding $V$ is a complex Hadamard matrix. One can also define the KD distribution in terms of frames $\Lambda_{ij}$, satisfying:

\begin{align}
    \Lambda_{ij} &= \frac{| a_i \rangle \langle b_j |}{\langle b_j | a_i \rangle}, & Q(\rho)_{ij} &= \frac{\text{Tr}(\Lambda_{ij}^\dagger \rho)}{\text{Tr}( \Lambda_{ij} \Lambda_{ij}^\dagger)},
\end{align}

\noindent where the presence of complex-valued $Q(\rho)_{ij}$ can be directly seen as a consequence of the non-Hermicity of $\Lambda_{ij}$. Finally, given a KD distribution one may reconstruct $\rho$ via the weighted sum:

\begin{equation} \label{eq:kdanddensity}
    \rho = \sum_{i j} \frac{| a_i \rangle \langle b_j |}{\langle b_j | a_i \rangle} Q(\rho)_{ij},
\end{equation}

\noindent which is only possible for informationally complete KD distributions, i.e., when $m_{\mathcal{A,B}} > 0$. 

\textit{Kirkwood--Dirac Positivity.} When a KD distribution takes on exclusively real non-negative values, it resembles a classical joint probability distribution. In those cases we say that it is KD-positive, or KD-classical. 

\begin{definition}[Total Non-Positivity]
    The total non-positivity of a KD distribution $Q(\rho)$ is defined as:

    \begin{equation} \label{eq:nonposdef}
    \mathcal{N}(Q(\rho)) = \sum_{i j} | Q(\rho)_{ij} | \geq 1.
\end{equation}

\noindent We call a KD distribution KD-positive if and only if $\mathcal{N}(Q(\rho)) = 1$, and non-positive otherwise. Similarly, if $Q(\rho)_{ij} \in \mathbb{R}$ for all $i, j$ we call the distribution KD-real, and non-real otherwise.
\end{definition}

KD non-positivity has been linked to non-classical phenomena in quantum non-contextuality, and advantages in quantum metrology. The bases $\mathcal{A, B}$ are themselves pure KD-positive states, for which $Q_{ij}(a_k) = \delta_{k i} | \langle a_k | b_j \rangle |^2$ (and similarly for $Q_{ij}(b_m)$). Recent work \cite{langrenez2024} has shown that for the vast majority of KD distributions (i.e., with probability 1, when $V$ is sampled via the Haar measure on $U(d)$), the basis states from $\mathcal{A, B}$ are the only KD-positive pure states. Nonetheless, for some choices of $V$ pure, non-basis KD-positive states exist, and there can also exist mixed KD-positive states lying outside of the convex set of pure KD-positive states \cite{Langrenez_2024_characterising}.

\textit{Support Uncertainties.} For a pure state $| \psi \rangle$ described on a KD distribution it is also useful to define the support uncertainties $n_\mathcal{A}, n_\mathcal{B}$, which count the number of non-zero inner products between $| \psi \rangle$ and the two reference bases:

\begin{align}
    n_\mathcal{A} &= | S_\mathcal{A} | = | \{ |a_i \rangle \in \mathcal{A} : \langle a_i | \psi \rangle \neq 0 \} |, \nonumber \\ n_\mathcal{B} &= | S_\mathcal{B} | = | \{ |b_j \rangle \in \mathcal{B} : \langle b_j | \psi \rangle \neq 0 \} |. \label{eq:supunc}
\end{align}

\noindent For informationally complete distributions, the product $n_\mathcal{A} n_\mathcal{B}$ gives the number of non-zero entries in the KD distribution $Q(\psi)_{ij}$, i.e., its support. In \cite{De_Bievre_2021}, it was shown that for all pure KD-positive states, the sum $n_\mathcal{A} + n_\mathcal{B}$ satisfies:

\begin{equation}
    n_\mathcal{A} + n_\mathcal{B} \leq d + 1,
\end{equation}

\noindent and that for MUBs, all pure KD-positive states also satisfy \cite{Xu_2024, Langrenez_2024_characterising}:

\begin{equation}
    n_{\mathcal{A}} n_{\mathcal{B}} = d.
\end{equation}

\textit{KD-positive States.} In Section \ref{section:kdstochasticunitaries} we will classify unitaries which preserve KD positivity, i.e. unitaries which always map KD-positive states to other KD-positive states (Definition \ref{def:kdpospreserving}). A necessary condition for a unitary to satisfy this property is that, up to phase, it permutes elements of the set of pure KD-positive states, which we refer to as $\mathcal{E}^\mathrm{pure}_{\mathrm{KD}+}$. A large factor in our classification is whether the set $\mathcal{C} = \mathcal{E}^\mathrm{pure}_{\mathrm{KD}+} \setminus (\mathcal{A} \cup \mathcal{B})$ of pure, non-reference basis KD-positive states is empty. In general, we will write $\mathcal{E}^\mathrm{pure}_{\mathrm{KD}+} = \mathcal{A} \cup \mathcal{B} \cup \mathcal{C}$ to cover both scenarios. Instances where $\mathcal{C} = \emptyset$ include Haar-random choices of $V$ \cite{Langrenez_2024_characterising}, as well as $V = \mathrm{DFT}_d$ for $d$ prime \cite{Xu_2024}. On the other hand, $\mathcal{C}$ is non-empty for $V = \mathrm{DFT}_d$ when $d$ is composite. In many KD distributions, the set of all KD-positive states $\mathcal{E}_{\mathrm{KD}+}$ is known to be the convex hull of $\mathcal{E}^\mathrm{pure}_{\mathrm{KD}+}$. This is true for Haar-random $V$ \cite{langrenez2024}, as well as $V = \mathrm{DFT}_d$ when $d = p^k$ is a prime power \cite{debievre2025,Langrenez_2024_characterising, Yang_2024}. Whenever it holds true that $\mathcal{E}_{\mathrm{KD}+} = \mathrm{conv}(\mathcal{E}^\mathrm{pure}_{\mathrm{KD}+})$, positivity preservation for pure states is a sufficient condition for positivity preservation for all states. In other cases, whenever $\mathcal{E}_{\mathrm{KD}+} \neq \mathrm{conv}(\mathcal{E}^\mathrm{pure}_{\mathrm{KD}+})$ and `exotic' impure, KD-positive states exist (such as for $V = \mathrm{DFT}_6$ or $V = H^{\otimes n}$ \cite{debievre2025}) there remains a possibility that unitaries which permute elements of $\mathcal{E}^\mathrm{pure}_{\mathrm{KD}+}$ map impure, KD-positive states from $\mathcal{E}_{\mathrm{KD}+} \setminus \mathrm{conv}(\mathcal{E}^\mathrm{pure}_{\mathrm{KD}+})$ to non-positive states. For all of our identified types of positivity-preserving unitaries (Lemma \ref{lemma:nonpospresunitaries} and Theorem \ref{thm:ustar}) we find that this is not the case.

\textit{KD Distributions as Vectors}. We will make use of vectorisation, which is the process of representing quantum states $\rho \in \mathbb{C}^{d \times d}$ as vectors in $\mathbb{C}^{d^2}$, acted on by larger matrices in $\mathbb{C}^{d^2 \times d^2}$ known as superoperators. For KD distributions, this representation was explicitly studied in \cite{schmid2024kirkwooddirac}. In practice, it allows us to represent any evolution of KD distributions as a linear transformation of the corresponding vectorised form. We proceed by stacking the rows of a KD distribution matrix $Q(\rho)$ into a single column vector:

\begin{equation}
    \boldsymbol{|} Q(\rho) \boldsymbol{\rangle \rangle} = \begin{pmatrix}
    Q_{11} & Q_{12} & \dots & Q_{1d} & Q_{21} & \dots & Q_{dd}
    \end{pmatrix}^T.
\end{equation}

\noindent It is also useful to define the dual row vector $\boldsymbol{\langle \langle 1|} = (1, 1, \dots, 1)$, composed of all $1$'s. Clearly, all KD distribution vectors satisfy $\boldsymbol{\langle \langle 1 |} Q(\rho) \boldsymbol{\rangle \rangle} = 1$. The total non-positivity of a KD distribution can then be written as the $\ell_1$-norm of $\boldsymbol{|} Q(\rho) \boldsymbol{\rangle \rangle}$:

\begin{equation}
    \mathcal{N}(Q(\rho)) = \| \boldsymbol{|} Q(\rho) \boldsymbol{\rangle \rangle} \|_1.
\end{equation}

\textit{KD Superoperators}. In the context of classical simulation, we will consider the evolution of a KD distribution $Q(\rho) \rightarrow Q(\mathcal{E}(\rho))$ under a quantum channel (a completely positive, trace-preserving map) $\mathcal{E}$ acting on $\rho$. In the larger space $\mathbb{C}^{d^2 \times d^2}$, this may be written as the multiplication of $\boldsymbol{|}Q(\rho) \boldsymbol{\rangle \rangle}$ by a $d^2 \times d^2$ matrix $\hat{\mathcal{E}}$: 

\begin{equation}
    \boldsymbol{ |} Q(\mathcal{E}(\rho))  \boldsymbol{ \rangle \rangle} = \hat{\mathcal{E}} \boldsymbol{|} Q(\rho) \boldsymbol{\rangle \rangle}.
\end{equation}

\noindent If $\mathcal{E}$ is composed of a set of Kraus operators $\{ K_\mu \}$ (such that $\sum_\mu K^\dagger_\mu K_\mu = \mathbb{1}_d$), then the quasiprobabilities $Q_{ij}$ evolve as follows:

\begin{align}
    Q_{ij} & \rightarrow \sum_\mu \langle b_j | a_i \rangle \times \langle a_i | K_\mu \rho K_\mu^\dagger | b_j \rangle \nonumber \\
    & = \sum_\mu \sum_{k l} \langle b_j | a_i \rangle \times \langle a_i | K_\mu | a_k \rangle \langle a_k | \rho | b_l \rangle \langle b_l | K_\mu^\dagger | b_j \rangle \nonumber \\
    & = \sum_{k l} \left( \sum_\mu  \frac{\langle b_j | a_i \rangle}{\langle b_l | a_k \rangle} \langle a_i | K_\mu | a_k \rangle  \langle b_l | K_\mu^\dagger | b_j \rangle  \right) \times Q_{kl} \nonumber \\
    & = \sum_{k l} \hat{\mathcal{E}}_{ij, kl}  Q_{kl}.
\end{align}

\noindent This allows us to identify the matrix elements $\hat{\mathcal{E}}_{ij, kl}$:

\begin{equation}
    \hat{\mathcal{E}}_{ij, kl} = \sum_\mu  \frac{\langle b_j | a_i \rangle}{\langle b_l | a_k \rangle} \langle a_i | K_\mu | a_k \rangle  \langle b_l | K_\mu^\dagger | b_j \rangle.
\end{equation}

\noindent One may verify that $\hat{\mathcal{E}}$ preserves the element sum of KD quasiprobability vectors. This is because its columns sum to 1:

\begin{align}
    \sum_{ij} \hat{\mathcal{E}}_{ij, kl} 
    &= \frac{1}{\langle b_l | a_k \rangle} \sum_{ij} \sum_\mu \langle b_l | K_\mu^\dagger | b_j \rangle \langle b_j | a_i \rangle \langle a_i | K_\mu | a_k \rangle \nonumber \\
    &= \frac{1}{\langle b_l | a_k \rangle} \sum_\mu \langle b_l | K_\mu^\dagger K_\mu |a_k \rangle \nonumber \\
    &= \langle b_l |a_k \rangle / \langle b_l | a_k \rangle = 1,
    \end{align}

\noindent from which it follows that $\boldsymbol{\langle \langle} \mathbf{1} \boldsymbol{|} \hat{\mathcal{E}} = \boldsymbol{\langle \langle} \mathbf{1} \boldsymbol{|} $, hence $\boldsymbol{\langle \langle} \mathbf{1} \boldsymbol{|} Q(\mathcal{E}(\rho)) \boldsymbol{\rangle \rangle} = \boldsymbol{\langle \langle} \mathbf{1} \boldsymbol{|} \hat{\mathcal{E}} \boldsymbol{|} Q(\rho) \boldsymbol{\rangle \rangle} = 1$. In light of this, one may view the superoperators $\hat{\mathcal{E}}$ as left quasi-stochastic matrices acting on KD quasiprobability vectors (recall that a left stochastic matrix is a square matrix with non-negative entries whose columns sum to 1). In particular, when $\hat{\mathcal{E}}$ is stochastic, the triangle inequality implies it can only decrease the $\ell_1$-norm of the KD distribution vector as $\| \hat{\mathcal{E}} \boldsymbol{|}Q(\rho) \boldsymbol{\rangle \rangle} \|_1 \leq \|  \boldsymbol{|}Q(\rho) \boldsymbol{\rangle \rangle} \|_1 $. For KD-positive inputs $\boldsymbol{|} Q(\rho) \boldsymbol{\rangle \rangle}$ it follows that the output distribution is also KD-positive, in full analogy to the evolution of a probability vector by a Markov chain. 

\textit{Unitary Channels.} If $\mathcal{E}$ is an action of a unitary $U$, we write the corresponding superoperator as $\hat{\mathcal{E}}_U$. In this case, the matrix elements simplify to:

\begin{equation}
    (\hat{\mathcal{E}}_U)_{ij, kl} = \frac{\langle b_j | a_i \rangle}{\langle b_l | a_k \rangle} \langle a_i | U | a_k \rangle  \langle b_l | U^\dagger | b_j \rangle. \label{eq:unitarysuperopelement}
\end{equation}

\noindent Furthermore, each unitary-induced superoperator $\hat{\mathcal{E}}_U$ must have an inverse given by $\hat{\mathcal{E}}^{-1}_U = \hat{\mathcal{E}}_{U^\dagger}$, which follows from the composition rule:

\begin{align}
    \hat{\mathcal{E}}_{U^\dagger} \hat{\mathcal{E}}_U \boldsymbol{|} Q(\rho) \boldsymbol{\rangle \rangle} &= \hat{\mathcal{E}}_{U^\dagger} \boldsymbol{|} Q( U \rho U^\dagger) \boldsymbol{\rangle \rangle} \nonumber \\
    & = \boldsymbol{|} Q(U^\dagger U \rho U^\dagger U ) \boldsymbol{\rangle \rangle} \nonumber \\
    &= \boldsymbol{|} Q(\rho ) \boldsymbol{\rangle \rangle}. \label{eq:unitaryinverse}
\end{align}

\noindent We may relate the elements of $\hat{\mathcal{E}}^{-1}_U$ to those of $\hat{\mathcal{E}}_{U}$:

\begin{align}
    (\hat{\mathcal{E}}_{U}^{-1})_{ij, kl} &= \frac{\langle b_j | a_i \rangle}{\langle b_l | a_k \rangle} \langle a_i | U^\dagger | a_k \rangle  \langle b_l | U | b_j \rangle \nonumber \\
    &= \frac{\langle b_j | a_i \rangle}{\langle b_l | a_k \rangle} \left( \langle a_k | U | a_i \rangle  \langle b_j | U^\dagger | b_l \rangle \right)^* \nonumber \\
    &= \frac{\langle b_j | a_i \rangle}{\langle b_l | a_k \rangle} \times \left( \frac{\langle b_j | a_i \rangle}{\langle b_l | a_k \rangle} (\hat{\mathcal{E}}_{U})_{kl, ij} \right)^* \nonumber \\
    &= \left| \frac{\langle b_j | a_i \rangle}{\langle b_l | a_k \rangle} \right|^2 (\hat{\mathcal{E}}_{U})_{kl, ij}^*. \label{eq:invsuperopelement}
    \end{align}

\noindent Thus, if $\mathcal{A, B}$ are mutually unbiased, then $\hat{\mathcal{E}}_U$ is always a unitary matrix on $\mathbb{C}^{d^2 \times d^2}$. In this case, the rows of $\hat{\mathcal{E}}_{U} = \hat{\mathcal{E}}_{U^\dagger}^\dagger$ are the complex conjugated columns of $\hat{\mathcal{E}}_{U^\dagger}$ (which sum to $1$), and therefore $\hat{\mathcal{E}}_{U}$ is a doubly quasi-stochastic matrix.

\textit{POVM Elements}. Vectorisation provides a recipe for representing and evolving KD quasiprobabilities through matrix multiplication in $\mathbb{C}^{d^2}$. To make this complete, we also need a way to calculate measurement probabilities, i.e. quantities $\text{Tr}(F U \rho U^\dagger)$, where $F$ is some POVM element. This is very simple and involves dual vectors $\boldsymbol{\langle \langle} F \boldsymbol{|} \in (\mathbb{C}^{d^2})^*$ which map KD vectors to Born rule probabilities. The dual vectors $\boldsymbol{\langle \langle} F \boldsymbol{|}$ have elements:

\begin{equation}
    F_{ij} = \frac{\langle b_j | F | a_i \rangle}{\langle b_j | a_i \rangle},
\end{equation}

\noindent so that measurement probabilities are given by:

\begin{align}
    \text{Tr}(F U \rho U^\dagger) &= \boldsymbol{\langle \langle} F \boldsymbol{|} \hat{\mathcal{E}}_U \boldsymbol{|} Q(\rho) \boldsymbol{\rangle \rangle} \\
    & = \sum_{ij, kl} F_{ij} (\hat{\mathcal{E}}_U)_{ij, kl} Q_{kl}. \label{eq:kdexpsum}
\end{align}

\noindent If $F$ is a projective measurement in the basis $\mathcal{A}$ or $\mathcal{B}$ (for example $F = | a_k \rangle \langle a_k |$), then $\boldsymbol{\langle \langle} F \boldsymbol{|}$ has elements:

\begin{equation}
    F_{ij} = \frac{ \langle b_j | a_k \rangle \langle a_k | a_i \rangle}{\langle b_j | a_i \rangle} = \delta_{ik},
\end{equation}

\noindent and Equation \eqref{eq:kdexpsum} reduces to the marginalisation of the quasiprobability distribution:

\begin{equation}
    \text{Tr}(F \rho ) = \sum_{ij} \delta_{ik} Q_{ij} = \sum_{j} Q_{kj} = \langle a_k | \rho | a_k \rangle.
\end{equation}

\textit{Experimental Estimation of KD Superoperator Elements.} In Appendix \ref{appendix:cycletest}, we offer a straightforward method for estimating the elements $(\hat{\mathcal{E}}_U)_{ij, kl}$ of a KD superoperator induced by a unitary $U$ on any informationally complete choice of $\mathcal{A, B}$. To do this, one can apply the cycle test algorithm \cite{Wagner_2024, oszmaniec2021measuring}, originally proposed for estimating Kirkwood--Dirac quasiprobabilities $Q(\rho)_{ij}$ of an unknown state.

\section{Classification of KD Positivity-Preserving Unitaries}
\label{section:kdstochasticunitaries}

We now classify unitaries which preserve KD positivity and total non-positivity, as well as completely characterise KD-stochastic unitaries. In Section \ref{section:kdsimulation} we discuss the implications of these results for the classical simulability of KD-positive states. Our classification is complete for all distributions where $\mathcal{C} = \emptyset$, randomly sampled transition matrices, and Fourier-conjugate bases with $d = p^k, pq$ (with $p, q$ prime). Here and throughout we will assume informational completeness (i.e. $m_{\mathcal{A, B}} > 0$), stating otherwise where relevant. We leave other instances of KD distributions for future work.

\begin{definition}[Positivity Preservation] \label{def:kdpospreserving}
    A unitary $U \in \mathcal{U}_{\mathcal{N}+}$ is said to be positivity-preserving if and only if for any density matrix $\rho$ such that $\mathcal{N}(Q(\rho)) = 1$, it is true that $\mathcal{N}(Q(U \rho U^\dagger)) = 1$, i.e. $U \in \mathcal{U}_{\mathcal{N}+}$ maps KD-positive states to KD-positive states.
\end{definition}

\begin{definition}[$\mathcal{N}$-preservation]
    A unitary $U \in \mathcal{U}_{\mathcal{N}}$ is said to be $\mathcal{N}$-preserving if and only if for any density matrix $\rho$, $\mathcal{N}(Q(U \rho U^\dagger)) = \mathcal{N}(Q(\rho))$, i.e. $\mathcal{U}_{\mathcal{N}}$ preserves the KD total non-positivity of $\rho$. If $U$ is $\mathcal{N}$-preserving, then it is positivity-preserving.
\end{definition}

\begin{definition}[Kirkwood--Dirac Stochasticity]
    A unitary $U \in \mathcal{U}_\mathrm{stoch}$ is said to be KD-stochastic if and only if its induced superoperator $\hat{\mathcal{E}}_U$ is a stochastic matrix. If $U$ is KD-stochastic, then by Theorem \ref{theorem:genperm} it is $\mathcal{N}$-preserving.
\end{definition}

Depending on the geometry of $\mathcal{E}^\mathrm{pure}_{\mathrm{KD}+}$ the three notions may, or may not be synonymous. Although an arbitrary stochastic superoperator cannot increase the $\ell_1$-norm of a complex vector (and so is positivity-preserving), Theorem \ref{theorem:genperm} further implies that $\mathcal{U}_\mathrm{stoch}$ can only contain superoperators which are permutations, and thus are $\mathcal{N}$-preserving. We will identify examples where no such (non-trivial) unitaries exist, as well as examples where the sets are both non-trivial and distinct. In all cases, they satisfy $\mathcal{U}_\mathrm{stoch} \subseteq \mathcal{U}_{\mathcal{N}} \subseteq \mathcal{U}_{\mathcal{N}_+}$. 

\textit{Methodology.} A necessary condition for a unitary $U$ to be positivity-preserving for all states $\rho \in \mathcal{E}_{\mathrm{KD}+}$ is that it is positivity-preserving on pure states $|\psi\rangle \in \mathcal{E}^\mathrm{pure}_{\mathrm{KD}+}$. Therefore, our approach will be to identify all possible families of unitaries satisfying this necessary condition, and (where appropriate) verify that they are positivity-preserving for $\rho \in \mathcal{E}_{\mathrm{KD}+} \setminus \mathrm{conv}(\mathcal{E}^\mathrm{pure}_{\mathrm{KD}+})$. We assume that the set of pure KD-positive states is finite -- to the best of our knowledge this is true for all informationally complete KD distributions. Then, any positivity-preserving unitary must form a generalised permutation of the set $\mathcal{A} \cup \mathcal{B} \cup \mathcal{C}$ of pure KD-positive states. 

\begin{definition}[Generalised Permutations]
    The set $M_d$ of unitary generalised permutations (or monomial unitary matrices) is defined as:
\begin{equation}
    M_d = \{ D P_\sigma \ : \ D = \mathrm{diag}(e^{i \theta_1}, \dots, e^{i \theta_d}), \ \sigma \in S_d \}.
\end{equation}
\end{definition}

\noindent We begin with the simplest case where $\mathcal{C} = \emptyset$, and the only pure KD-positive states are the basis vectors from $\mathcal{A}$ and $\mathcal{B}$. For such distributions, we find that $\mathcal{U}_{\mathcal{N}}$ can only consist of two types of generalised permutations, which we call type \textbf{I} and \textbf{II}. Such unitaries are then shown to be $\mathcal{N}$-preserving, and so preserve positivity on any $\rho \in \mathcal{E}_{\mathrm{KD}+}$ regardless of the existence of exotic, impure KD-positive states. Thus, for any distribution where $\mathcal{C} = \emptyset$ our classification of $\mathcal{U}_{\mathcal{N}} = \mathcal{U}_{\mathcal{N}+}$ into types \textbf{I, II} is exhaustive. 

\begin{lemma} \label{lemma:pospresunitaries}
If $\mathcal{C} = \emptyset$, i.e. $\mathcal{E}^\mathrm{pure}_{\mathrm{KD}+} = \mathcal{A} \cup \mathcal{B}$, then any positivity-preserving unitary $U$ either satisfies $U(\mathcal{A}) = \mathcal{A}$ and $U(\mathcal{B}) = \mathcal{B}$ (type \textbf{I}), or $U(\mathcal{A}) = \mathcal{B}$ and $U(\mathcal{B}) = \mathcal{A}$ (type \textbf{II}). The set of all such unitaries is given by:

\begin{equation}
   \mathcal{U}_{\mathcal{N}_+}=  \underbrace{\left( M_d \cap V M_d V^\dagger \right)}_{{\mathrm{Type}  \ \mathbf{I}}} \cup \underbrace{\left( VM_d\cap M_dV^\dagger \right)}_{{\mathrm{Type} \ \mathbf{II}}}
\end{equation}

\end{lemma}

\begin{proof}
    Since the only pure KD-positive states are the basis states, any positivity-preserving map $U$ must be a generalised permutation of the vectors in $\mathcal{A} \cup \mathcal{B}$ up to phase. Prima facie, there are three possibilities: type \textbf{I}, where $U$ is a generalised permutation on $\mathcal{A}$ and $\mathcal{B}$, in which case $U \in M_d$ is a (possibly distinct) monomial unitary on both bases, hence $U \in M_d \cap V M_d V^\dagger$. For type \textbf{II}, $U$ is a bijection between $\mathcal{A}$ and $\mathcal{B}$, in which case $U$ can be written as $U = V D P_\sigma$ on the $\mathcal{A}$ basis, and as $U = D' P_{\sigma'} V^\dagger$ on the $\mathcal{B}$ basis, hence $U \in VM_d\cap M_dV^\dagger$. The third possibility is that $U$ maps some vectors from $\mathcal{A}$ to $\mathcal{A}$, and some to $\mathcal{B}$ (and similarly for $\mathcal{B}$). This is not possible. Suppose $U|a_i\rangle = e^{i \theta_i} |a_{i'} \rangle$ and $U|a_j\rangle = e^{i \theta_j} |b_{j'} \rangle$ for some $i \neq j$. This would imply that:

    \begin{align}
        \langle a_i | U^\dagger U | a_j \rangle &= e^{i (\theta_j - \theta_i)} \langle a_{i'}| b_{j'} \rangle \\ &= e^{i (\theta_j - \theta_i)} \langle a_{i'}| V | a_{j'} \rangle \neq 0,
    \end{align} 

    \noindent which is a contradiction, since $U$ is unitary and $\mathcal{A}$ is an orthonormal basis. Thus, positivity-preserving unitaries must be of the type \textbf{I} or \textbf{II}.  
\end{proof}

\begin{lemma} \label{lemma:nonpospresunitaries}
Positivity-preserving unitaries of type \textbf{I} or \textbf{II} are also $\mathcal{N}$-preserving. In other words, if $\mathcal{U}_{\mathcal{N}_+} = \bigl(M_d\cap V M_d V^\dagger\bigr)\;\cup\;\bigl(VM_d\cap M_dV^\dagger\bigr)$,
then $\mathcal{U}_{\mathcal{N}_+} = \mathcal{U}_{\mathcal{N}}$.
\end{lemma}
\begin{proof}
    Clearly, by definition $\mathcal{U}_{\mathcal{N}} \subseteq \mathcal{U}_{\mathcal{N}_+}$. We will show that $\mathcal{U}_{\mathcal{N}_+} \subseteq \mathcal{U}_{\mathcal{N}}$. Suppose $U \in \mathcal{U}_{\mathcal{N}_+}$. We can consider types \textbf{I} and \textbf{II} separately. In the first case, $U$ acts as:

    \begin{align}
    U|a_i\rangle &= e^{i \theta_i} |a_{\sigma(i)} \rangle, & U|b_j\rangle &= e^{i \phi_j} |b_{\tau(j)} \rangle.
    \end{align}

    \noindent Hence,
    \begin{align}
        U^\dagger |a_i\rangle \langle a_i | U &= |a_{\sigma^{-1}(i)} \rangle \langle a_{\sigma^{-1}(i)} |, \nonumber \\
        U^\dagger |b_j\rangle \langle b_j | U &= |b_{\tau^{-1}(j)} \rangle \langle b_{\tau^{-1}(j)} |.
    \end{align}
    
\noindent Hence, $\rho \rightarrow U \rho U^\dagger$ induces the following transformation on the KD distribution:

\begin{align}
    Q_{ij}(U \rho U^\dagger) &= \langle b_j | U U^\dagger |a_i \rangle \langle a_i | U \rho U^\dagger | b_j \rangle \nonumber \\
    & = \langle b_{\tau^{-1}(j)} |a_{\sigma^{-1}(i)} \rangle \langle a_{\sigma^{-1}(i)} | \rho | b_{\tau^{-1}(j)} \rangle \nonumber \\
    & = Q_{\sigma^{-1}(i), \tau^{-1}(j)}(\rho).
\end{align}

\noindent This implies that $\mathcal{N}(Q(U \rho U^\dagger)) = \mathcal{N}(Q(\rho))$ for all $\rho$. For unitaries of type \textbf{II}, we may write:

\begin{align}
    U|a_i\rangle &= e^{i \theta_i} |b_{\sigma(i)} \rangle, & U|b_j\rangle &= e^{i \phi_j} |a_{\tau(j)} \rangle.
    \end{align}

    \noindent Hence,
    \begin{align}
        U^\dagger |a_i\rangle \langle a_i | U &= |b_{\tau^{-1}(i)} \rangle \langle b_{\tau^{-1}(i)} |, \nonumber \\
        U^\dagger |b_j\rangle \langle b_j | U &= |a_{\sigma^{-1}(j)} \rangle \langle a_{\sigma^{-1}(j)} |,
    \end{align}

\noindent and so:
\begin{align}
    Q_{ij}(U \rho U^\dagger) &= \langle b_j | U U^\dagger |a_i \rangle \langle a_i | U \rho U^\dagger | b_j \rangle \nonumber \\
    & = \langle a_{\sigma^{-1}(j)} |b_{\tau^{-1}(i)} \rangle \langle b_{\tau^{-1}(i)} | \rho | a_{\sigma^{-1}(j)} \rangle \nonumber \\
    & = Q^*_{\sigma^{-1}(j), \tau^{-1}(i)}(\rho).
\end{align}

\noindent This again implies that the total non-positivity is preserved for any state. Thus, if $U \in \mathcal{U}_{\mathcal{N}_+}$ is of type \textbf{I} or \textbf{II}, it also preserves the total non-positivity of any KD distribution. If these are the only types of unitaries in $\mathcal{U}_{\mathcal{N}_+}$, then any $\mathcal{N}$-preserving unitary must also be positivity-preserving, and so $\mathcal{U}_{\mathcal{N}_+} = \mathcal{U}_{\mathcal{N}}$.

\end{proof}

\begin{remark}
    If $U \in \mathcal{U}_{\mathcal{N}+}$ is a type \textbf{I} or \textbf{II} unitary, then it must also form a generalised permutation on $\mathcal{C}$, and be $\mathcal{N}$-preserving for all $\rho \in \mathcal{E}_{\mathrm{KD}+}$. 
\end{remark}
\begin{proof}
    We know from Lemma \ref{lemma:nonpospresunitaries} that if $U$ is of type \textbf{I} or \textbf{II}, then $U \in \mathcal{U}_{\mathcal{N}}$ and thus maps all KD-positive states to KD-positive states. The only way for this to be true is if $U$ also forms a generalised permutation on $\mathcal{C}$. Since type \textbf{I} and \textbf{II} unitaries always permute entries of the KD distribution, they also preserve the total non-positivity of any impure state. 
\end{proof}

When $\mathcal{C} = \emptyset$, any positivity-preserving unitary is also $\mathcal{N}$-preserving. Furthermore, the action of any type \textbf{I} or \textbf{II} unitary on KD distributions is said to be covariant, meaning that it corresponds to a relabelling and conjugation of the matrix elements $Q_{ij}$. In particular, in Section \ref{section:kdsimulation} we show this implies that the evolution of positive KD distributions under any quantum circuit composed of such unitaries can be classically efficiently simulated. 

\textit{Fourier-conjugate Bases.} For a variety of interesting choices of transition matrices it holds that $\mathcal{C} \neq \emptyset$, including in the case of the discrete Fourier transform. We now prove that when $V = \mathrm{DFT}_d$ in dimension $d = p^k$ with $p$ prime, the structure of the set $\mathcal{C}$ of non-basis pure KD-positive states still forces positivity-preserving unitaries to only be of type \textbf{I} or \textbf{II}. As such unitaries act covariantly, this will imply that any positivity-preserving unitary can be efficiently simulated.

\begin{theorem} \label{thm:primefourier}
    For $V = \mathrm{DFT}_d$ in dimension $d = p^k$ with $p$ prime, $\mathcal{U}_{\mathcal{N}_+}$ consists of type \textbf{I} and \textbf{II} unitaries only. Consequently, $\mathcal{U}_{\mathcal{N}_+} = \mathcal{U}_{\mathcal{N}}$.
\end{theorem}
\begin{proof}
If $k=1$, then $\mathcal{C} = \emptyset$ and the result is immediate from Lemma \ref{lemma:pospresunitaries}. For $k > 1$, it has been shown that the set $\mathcal{C}$ of non-basis pure KD-positive states decomposes as follows \cite{Xu_2024, De_Bievre_2021}:

\begin{equation}
    \mathcal{C} = \bigcup_{uv = d} \mathcal{C}_{u, v},
\end{equation}

\noindent where each $\mathcal{C}_{u, v} = \{ |m, s \rangle_{u, v} \}$ is an orthonormal basis on $\mathbb{C}^d$ defined as:

\begin{equation}
    |m, s \rangle_{u, v} = \frac{1}{\sqrt{v}} \sum_{k = 0}^{v-1} e^{\frac{2 \pi i}{v} sk} | a_{ku + m} \rangle, \quad m \in \mathbb{Z}_u, s \in \mathbb{Z}_v .
\end{equation}

\noindent We will show that despite $\mathcal{C}$ being non-empty for $d = p^k$, any generalised permutation on the set of KD-positive states $\mathcal{A} \cup \mathcal{B} \cup \mathcal{C}$ must be of either type \textbf{I} or \textbf{II}. First, we calculate all possible inner products between $|m, s \rangle_{u, v}$ and $| x, y \rangle_{a, b}$:

\begin{equation}
\langle m, s | x, y \rangle = \frac{1}{\sqrt{vb}} \sum_{k=0}^{v-1}\sum_{l=0}^{b-1} e^{\frac{-2 \pi i}{v} sk} e^{\frac{2 \pi i}{b} yl} \delta_{ku+m,\ l a+x}.
\end{equation}

\noindent From the delta function, the double sum only receives contributions from pairs $(k, l)$ satisfying $ku - l a = x - m$. This is a linear Diophantine equation, whose solutions exist if and only if $g | (x - m) $, where $g = \gcd(u, a)$. We therefore assume $x = m \bmod g$. We pick a solution $(k_0, l_0)$. Then, $(k, l) = (k_0 + ra/g, l_0 + ru/g)$ also form solutions. Since $u/g$ and $a/g$ are coprime, we have $uv/g = ab/g$, which implies that $v = ha/g$ and $b = hu/g$ where $h = \gcd(v, b)$. Since $a/g = v/h$ and $u/g = b/h$, there are $h = \gcd(v, b)$ unique solutions in the allowed ranges. Hence, the inner product can be rewritten as:

\begin{align}
\langle m,s | x,y \rangle &= \frac{1}{\sqrt{vb}} \sum_{r=0}^{h-1} e^{\frac{-2\pi i}{v} s(k_0 +r \frac{a}{g})}\, e^{\frac{2\pi i}{b} y(l_0+r \frac{u}{g})} \\
&=
\frac{e^{2\pi i \left( \frac{y l_0 }{b} -  \frac{s k_0 }{v} \right)}}{\sqrt{vb}}
\sum_{r=0}^{h-1}
e^{2\pi i r\left(\frac{yu}{gb}-\frac{sa}{gv} \right)}.
\end{align}

\noindent From $uv = ab$, we have $\frac{yu}{gb}-\frac{sa}{gv} = \frac{a}{gv}(y - s)$, where $a / gv = 1/h$. Hence, the inner product is given by:

\begin{align}
\langle m,s | x,y \rangle &=
\frac{e^{2\pi i \left( \frac{y l_0 }{b} -  \frac{s k_0 }{v} \right)}}{\sqrt{vb}}
\sum_{r=0}^{h-1}
e^{\frac{2\pi i}{h} r (y - s)} \label{eq:dft_inner_product}\\
&= \begin{cases} 
\frac{h}{\sqrt{vb}} e^{2\pi i \left( \frac{y l_0 }{b} -  \frac{s k_0 }{v} \right)}, & x = m \bmod g, \\
& y = s \bmod h. \\ 0, & \mathrm{otherwise.} \end{cases}
\end{align}

\noindent Now, suppose that $d = p^k$. For two KD-positive bases $\mathcal{C}_{u, v}$ and $\mathcal{C}_{a, b}$  with $ab = uv = d$, the non-zero overlaps satisfy $|\langle m,s | x,y \rangle|^2 = \gcd(b, v)^2 / bv$. Since every divisor of $d$ is now a power of $p$, we may write $b = p^r$ and $v = p^s$, where $r, s \in \{0, \dots, k\}$. Then, $|\langle m,s | x,y \rangle|^2  = 1/p^{|r - s|}$. For mutually unbiased bases, we require $|r - s| = k$, which can only occur when $(r,s ) = (k, 0)$ or $(r, s) = (0, k)$. This implies that $\mathcal{A, B}$ are the only mutually unbiased pair of KD-positive bases and that for any state $|c_i\rangle \in \mathcal{C}$ and any other KD-positive state $|\phi \rangle \in \mathcal{E}_\mathrm{KD+}^\mathrm{pure}$, $|\langle c_i| \phi \rangle|^2 \neq 1/d$. Since $U$ preserves inner products, its images $U(\mathcal{A})$ and $U(\mathcal{B})$ must also form a MUB pair. Thus, if $U$ were to map any element of $\mathcal{A}$ to an element of $\mathcal{C}$, then $|\langle b_j | U^\dagger U|a_i \rangle|^2 = |\langle \phi |c_i \rangle|^2 \neq 1/d$, violating the unitarity of $U$. This narrows down $U$ to be of type \textbf{I} or \textbf{II} only. From Lemma \ref{lemma:nonpospresunitaries}, it follows that $\mathcal{U}_{\mathcal{N}_+} = \mathcal{U}_{\mathcal{N}}$.
\end{proof}

Interestingly, Theorem \ref{thm:primefourier} does not apply for composite dimensions $d$, where additional mutually unbiased pairs $\mathcal{C}_{u, v}$ and $\mathcal{C}_{a, b}$ can exist. Consequently, we can identify unitaries which permute $\mathcal{A} \cup \mathcal{B} \cup \mathcal{C}$ but are not of type \textbf{I} or \textbf{II}. One class of such unitaries turns out to belong in $\mathcal{U}_{\mathcal{N}_+} \setminus \mathcal{U}_{\mathcal{N}}$, meaning they are positivity-preserving but not $\mathcal{N}$-preserving on pure states. To demonstrate this, take $d=6$. From the above decomposition of $\mathcal{C}$, the full set of pure, KD-positive states is given by $\mathcal{E}^\mathrm{pure}_{\mathrm{KD}+} = \mathcal{A} \cup \mathcal{B} \cup \mathcal{C}_{2, 3} \cup \mathcal{C}_{3, 2}$. The transition matrix is $V = \mathrm{DFT}_6$, with elements $V_{ij} = \omega^{ij}/\sqrt{6}$, where $\omega = e^{\pi i / 3}$. Consider the unitary $U_\star$ defined by:

\begin{equation}
U_\star = \frac{1}{\sqrt{3}} \begin{pmatrix} 1 & 0 & 1 & 0 & 1 & 0 \\ 0 & \omega^2 & 0 & 1 & 0 & \omega^4 \\ 1 & 0 & \omega^2 & 0 & \omega^4 & 0 \\ 0 & 1 & 0 & 1 & 0 & 1 \\ 1 & 0 & \omega^4 & 0 & \omega^2 & 0 \\ 0 & \omega^4 & 0 & 1 & 0 & \omega^2 \end{pmatrix}.
\end{equation}

\noindent When applied to $\mathcal{B}$, its image (in the $\mathcal{A}$ basis) is given by the rows of $U_\star V$:

\begin{equation}
U_\star V = \frac{1}{\sqrt{2}} \begin{pmatrix} 1 & 0 & 0 & 1 & 0 & 0 \\ 0 & -1 & 0 & 0 & 1 & 0 \\ 0 & 0 & 1 & 0 & 0 & 1 \\ 1 & 0 & 0 & -1 & 0 & 0 \\ 0 & 1 & 0 & 0 & 1 & 0 \\ 0 & 0 & 1 & 0 & 0 & -1 \end{pmatrix}.
\end{equation} 

\noindent Direct calculation thus shows that $U_\star(\mathcal{A}) = \mathcal{C}_{2, 3}$, $U_\star(\mathcal{B}) = \mathcal{C}_{3, 2}$, $U_\star(\mathcal{C}_{2, 3}) = \mathcal{A}$ and $U_\star(\mathcal{C}_{3, 2}) = \mathcal{B}$. Hence, $U_\star$ is a generalised permutation on the set of pure KD-positive states. However, it is not of type \textbf{I} or \textbf{II}. In addition, its action on the KD distribution does not preserve the total non-positivity. To see this, consider the state $| \psi \rangle = (|a_1\rangle + |a_2\rangle + |a_3\rangle)/\sqrt{3}$, which is non-positive with $\mathcal{N}(Q(|\psi\rangle)) = 4/3$. The state $U_\star|\psi\rangle$ is also non-positive, but with $\mathcal{N}(Q(U_\star|\psi\rangle)) = 7/3$. Hence, $U_\star$ does not preserve the total non-positivity of the KD distribution, and so $U_\star \notin \mathcal{U}_{\mathcal{N}}$. 

For this choice of KD distribution it has been shown that positive, mixed states outside $\mathrm{conv}(\mathcal{E}^\mathrm{pure}_{\mathrm{KD}+})$ exist \cite{debievre2025}. We now show that $U_\star$ preserves positivity for \textit{all} states in $\mathcal{E}_{\mathrm{KD}+}$, including ones outside the convex hull of $\mathcal{E}^\mathrm{pure}_{\mathrm{KD}+}$, and so $U_\star \in \mathcal{U}_{\mathcal{N}+}$.

\begin{theorem} \label{thm:ustar}
    For $V = \mathrm{DFT}_{pq}$, the type \textbf{III} unitary $U_\star$ maps all KD-positive states to KD-positive states. 
\end{theorem}
\begin{proof}
    Here we prove the $d=6$ case for our given example of $U_\star$ explicitly. For any $d=pq$, the unitary $U_\star$ can be shown to always exist (we show this next, in Lemma \ref{cor:compositefourier}), and the proof that this choice of $U_\star$ is $\mathcal{N}$-preserving for all KD-real states is given in Appendix \ref{app:type3proofs}. We will label states from $\mathcal{C}_{2, 3}$ as $|c_i \rangle = V_c |a_i\rangle$ and states from $\mathcal{C}_{3, 2}$ as $|d_i \rangle = V_d|a_i\rangle$, where:
    \begin{align}
        V_c &= \frac{1}{\sqrt{3}} \begin{pmatrix} 1 & 1 & 1 & 0 & 0 & 0 \\ 0 & 0 & 0 & 1 & 1 & 1 \\
        1 & \omega^2 & \omega^4 & 0 & 0 & 0 \\
        0 & 0 & 0 & 1 & \omega^2 & \omega^4 \\
        1 & \omega^4 & \omega^2 & 0 & 0 & 0 \\
        0 & 0 & 0 & 1 & \omega^4 & \omega^2 
        \end{pmatrix}, \\
        V_d &= \frac{1}{\sqrt{2}} \begin{pmatrix} 1 & 1 & 0 & 0 & 0 & 0 \\ 0 & 0 & 1 & 1 & 0 & 0 \\
        0 & 0 & 0 & 0 & 1 & 1 \\
        1 & -1 & 0 & 0 & 0 & 0 \\
        0 & 0 & 1 & -1 & 0 & 0 \\
        0 & 0 & 0 & 0 & 1 & -1
        \end{pmatrix}.
    \end{align}
    
    We make use of a recent result of \cite{Xu_2025}, which shows that any KD-real state $\rho$ (including $\rho \in \mathcal{E}_{\mathrm{KD}+}$) can be decomposed as a linear combination of pure states from $\mathcal{E}^\mathrm{pure}_{\mathrm{KD}+}$, with real coefficients:

\begin{equation} \label{eq:lineardecomposition}
    \rho = \sum_{i=1}^6 \left( \alpha_i |a_i \rangle \langle a_i| + \beta_i |b_i \rangle \langle b_i| + \gamma_i |c_i \rangle \langle c_i| +  \varepsilon_i |d_i \rangle \langle d_i| \right),
\end{equation}

\noindent where the (real) coefficients sum as:

\begin{equation} \label{eq:coefficientsum}
\sum_{i=1}^6 (\alpha_i + \beta_i + \gamma_i + \varepsilon_i) = 1.
\end{equation} 

\noindent By linearity (c.f. Equation \eqref{eq:kddef}), we may write any $Q(\rho)$ as a linear combination of KD distributions of pure states. Explicitly, we get $Q(\rho) = \sum_{i} (\alpha_i Q(|a_i\rangle ) + \beta_i Q(|b_i\rangle) + \gamma_i Q(|c_i\rangle) + \varepsilon_i Q(|d_i\rangle))$. Even more explicitly, we may write $Q(\rho)$ as a sum of four matrices:

\begin{align}
    &\frac{1}{6} \begin{pmatrix}
    \alpha_1 & \alpha_1 & \alpha_1 & \alpha_1 & \alpha_1 & \alpha_1 \\
    \alpha_2 & \alpha_2 & \alpha_2 & \alpha_2 & \alpha_2 & \alpha_2 \\
    \alpha_3 & \alpha_3 & \alpha_3 & \alpha_3 & \alpha_3 & \alpha_3 \\
    \alpha_4 & \alpha_4 & \alpha_4 & \alpha_4 & \alpha_4 & \alpha_4 \\
    \alpha_5 & \alpha_5 & \alpha_5 & \alpha_5 & \alpha_5 & \alpha_5 \\
    \alpha_6 & \alpha_6 & \alpha_6 & \alpha_6 & \alpha_6 & \alpha_6
    \end{pmatrix} + \frac{1}{6}
    \begin{pmatrix}
    \beta_1 & \beta_2 & \beta_3 & \beta_4 & \beta_5 & \beta_6 \\
    \beta_1 & \beta_2 & \beta_3 & \beta_4 & \beta_5 & \beta_6 \\
    \beta_1 & \beta_2 & \beta_3 & \beta_4 & \beta_5 & \beta_6 \\
    \beta_1 & \beta_2 & \beta_3 & \beta_4 & \beta_5 & \beta_6 \\
    \beta_1 & \beta_2 & \beta_3 & \beta_4 & \beta_5 & \beta_6 \\
    \beta_1 & \beta_2 & \beta_3 & \beta_4 & \beta_5 & \beta_6 
    \end{pmatrix} \\
    &+ \frac{1}{6}
    \begin{pmatrix}
    \gamma_1 & \gamma_2 & \gamma_3 & \gamma_1 & \gamma_2 & \gamma_3 \\
    \gamma_4 & \gamma_5 & \gamma_6 & \gamma_4 & \gamma_5 & \gamma_6 \\
    \gamma_1 & \gamma_2 & \gamma_3 & \gamma_1 & \gamma_2 & \gamma_3 \\
    \gamma_4 & \gamma_5 & \gamma_6 & \gamma_4 & \gamma_5 & \gamma_6 \\
    \gamma_1 & \gamma_2 & \gamma_3 & \gamma_1 & \gamma_2 & \gamma_3 \\
    \gamma_4 & \gamma_5 & \gamma_6 & \gamma_4 & \gamma_5 & \gamma_6 
    \end{pmatrix} + \frac{1}{6}
    \begin{pmatrix}
    \varepsilon_1 & \varepsilon_2 & \varepsilon_1 & \varepsilon_2 & \varepsilon_1 & \varepsilon_2 \\
    \varepsilon_3 & \varepsilon_4 & \varepsilon_3 & \varepsilon_4 & \varepsilon_3 & \varepsilon_4 \\
    \varepsilon_5 & \varepsilon_6 & \varepsilon_5 & \varepsilon_6 & \varepsilon_5 & \varepsilon_6 \\
    \varepsilon_1 & \varepsilon_2 & \varepsilon_1 & \varepsilon_2 & \varepsilon_1 & \varepsilon_2 \\
    \varepsilon_3 & \varepsilon_4 & \varepsilon_3 & \varepsilon_4 & \varepsilon_3 & \varepsilon_4 \\
    \varepsilon_5 & \varepsilon_6 & \varepsilon_5 & \varepsilon_6 & \varepsilon_5 & \varepsilon_6
    \end{pmatrix}.
\end{align}

\noindent Each entry $Q(\rho)_{ij}$ is one of $36$ linear combinations of the coefficients $\alpha_x, \beta_y, \gamma_z, \varepsilon_w$ for some $x, y, z, w \in \llbracket 6 \rrbracket$. Assume that $\rho$ is an impure, KD-positive state not in $\mathrm{conv}(\mathcal{E}^\mathrm{pure}_{\mathrm{KD}+})$. Although the individual coefficients in Equation \eqref{eq:lineardecomposition} are not necessarily positive, from the KD positivity of $\rho$ it follows that each entry combination corresponding to a quasiprobability satisfies $Q(\rho)_{ij} \geq 0$. Now, recall that $U_\star$ is a generalised permutation on the set of pure KD-positive states. We can directly compute $U_\star \rho U_\star^\dagger$ by conjugating each pure state in the decomposition of \eqref{eq:lineardecomposition}, and then taking the sum. This gives us a new linear combination of pure state KD distributions. Thus, $Q(U_\star \rho U_\star^\dagger)$ is given by:

\begin{align} & \frac{1}{6}
     \begin{pmatrix}
     \alpha_1 & \alpha_3 & \alpha_5 & \alpha_1 &\alpha_3   & \alpha_5 \\
     \alpha_4 & \alpha_6 & \alpha_2 & \alpha_4 & \alpha_6 & \alpha_2 \\
     \alpha_1 & \alpha_3 & \alpha_5 & \alpha_1 &\alpha_3   & \alpha_5 \\
     \alpha_4 & \alpha_6 & \alpha_2 & \alpha_4 & \alpha_6 & \alpha_2 \\
     \alpha_1 & \alpha_3 & \alpha_5 & \alpha_1 &\alpha_3   & \alpha_5 \\
     \alpha_4 & \alpha_6 & \alpha_2 & \alpha_4 & \alpha_6 & \alpha_2 
    \end{pmatrix} +\frac{1}{6}
    \begin{pmatrix}
     \beta_1 & \beta_4 & \beta_1 & \beta_4 & \beta_1 & \beta_4 \\
     \beta_5 & \beta_2 & \beta_5 & \beta_2 & \beta_5 & \beta_2 \\
     \beta_3 & \beta_6 & \beta_3 & \beta_6  & \beta_3 & \beta_6  \\
     \beta_1 & \beta_4 & \beta_1 & \beta_4 & \beta_1 & \beta_4 \\
     \beta_5 & \beta_2 & \beta_5 & \beta_2 & \beta_5 & \beta_2 \\
     \beta_3 & \beta_6  & \beta_3 & \beta_6  & \beta_3 & \beta_6  
    \end{pmatrix} \\
    &+ \frac{1}{6}
    \begin{pmatrix}
     \gamma_1 & \gamma_1 & \gamma_1 & \gamma_1 & \gamma_1 & \gamma_1 \\
     \gamma_5 & \gamma_5 & \gamma_5 & \gamma_5 & \gamma_5 & \gamma_5 \\
     \gamma_3 & \gamma_3 & \gamma_3 & \gamma_3 & \gamma_3 & \gamma_3 \\
     \gamma_4 & \gamma_4 & \gamma_4 & \gamma_4 & \gamma_4 & \gamma_4 \\
     \gamma_2 & \gamma_2 & \gamma_2 & \gamma_2 & \gamma_2 & \gamma_2 \\
     \gamma_6 & \gamma_6 & \gamma_6 & \gamma_6 & \gamma_6 & \gamma_6 
    \end{pmatrix} + \frac{1}{6}
    \begin{pmatrix}
     \varepsilon_1 & \varepsilon_6 & \varepsilon_3 & \varepsilon_2 & \varepsilon_5 & \varepsilon_4 \\
     \varepsilon_1 & \varepsilon_6 & \varepsilon_3 & \varepsilon_2 & \varepsilon_5 & \varepsilon_4 \\
     \varepsilon_1 & \varepsilon_6 & \varepsilon_3 & \varepsilon_2 & \varepsilon_5 & \varepsilon_4 \\
     \varepsilon_1 & \varepsilon_6 & \varepsilon_3 & \varepsilon_2 & \varepsilon_5 & \varepsilon_4  \\
     \varepsilon_1 & \varepsilon_6 & \varepsilon_3 & \varepsilon_2 & \varepsilon_5 & \varepsilon_4  \\
     \varepsilon_1 & \varepsilon_6  & \varepsilon_3  &\varepsilon_2  &\varepsilon_5  &\varepsilon_4  
    \end{pmatrix}.
\end{align}

\noindent Direct observation shows that of the $36$ possible combinations of coefficients $\alpha_x, \beta_y, \gamma_z, \varepsilon_w$ that appear in the entries of $Q(\rho)$, each appears exactly once in $Q(U_\star \rho U_\star^\dagger)$. Hence, each entry of $Q(U_\star \rho U_\star^\dagger)$ is a permuted entry of $Q(\rho)$. Therefore, $U_\star$ maps $\rho$ to a KD-positive state. Since $\rho$ was an arbitrary KD-positive state, we conclude that $U_\star$ maps all KD-positive states to KD-positive states. We additionally conclude that $U_\star$ is covariant on all KD-real states, satisfying $Q(U_\star \rho U_\star^\dagger)_{ij} = Q(\rho)_{\gamma^{-1}(ij)}$ for some permutation $\gamma^{-1}$ of the indices. This permutation can be explicitly read off by comparing the above expressions for $Q(\rho)$ and $Q(U_\star \rho U_\star^\dagger)$. Notably, it is not a product of row and column permutations. A \texttt{Mathematica} notebook containing our calculations is provided in the Supplemental Material \cite{SupplementalMaterial}.
\end{proof}

Next, we prove that $U_\star$ always exists for $V = \mathrm{DFT}_{pq}$ with $p, q$ being distinct primes:

\begin{lemma} \label{cor:compositefourier}
For $V = \mathrm{DFT}_d$ in dimension $d = pq$ with $p, q$ distinct primes, there always exist type \textbf{III} unitaries $U_\star \in \mathcal{U}_{\mathcal{N}_+}$, such that $U_\star(\mathcal{A}) = \mathcal{C}_{p, q}$, $U_\star(\mathcal{C}_{p, q}) = \mathcal{A}$, $U_\star(\mathcal{B}) = \mathcal{C}_{q, p}$ and $U_\star(\mathcal{C}_{q, p}) = \mathcal{B}$. Furthermore, action by $U_\star$ is not covariant on the KD distribution in general, and types \textbf{I}, \textbf{II} and \textbf{III} exhaust the set $\mathcal{U}_{\mathcal{N}_+}$.
\end{lemma}

\begin{proof}
In this case, $\mathcal{E}_{\mathrm{KD}+}^\mathrm{pure} = \mathcal{A} \cup \mathcal{B} \cup \mathcal{C}_{p, q} \cup \mathcal{C}_{q, p}$. Taking $(a,b)=(q,p)$ and $(u,v)=(p,q)$ in Equation \eqref{eq:dft_inner_product}, we have that $|\langle m, s | x, y \rangle|^2 = \gcd(q, p)^2 / qp = 1/d$. Hence, the bases $\mathcal{C}_{p, q}$ and $\mathcal{C}_{q, p}$ are mutually unbiased. Since $p$ and $q$ are coprime, we can associate each index $i \in \mathbb{Z}_d$ with a pair of indices $(i_p, i_q) \in \mathbb{Z}_p \times \mathbb{Z}_q$ as follows:

\begin{align}
(i_p,i_q) & =(i \mod p, \ i \mod q), \label{eq:CRT} \\
i & = i_p \, q \bar{q} + i_q \, p \bar{p} \mod{pq},
\end{align}

\noindent where $\bar{p} = p^{-1} \bmod q$ and $\bar{q} = q^{-1} \bmod p$. This mapping is bijective by the Chinese remainder theorem. We can therefore define a unitary permutation $P$, which decomposes $\mathbb{C}^d$ into a tensor product $\mathbb{C}^p \otimes \mathbb{C}^q$:

\begin{align}
P|a_i\rangle &= |i_p \rangle \otimes |i_q\rangle, & P^\dagger (|i_p \rangle \otimes |i_q\rangle) &= |a_{i_p \, q \bar q + i_q \, p \bar p} \rangle,
\end{align}

\noindent where $|i_p\rangle \in \mathbb{C}^p$ and $|i_q\rangle \in \mathbb{C}^q$. We can define $U_\star$ as follows:

\begin{equation}
U_\star = P^\dagger (\mathbb{1}_p \otimes \mathrm{DFT}_q) P. \label{eq:ustargeneral}
\end{equation}

\noindent Its action on the $\mathcal{A}$ basis is:

\begin{align}
U_\star | a_i \rangle
&= P^\dagger (|i_p \rangle \otimes \mathrm{DFT}_q | i_q \rangle ) \\
&= P^\dagger \left( |i_p \rangle \otimes \frac{1}{\sqrt{q}} \sum_{s_q=0}^{q-1} e^{\frac{2 \pi i}{q} s_q i_q} |s_q \rangle \right) \\
&= \frac{1}{\sqrt{q}} \sum_{s_q=0}^{q-1} e^{\frac{2 \pi i}{q} s_q i_q} | a_{i_p \, q \bar q + s_q \, p \bar p} \rangle.
\end{align}

\noindent Now, write $n = i_p \, q \bar{q} + s_q \, p \bar{p} \in \mathcal{I}$ as the indices of the $\mathcal{A}$ basis states entering the sum. For all $n \in \mathcal{I}$, we have that $n \bmod p = i_p$ is fixed, so as $s_q$ runs over all of $\mathbb{Z}_q$ it always contributes a multiple of $p$. Therefore, we may write $\mathcal{I} = \{ i_p, i_p + p, \dots, i_p + (q-1)p \}$. We therefore re-index the sum by writing $n = i_p + kp$, where $s_q = n \bmod q = (i_p + kp) \bmod q$ enters the exponent:

\begin{align}
U_\star |a_i \rangle &= \frac{1}{\sqrt{q}} \sum_{k=0}^{q-1} e^{\frac{2 \pi i}{q} i_q (i_p + kp)} |a_{kp + i_p} \rangle
\\
&= e^{\frac{2 \pi i}{q} i_q i_p} \frac{1}{\sqrt{q}} \sum_{k=0}^{q-1} e^{\frac{2 \pi i}{q} pi_q \cdot k} |a_{kp + i_p} \rangle
\\ &= e^{\frac{2 \pi i}{q} i_q i_p} |i_p, p \, i_q \rangle_{p,q}
\end{align}

\noindent Therefore, $U_\star(\mathcal A)=\mathcal C_{p,q}$. Similarly, we can show that $U_\star(\mathcal B) = \mathcal C_{q, p}$. First we calculate the action of $P$ on the $\mathcal{B}$ states:

\begin{align}
    P | b_j \rangle &= \frac{1}{\sqrt{pq}} \sum_{l=0}^{pq-1} e^{\frac{2 \pi i}{pq} jl} P | a_l \rangle \\
    &= \frac{1}{\sqrt{pq}} \sum_{l=0}^{pq-1} e^{\frac{2 \pi i}{pq} jl} |l_p\rangle \otimes |l_q\rangle \\
    &= \frac{1}{\sqrt{pq}} \sum_{l_p=0}^{p-1} \sum_{l_q=0}^{q-1} e^{\frac{2 \pi i}{pq} (j_p \, q \bar{q} + j_q \, p \bar{p})(l_p \, q \bar{q} + l_q \,p \bar{p})} |l_p\rangle \otimes |l_q\rangle \\
     &= \frac{1}{\sqrt{pq}} \sum_{l_p=0}^{p-1} \sum_{l_q=0}^{q-1} e^{\frac{2 \pi i}{p} \bar{q} j_p l_p} e^{\frac{2 \pi i}{q} \bar{p} j_q l_q} |l_p\rangle \otimes |l_q\rangle \\
     &= \mathrm{DFT}_p |\bar{q} \, j_p \rangle \otimes \mathrm{DFT}_q |\bar{p} \, j_q \rangle.  \label{eq:Ustarb} \\
\end{align}

\noindent Here, between the third and fourth line the cross-terms in the exponent are integers. Hence, 

\begin{align}
U_\star | b_j \rangle &= P^\dagger (\mathrm{DFT}_p |\bar{q} \, j_p \rangle \otimes \mathrm{DFT}^2_q |\bar{p} \, j_q \rangle) \\
& = P^\dagger (\mathrm{DFT}_p |\bar{q} \, j_p \rangle \otimes |- \bar{p} \, j_q \rangle) \\
& =P^\dagger \left( \frac{1}{\sqrt{p}} \sum_{s_p=0}^{p-1} e^{\frac{2 \pi i}{p} s_p \bar{q} j_p} |s_p\rangle \otimes |- \bar{p} \, j_q \rangle \right)\\
& =  \frac{1}{\sqrt{p}} \sum_{s_p=0}^{p-1} e^{\frac{2 \pi i}{p} s_p \bar{q} j_p} |a_{ s_p \, q \bar{q} - \bar{p} \, j_q \, p \bar{p}} \rangle,
\end{align}

\noindent where we used $\mathrm{DFT}^2_d |i\rangle = |-i \bmod d\rangle$ for all $i \in \mathbb{Z}_d$. Again, we re-index the sum, which runs over a set $\mathcal{J}$ consisting of integers $n = s_p \, q \bar{q} - \bar{p} \, j_q \, p \bar{p}$ with $n \bmod q = - \bar{p} \, j_q \bmod q$ fixed. Thus, we take $n = kq + (- \bar{p} \, j_q \bmod q)$ and substitute $s_p = (kq + (- \bar{p} \, j_q \bmod q)) \bmod p$:

\begin{align}
U_\star | &b_j \rangle = \frac{1}{\sqrt{p}} \sum_{k=0}^{p-1} e^{\frac{2 \pi i}{p} \bar{q} j_p (kq + (- \bar{p} \, j_q \bmod q) )} |a_{ kq + (- \bar{p} \, j_q \bmod q) } \rangle \\
&= e^{\frac{2\pi i}{p} \bar{q}j_p (- \bar{p} \, j_q \bmod q)} \frac{1}{\sqrt{p}} \sum_{k=0}^{p-1} e^{\frac{2 \pi i}{p}  j_p \cdot k } |a_{ kq + (- \bar{p} \, j_q \bmod q) } \rangle \\
& = e^{\frac{2\pi i}{p} \bar{q}j_p (- \bar{p} \, j_q \bmod q)} | - \bar{p} \, j_q, j_p \rangle_{q, p}.
\end{align}

\noindent  It follows that $U_\star(\mathcal{B})=\mathcal{C}_{q,p}$. Moreover, $U_\star^2 = P^\dagger (\mathbb{1}_p \otimes \mathrm{DFT}_q^2)P$, so $U_\star^2$ is a permutation on $\mathcal{A}$ and $\mathcal{B}$. Hence, $U_\star(\mathcal C_{p, q}) = U_\star^2(\mathcal A) = \mathcal A$ and $U_\star(\mathcal C_{q, p}) = U_\star^2(\mathcal B) = \mathcal B$. Therefore, $U_\star$ is a positivity-preserving generalised permutation on the set of all pure KD-positive states. Appendix \ref{app:type3proofs} additionally shows it is positivity-preserving on any KD-positive state under $V = \mathrm{DFT}^{\otimes n}_{pq}$.

Next, we show that the action of $U_\star$ is not covariant with respect to non-real KD distributions. Instead, the resulting KD distribution $Q(U_\star \rho U_\star^\dagger)$ can be expressed as a complex linear combination over elements of $Q(\rho)$. First, we perform a global change of basis. We change into the $P(\mathcal{A})$ basis $|i_p, i_q \rangle = |i_p\rangle \otimes |i_q\rangle$, and the $P(\mathcal{B})$ basis $|j_p, j_q \rangle = \mathrm{DFT}_p |\bar{q} \, j_p\rangle \otimes \mathrm{DFT}_q |\bar{p} \, j_q\rangle$. The quantum state becomes $\tilde{\rho} = P \rho P^\dagger$. This gives an identical KD distribution, with entries indexed by pairs of indices $(i_p, i_q)$ and $(j_p, j_q)$ instead:

\begin{align}
    Q_{i_p, i_q ; j_p, j_q} &= \langle b_j | P^\dagger P | a_i \rangle \langle a_i | P^\dagger P \rho P^\dagger P | b_j \rangle \\
    &= \langle j_p, j_q | i_p, i_q \rangle \langle i_p , i_q | \tilde{\rho} | j_p , j_q \rangle
\end{align}

\noindent Following the global change of basis, $U_\star$ is given by $P U_\star P^\dagger = \mathbb{1}_p \otimes \mathrm{DFT}_q$. Hence $ Q_{i_p, i_q ; j_p, j_q} $ transforms as:

\begin{align}
    Q_{i_p, i_q ; j_p, j_q} &\rightarrow \langle j_p, j_q | i_p, i_q \rangle \times \\
    &\langle i_p , i_q | (\mathbb{1}_p \otimes \mathrm{DFT}_q) \tilde{\rho} (\mathbb{1}_p \otimes \mathrm{DFT}^\dagger_q) | j_p , j_q \rangle.
\end{align}

\noindent The second term may be expanded as:
\begin{align}
    &\langle i_p , i_q | (\mathbb{1}_p \otimes \mathrm{DFT}_q) \tilde{\rho} (\mathbb{1}_p \otimes \mathrm{DFT}^\dagger_q) | j_p , j_q \rangle  \\
    &= \frac{1}{q}\sum_{s_q, t_q=0}^{q-1} e^{\frac{2 \pi i}{q} (i_q \cdot s_q - \bar{p} j_q \cdot \bar{p} t_q)} \langle i_p, s_q | \tilde{\rho} | j_p, t_q \rangle 
\end{align}

\noindent Where we used the fact that $\langle j_p, t_q | \mathbb{1}_p \otimes \mathrm{DFT}^\dagger_q |j_p, j_q \rangle = \langle \bar{p} \, t_q | \mathrm{DFT}_q^\dagger | \bar{p} \, j_q \rangle $. From substituting this in along with $\langle i_p, s_q | \tilde{\rho} | j_p, t_q \rangle = Q_{i_p, s_q; j_p, t_q} / \langle j_p, t_q | i_p, s_q \rangle $, expanding out the ratio $\langle j_p, j_q | i_p, i_q \rangle / \langle j_p, t_q | i_p, s_q \rangle$ and factorising, we obtain:

\begin{equation}
    Q_{i_p, i_q ; j_p, j_q}  \rightarrow \frac{1}{q}\sum_{s_q, t_q=0}^{q-1} e^{\frac{2 \pi i}{q} (i_q + \bar{p} t_q) \cdot (s_q - \bar{p} j_q)} Q_{i_p, s_q; j_p, t_q} \label{eq:ustartransform}
\end{equation}

\noindent Which in general is not a covariant transformation. Furthermore, for a generic choice of $Q(\rho)$ there is no reason to expect that $\mathcal{N}(Q(U_\star \rho U_\star^\dagger)) = \mathcal{N}(Q(\rho))$, and indeed we have found examples where this is not the case. Hence, $U_\star \notin \mathcal{U}_{\mathcal{N}}$.

Finally, we show that no positivity-preserving unitaries can exist beyond types \textbf{I}, \textbf{II} and \textbf{III}. Inner product preservation under $U$ is again the key constraint. Each set $\mathcal{A},\mathcal{B}, \mathcal{C}_{p,q}, \mathcal{C}_{q,p}$ forms an orthonormal basis. The inner products between states across different sets are as follows:

\begin{tabular}{c |c } 
Basis Pair & \ Possible Overlap Magnitudes \\ [2pt]  \hline 
\rule{0pt}{\normalbaselineskip} \ $(\mathcal{A}, \mathcal{B})$, $(\mathcal{C}_{p,q}, \mathcal{C}_{q,p})$ \ & $1/\sqrt{pq}$ \\ [2pt]
\ $(\mathcal{A}, \mathcal{C}_{p,q})$, $(\mathcal{B}, \mathcal{C}_{q,p})$ \ & 0, $1/\sqrt{q}$ \\ [2pt]
\ $(\mathcal{A}, \mathcal{C}_{q,p})$, $(\mathcal{B}, \mathcal{C}_{p,q})$ \ & 0, $1/\sqrt{p}$ \\ [2pt]
\end{tabular}

\

\noindent Without loss of generality, suppose that the image $U(\mathcal{A})$ is `split', so that $U$ maps $|a_1 \rangle$ and $|a_2 \rangle$ to two different basis sets. If these two sets are mutually unbiased -- i.e. $\mathcal{A}$ and $\mathcal{B}$, or $\mathcal{C}_{q,p}$ and $\mathcal{C}_{p,q}$, then this implies $|\langle a_1 | U^\dagger U |a_2 \rangle| \neq 0$. On the other hand, if these sets are not mutually unbiased, then no matter which set $U$ maps $|b_i \rangle \in \mathcal{B}$ to, either $|\langle a_1 | U^\dagger U |b_i \rangle| \neq 1/\sqrt{d}$ or $|\langle a_2 | U^\dagger U |b_i \rangle| \neq 1/\sqrt{d}$. In all cases, this implies that $U$ does not preserve the inner product, so (by the same reasoning as for $\mathcal{A}$) the image of each basis set cannot be split across distinct basis sets. In effect, only types \textbf{I}, \textbf{II} and \textbf{III} unitaries are possible. We also note that by symmetry, the above lemma straightforwardly applies to the case of $U'_\star = P^\dagger(\mathrm{DFT}_p \otimes \mathbb{1}_{q})P$, which is another instance of a type \textbf{III} unitary. 

\end{proof}

\noindent Although in general the action of $U_\star$ on non-real KD distributions is not covariant, it is still more straightforward to describe over generic unitaries -- for which a full construction of the superoperator $\hat{\mathcal{E}}_U$ is typically required. Regardless of this fact, the presence of such unitaries in any KD-based classical simulation algorithm poses an additional difficulty. We discuss the implications of this result in Section \ref{section:kdsimulation}.

\textit{KD-stochastic unitaries}. We now turn to the question of identifying unitaries $U$ which induce stochastic superoperators $\hat{\mathcal{E}}_U$ in the vectorised picture. Stochastic matrices have the property of preserving the $\ell_1$-norm of KD-positive inputs. However, we have seen that positivity-preserving unitaries do not need to induce a stochastic $\hat{\mathcal{E}}_U$ to preserve positivity. For example, if the transition matrix $V$ is Hermitian, then the unitary $U = V$ has a non-stochastic KD superoperator, but maps the KD matrix $Q(\rho)$ to its Hermitian conjugate:

\begin{equation}
    Q(\rho)_{ij} \rightarrow \langle b_j | U^\dagger U |a_i \rangle \langle a_i | U \rho U^\dagger | b_j \rangle = Q^*_{ji}(\rho).
\end{equation}

\noindent This corresponds to a $\mathcal{N}$-preserving unitary of type \textbf{II}. In contrast, for the odd-dimensional discrete Wigner distribution such a situation does not occur, as all positivity-preserving gates induce stochastic superoperators \cite{Pashayan_2015, Gross_2006}. We now show that KD-stochastic unitaries are precisely of type \textbf{I}.

\begin{theorem} \label{theorem:genperm}
    For a KD distribution defined on the bases $\mathcal{A} = \{ |a_i \rangle \}$ and $\mathcal{B} = \{ |b_j \rangle \}$, a unitary $U \in \mathcal{U}_{\mathrm{stoch}}$ if and only if $U \in M_d \cap V M_d V^\dagger$. In other words, $U$ satisfies:

    \begin{align*}
    U| a_i \rangle &= e^{i \theta_i} | a_{\sigma(i)} \rangle, & U| b_j \rangle &= e^{i \phi_j} | b_{\tau(j)} \rangle,  
    \end{align*}
    where $\sigma, \tau \in S_d$. It follows that $\mathcal{U}_{\mathrm{stoch}}$ consists of $\mathcal{N}$-preserving type \textbf{I} unitaries. 
\end{theorem}
\begin{proof}
    Suppose $U \in M_d \cap V M_d V^\dagger$, i.e. $U$ satisfies $U| a_i \rangle = e^{i \theta_i} | a_{\sigma(i)} \rangle$ and $U| b_j \rangle = e^{i \phi_j} | b_{\tau(j)} \rangle$. Then, 

    \begin{align*}
        (\hat{\mathcal{E}}_U)_{ij, kl} &= \frac{\langle b_j | a_i \rangle}{\langle b_l | U^\dagger U | a_k \rangle} e^{i \theta_k} \langle a_i |a_{\sigma(k)} \rangle e^{-i \phi_l} \langle b_{\tau(l)} | b_j \rangle  \\
        &= \delta_{i, \sigma(k)} \delta_{j, \tau(l)} \times  \frac{\langle b_j | a_i \rangle}{\langle b_{\tau(l)} | a_{\sigma(k)} \rangle} \times \frac{e^{i (\theta_k - \phi_l)}}{e^{i (\theta_k - \phi_l)}} \\
        & = \delta_{i, \sigma(k)} \delta_{j, \tau(l)},
    \end{align*}

        \noindent so $\hat{\mathcal{E}}_U \in S_d \times S_d$ is the Kronecker product of two permutation matrices.

        For the other direction, suppose the induced superoperator $\hat{\mathcal{E}}_U$ is stochastic. Then, all of its entries are non-negative. By Equation \eqref{eq:unitaryinverse}, it will always have a quasi-stochastic inverse $\hat{\mathcal{E}}_{U}^{-1}$. By Equation \eqref{eq:invsuperopelement}, this inverse will only contain non-negative entries and thus also be stochastic. A basic fact of linear algebra \cite{Ding01032013} is that if a matrix and its inverse are both stochastic, then that matrix must be a permutation. Therefore, $\hat{\mathcal{E}}_U$ must be a permutation. Hence, for each pair of indices $i, j$ there is a unique pair of indices $i^*, j^*$ such that $(\hat{\mathcal{E}}_U)_{ij, i^* j^*} \neq 0$. By Equation \eqref{eq:unitarysuperopelement}, $\langle a_i | U | a_{i^*} \rangle \langle b_j | U^\dagger | b_{j^*} \rangle \neq 0$, therefore $\langle a_i | U | a_{i^*} \rangle \neq 0$ and $\langle b_j | U^\dagger | b_{j^*} \rangle \neq 0$. Holding $i, j, j^*$ fixed we deduce that $\langle a_i | U | a_{k} \rangle = 0$ for $k \neq i^*$, and holding $i, j, i^*$ fixed we deduce that $\langle b_j | U^\dagger | b_{l} \rangle = 0$ for $l \neq j^*$. Hence, $U$ is a generalised permutation on the bases $\mathcal{A}$ and $\mathcal{B}$.
\end{proof}

\textit{Randomly Sampled Bases}. Given the stringent conditions for their existence, whether $\mathcal{U}_{\mathcal{N}}$, $\mathcal{U}_{\mathcal{N}+}$ and $\mathcal{U}_{\mathrm{stoch}}$ contain non-trivial unitaries is highly sensitive to the symmetries of the transition matrix $V$. Consequently, for most KD distributions this is not the case.

\begin{remark} \label{remark:haarrandom}
The vast majority of Kirkwood--Dirac distributions do not admit non-trivial positivity-preserving, $\mathcal{N}$-preserving or KD-stochastic unitaries. 
\end{remark}
\begin{proof} (Sketch)
We already know from \cite{langrenez2024} that for a generic choice of $V$ (with probability one, under the Haar measure on $U(d)$), $\mathcal{E}_{\mathrm{KD}+} = \mathrm{conv}(\mathcal{A} \cup \mathcal{B})$, and $\mathcal{C} = \emptyset$. Therefore, only type \textbf{I} and \textbf{II} can exist. In the first case, $U \in M_d \cap V M_d V^\dagger$. Elements of this set are ones which satisfy $U \in M_d$ and $V^\dagger U V \in M_d$. Hence, for some $U = D_\alpha P_\sigma$ there must exist some $D_\beta P_\tau$ such that $V^\dagger D_\alpha P_\sigma V = D_\beta P_\tau$. Therefore,

\begin{equation} \label{eq:genpermcondition}
D_\alpha P_\sigma V = V D_\beta P_\tau \rightarrow V_{\sigma(i) \tau(j)} = \alpha_{\sigma(i)} \beta^*_{\tau(j)} V_{ij},
\end{equation}

\noindent where $P_\sigma$ and $P_\tau$ are permutation matrices, and the above holds for all $i, j$. Hence, $|V_{\sigma(i) \tau(j)}| = |V_{ij}|$, implying that the modulus matrix $|V|$ must be invariant under some row and column permutation $\sigma$, $\tau$. Excluding the identity permutations, we therefore require at least one pair of distinct entries of $V$ to have the same modulus (this is a necessary, though not sufficient condition). Under the Haar measure on $U(d)$ the probability of this occurring is zero. Hence, for a generic choice of $V$, no non-trivial type \textbf{I} unitaries exist. Similarly, for non-trivial type \textbf{II} unitaries in $V M_d \cap M_d V^\dagger$ to exist, the necessary condition is that $|V_{i,\tau(j)}| = |V_{j,\sigma^{-1}(i)}|$ under some non-identity pair $\sigma, \tau$. Again, under the Haar measure this happens with probability zero. Therefore, for a randomly sampled $V$, with probability one we have $\mathcal{U}_{\mathcal{N}} = \mathcal{U}_{\mathcal{N}_+}= \mathcal{U}_{\mathrm{stoch}} = \{e^{i \theta} \mathbb{1}\}$.
\end{proof}

\textit{Examples of Type \textbf{I} and \textbf{II} Unitaries.} Although type \textbf{I} and \textbf{II} unitaries are rare, they do exist for symmetric choices of $V$. For example, in the Fourier-conjugate distribution in any dimension, type \textbf{I} unitaries are given by the Weyl--Heisenberg group \cite{debievre2025, Spriet2026} (see Appendix \ref{section:kdconvolutions} for a definition), whereas type \textbf{II} unitaries are given by a product of the Weyl--Heisenberg unitaries and the $\mathrm{DFT}$ matrix. In multi-qubit systems (where $\mathcal{A}$ and $\mathcal{B}$ are the computational and Hadamard-transformed bases), the positivity-preserving group is generated by single-qubit Pauli gates and CNOT for type \textbf{I}, and $U = H^{\otimes n}$ for type \textbf{II} \cite{Thio_2026}. Notably, single-qubit Hadamard gates are of neither type and can be shown to not preserve positivity. For non-Fourier distributions, in the spin-$1$ distribution presented in \cite{Langrenez_2024_characterising} with an involutory transition matrix (referred to as $U^\star$), one can identify type \textbf{I} unitaries with any $3 \times 3$ permutation matrix on the $\mathcal{A}$ basis, and type \textbf{II} unitaries with the product of any permutation and $U^\star$. As the states in the set $\mathcal{C}$ for the spin-$1$ system are not pairwise orthogonal, by a similar argument to Theorem \ref{thm:primefourier} we can show that no other types of positivity-preserving unitaries can exist. Finally, we remark that our classification is not exhaustive, since there may be other positivity-preserving, generalised permutations on $\mathcal{A} \cup \mathcal{B} \cup \mathcal{C}$ not of the three types we have identified. For example, it is in principle possible that such unitaries map some subset of $\mathcal{A}$ to a subset of $\mathcal{B}$, and some other subset of $\mathcal{A}$ to a subset of $\mathcal{C}$. In our search we have not found any such examples -- whether they could exist for an arbitrary choice of $V$ can usually be ruled out by symmetry arguments (namely the preservation of inner products between all the positive pure states under $U$, as we have argued in Theorem \ref{thm:primefourier}). These could be used to fully characterise $\mathcal{U}_{\mathcal{N}+}$ for $\mathrm{DFT}_d$ in the general case, which we leave for future work.

\textit{Completeness of Our Classification.} Collecting the results of this section, we formulate a general method for classifying positivity-preserving unitaries for arbitrary distributions: if $\mathcal{C} = \emptyset$, then only types \textbf{I, II} can exist (Lemma \ref{lemma:pospresunitaries}). As they are always $\mathcal{N}$-preserving (Lemma \ref{lemma:nonpospresunitaries}), this concludes the classification. If $\mathcal{C} \neq \emptyset$, one must check whether generalised permutations on $\mathcal{E}_\mathrm{KD+}^\mathrm{pure}$ beyond types \textbf{I}, \textbf{II} exist. Typically, they can be ruled out using symmetry arguments. If they are found, and $\mathcal{E}_\mathrm{KD+} = \mathrm{conv}(\mathcal{E}_\mathrm{KD+}^\mathrm{pure} )$, then this concludes the classification. Otherwise, one must also check if they preserve positivity of $\rho \in \mathcal{E}_\mathrm{KD+} \setminus \mathrm{conv}(\mathcal{E}_\mathrm{KD+}^\mathrm{pure}) $. Overall, our classification is complete for three cases: for the Haar-random choice of $V$, our method terminates at the first step. For $V= \mathrm{DFT}_{p^k}$, it terminates at the second. Finally, for $V= \mathrm{DFT}_{pq}$, it terminates at the third. 

\section{Estimating Born Rule Probabilities via KD Distributions}
\label{section:kdsimulation}

We now apply the results of Section \ref{section:kdstochasticunitaries} to the problem of estimating Born rule outcome probabilities via KD distributions. Throughout this section, `efficient simulation' means additive-error estimation of
Born outcome probabilities, with sample complexity polynomial in $(1/\epsilon, \log(1 / \delta))$, and the circuit size, assuming efficient access to the relevant KD quasiprobabilities and superoperator elements. To do this, we modify the algorithm of \cite{Pashayan_2015}, which makes use of a Monte Carlo--style method to estimate the outcome probability via Equation \eqref{eq:kdexpsum}. Consider an $n$-qudit input state $\rho$, a quantum circuit $U$, and a measurement operator $F$. We first decompose $U$ into a product of one- and two-qudit gates $U = U_N \cdots U_1$, and write pairs of indices (summed over in \eqref{eq:kdexpsum}) as $(i_0 j_0) = I_0, \dots, (i_N j_N) = I_N$. The probability of the measurement outcome is given by:

\begin{align} \text{Tr}(F U \rho U^\dagger) = \sum_{I_0, \dots, I_N} & F_{I_N} (\hat{\mathcal{E}}_{U_N})_{I_{N}, I_{N-1}} \cdots \label{eq:bornruleest} \\ & \dots (\hat{\mathcal{E}}_{U_2})_{I_2, I_1}  (\hat{\mathcal{E}}_{U_1})_{I_1, I_0} Q(\rho)_{I_0}. 
\end{align}

\textit{Assumptions.} We assume that the objects $\boldsymbol{|}Q \boldsymbol{\rangle \rangle}, \boldsymbol{\langle \langle} F \boldsymbol{|}$, and $\hat{\mathcal{E}}_{U_m}$ are efficiently describable (for example, when the state preparation and measurement is conducted in a product state basis and each $U_m$ acts on at most two qudits), and that individual elements from their rows and columns can be sampled efficiently under some probability distribution. We can use this decomposition to classically estimate Born rule outcome probabilities. We will make use of several norms:

\begin{align*}
    \|\boldsymbol{|} Q \boldsymbol{\rangle \rangle} \|_1 &= \sum_I |Q(\rho)_{I}| = \mathcal{N}(Q(\rho)) \\
    \| \boldsymbol{\langle \langle} F \boldsymbol{|} \|_1 &= \sum_I |F_I| = \mathcal{N}(F) \\
    \| \boldsymbol{\langle \langle} F \boldsymbol{|} \|_{\infty} &= \max_{I} | F_I |  \\
    \| (\hat{\mathcal{E}}_U)_{\cdot, I} \|_1 &= \sum_{J} |(\hat{\mathcal{E}}_U)_{J, I} |  \\
    \| \hat{\mathcal{E}}_U \|_1 &= \max_I \sum_{J} |(\hat{\mathcal{E}}_U)_{J, I} | = \max_I \| (\hat{\mathcal{E}}_U)_{\cdot, I} \|_1,
\end{align*}

\noindent where the first two are KD total non-positivities, the third is the largest absolute element of the POVM dual vector $\boldsymbol{\langle \langle} F \boldsymbol{|}$, the fourth is the $\ell_1$-norm of the column $I$ of $\hat{\mathcal{E}}_U$, and the last is the induced $\ell_1$-norm on $\hat{\mathcal{E}}_U$. 

\textit{The Probability Estimation Protocol.} We proceed as follows: first, take the absolute values of the quasiprobabilities in $\boldsymbol{|} Q \boldsymbol{\rangle \rangle}$, and randomly sample an index pair $(i_0, j_0) = I_0$ with probability $\text{Pr}(I_0) = |Q(\rho)_{I_0}| / \mathcal{N}(Q(\rho))$. Then, randomly sample the index pair $I_1$ with probability $\text{Pr}(I_1) = |(\hat{\mathcal{E}}_{U_1})_{I_1, I_0}| / \| (\hat{\mathcal{E}}_{U_1})_{\cdot, I_0} \|_1 $. Repeat this up to $I_{N-1}$, and then sample $I_N$ with probability $\text{Pr}(I_N) = |(\hat{\mathcal{E}}_{U_N})_{I_N, I_{N-1}}| / \| (\hat{\mathcal{E}}_{U_N})_{\cdot, I_{N-1}} \|_1 $.  After sampling $I_0, I_1, \dots, I_{N}$, also compute the complex phases $\text{Ph}(Q_{I_0}) = Q(\rho)_{I_0} / |Q_{I_0}|$, and $\text{Ph}((\hat{\mathcal{E}}_{U_k})_{I_k, I_{k-1}}) = (\hat{\mathcal{E}}_{U_k})_{I_k, I_{k-1}} / |(\hat{\mathcal{E}}_{U_k})_{I_k, I_{k-1}}|$ for $k\in \llbracket N \rrbracket$. Finally, calculate the complex random variable $Z(I_0, \dots, I_N)$ and its real part:

\begin{widetext}
\begin{align}
X(I_0, \dots, I_N) &= \text{Re}\left(Z(I_0, \dots, I_N) \right) \nonumber
\\ &= \text{Re} \left(  F_{I_N} \times \left[ \prod_{k = 1}^{N} \| (\hat{\mathcal{E}}_{U_{k}})_{\cdot, I_{k-1}} \|_1 \cdot \text{Ph}((\hat{\mathcal{E}}_{U_k})_{I_k, I_{k-1}}) \right]  \times  \mathcal{N}(Q(\rho)) \cdot \text{Ph}(Q_{I_0}) \right). \label{eq:xexpression}
\end{align}
\end{widetext}

\noindent This is then repeated many times, with the average $\tilde{X}$ forming an estimate of $\text{Tr}(F U \rho U^\dagger)$. In expectation (over the choices of random index pairs):

\begin{align}
    \mathbb{E}(Z(I_0, \ldots, I_N)) & = \text{Tr}(F U \rho U^\dagger), \nonumber \\
    \mathbb{E}(X(I_0, \ldots, I_N)) &= \text{Tr}(F U \rho U^\dagger).
\end{align}

\noindent Now, define the total induced non-positivity as the product of induced $\ell_1$-norms of the intermediate superoperators $\hat{\mathcal{E}}_{U_k}$ and the largest absolute element of $\boldsymbol{\langle \langle} F \boldsymbol{|}$:

\begin{equation}
    \mathcal{N}_I = \left[ \prod_{k = 1}^{N} \| \hat{\mathcal{E}}_{U_k} \|_1  \right] \times \| \boldsymbol{\langle \langle} F \boldsymbol{|} \|_{\infty}
\end{equation}

\noindent Each $Z(I_0, \ldots, I_N)$ is then bounded to have magnitude:

\begin{equation*}
 |Z(I_0, \ldots, I_N) | \leq \mathcal{N}(Q(\rho)) \times \mathcal{N}_I.
\end{equation*}

\noindent As $Z(I_0, \dots, I_N)$ is a complex random variable, in order to bound the number of samples required we can consider its real part, also bounded as $|X(I_0, \dots, I_N)| \leq \mathcal{N}(Q(\rho)) \mathcal{N}_I$, and use the Hoeffding inequality (in the same way as was done in \cite{Pashayan_2015}) to show that estimating the Born rule probabilities up to precision $\epsilon$ (and success probability $1-\delta$) requires $s(\epsilon, \delta)$ samples, where:

\begin{equation}
    s(\epsilon, \delta) = \frac{2}{\epsilon^2} \mathcal{N}(Q(\rho))^2 \mathcal{N}_I^2 \ln(2 / \delta).
\end{equation}

\noindent Clearly, the runtime of the estimation algorithm is bounded by $\mathcal{O}(\mathcal{N}(Q(\rho))^2 \mathcal{N}_I^2)$, and thus the efficiency of the estimation procedure is determined by the total non-positivity $\mathcal{N}(Q(\rho))$ of the input state, multiplied by the total induced non-positivity $\mathcal{N}_I$. In the case of the Wigner distribution, the input and induced non-positivities are independent. For example, when a Clifford circuit satisfies $\mathcal{N}_I = \mathcal{O}(1)$ the additional sample complexity of estimation on a non-positive input is multiplicative in $\mathcal{N}(Q(\rho))^2$. We will show that for the KD distribution, this happens by default with type \textbf{I} unitaries. For type \textbf{II} unitaries, the same holds once their covariance is accounted for. For type \textbf{III} unitaries, the induced non-positivity implicitly depends on whether the input distribution contains complex entries.

\textit{Incorporating Type \textbf{I, II} Unitaries.} We now show how to accommodate KD positivity-preserving unitaries in the algorithm. For type \textbf{I} unitaries, we have already established via Theorem \ref{theorem:genperm} that the induced superoperator $\hat{\mathcal{E}}_U$ is a permutation matrix. Hence, for such unitaries the induced non-positivity is $\| \hat{\mathcal{E}}_U \|_1 = 1$, and no modification is necessary. For type \textbf{II} unitaries, in general $\| \hat{\mathcal{E}}_U \|_1 > 1$. However, Lemma \ref{lemma:nonpospresunitaries} asserts that for any state $\rho$, $Q(U \rho U^\dagger)_{ij} = Q^*_{\sigma^{-1}(j), \tau^{-1}(i)}(\rho)$ for some $\sigma, \tau \in S_d$. Suppose that the circuit $U$ contains a type \textbf{II} unitary $U_k$ at step $k$. We may write $V = U_N \cdots U_{k+1}$ and $W = U_{k-1} \cdots U_1$, so that $U = V U_k W$. Also writing $(\sigma^{-1}(j), \tau^{-1}(i)) = \gamma^{-1}(I)$ and $Q(U \rho U^\dagger)_I = Q^*(\rho)_{\gamma^{-1}(I)}$ as a shorthand, Equation \eqref{eq:bornruleest} becomes:

\begin{align}
    \text{Tr}&(F (V U_k W)  \rho (V U_k W)^\dagger) \label{eq:bornruleest2} \\ &= \sum_{H, I_k, J, L} F_{H} (\hat{\mathcal{E}}_V)_{H, I_k}(\hat{\mathcal{E}}_{U_k})_{I_k, J} (\hat{\mathcal{E}}_W)_{J, L}Q(\rho)_L \\ &=
    \sum_{H, I_k, J} F_{H} (\hat{\mathcal{E}}_V)_{H, I_k}(\hat{\mathcal{E}}_{U_k})_{I_k, J}Q( W\rho W^\dagger)_J \\
    &=
    \sum_{H, I_k} F_{H} (\hat{\mathcal{E}}_V)_{H, I_k} Q( U_k W\rho W^\dagger U_k^\dagger)_{I_k} \\
    &=
    \sum_{H, I_k} F_{H} (\hat{\mathcal{E}}_V)_{H, I_k} Q( W\rho W^\dagger)^*_{\gamma^{-1}(I_k)} \\
    &=
    \sum_{H, I_k, L} F_{H} (\hat{\mathcal{E}}_V)_{H, I_k} (\hat{\mathcal{E}}_W)^*_{\gamma^{-1}(I_k), L}Q(\rho)^*_L \\
    &=
    \sum_{H, J, L} F_{H} (\hat{\mathcal{E}}_V)_{H, \gamma(J)}(\hat{\mathcal{E}}_W)^*_{J, L}Q(\rho)^*_L
\end{align}

\noindent Hence, the algorithm may be modified as follows. Whenever a type \textbf{II} unitary $U$ is encountered at step $k$, proceed deterministically, setting $I_k = \gamma(I_{k-1})$, $\| (\hat{\mathcal{E}}_{U_{k}})_{\cdot, I_{k-1}} \|_1=1$ and $\text{Ph}((\hat{\mathcal{E}}_{U_k})_{I_k, I_{k-1}}) = 1$. Furthermore, complex conjugate all phases as $\text{Ph}(Q_{I_0}) \rightarrow \text{Ph}(Q_{I_0})^*$ and $\text{Ph}((\hat{\mathcal{E}}_{U_m})_{I_m, I_{m-1}}) \rightarrow \text{Ph}((\hat{\mathcal{E}}_{U_m})_{I_m, I_{m-1}})^*$ for all $m < k$. The rest of the algorithm proceeds as normal, with $\tilde{X}$ forming the estimate for $\text{Tr}(F U \rho U^\dagger)$. This modification directly accounts for the covariant transformations under type \textbf{II}, ensuring that the induced $\ell_1$-norm of their superoperators does not contribute to $\mathcal{N}_I$.

\textit{Incorporating Type \textbf{III} Unitaries.} For type \textbf{III} unitaries, there is an additional complication. As remarked in Lemma \ref{cor:compositefourier}, such unitaries do not generally induce a covariant transformation on the KD distribution, and thus the induced non-positivity satisfies $\| \hat{\mathcal{E}}_U \|_1 > 1$. On the other hand, as shown in Theorem \ref{thm:ustar}, the type \textbf{III} unitary $U_\star$ exhibits covariance on KD-real inputs, permuting their entries as $Q(U_\star \rho U_\star^\dagger)_{ij} = Q(\rho)_{\gamma^{-1}(i j)}$ for some $\gamma^{-1} \in S_{d^2}$. The exact form of $\gamma$ is given in Appendix \ref{app:type3proofs}: the mapping is obtained via CRT decompositions of $(i, j)$ and modular arithmetic, which is computationally efficient to implement. Thus, if the state $(U_{k-1} \cdots U_1)\rho (U_{k-1} \cdots U_1)^\dagger$ is KD-real, type \textbf{III} unitaries can be incorporated into the estimation algorithm in a similar way to type \textbf{II} unitaries, i.e. by deterministically setting $\| (\hat{\mathcal{E}}_{U_{k}})_{\cdot, I_{k-1}} \|_1 = \text{Ph}((\hat{\mathcal{E}}_{U_k})_{I_k, I_{k-1}}) = 1$ and $I_k = \gamma(I_{k-1})$ (c.f. Equation \eqref{eq:bornruleest2}) whenever they appear. On the other hand, if the input at step $k$ is non-real, covariance fails and this modification is not possible. Instead, sampling from the columns of $\hat{\mathcal{E}}_U$ becomes necessary, contributing a multiplicative factor to $\mathcal{N}_I$ for each instance of a type \textbf{III} unitary. This leads to a sharp increase in sample complexity between KD-real and non-real distributions. Consider for example a positivity-preserving circuit containing type \textbf{III} unitaries. If the input state is KD-positive (and thus KD-real), then the total induced non-positivity $\mathcal{N}_I$ is at most $\mathcal{O}(1)$, and the output probabilities can be efficiently estimated. However, if any $Q(\rho)_{ij} \in \mathbb{C}$ then in the worst case $\mathcal{N}_I$ grows exponentially large in the number of type \textbf{III} unitaries, and efficient classical estimation fails (such circuits may still be efficiently simulable with other methods, however). In this sense, the presence of type \textbf{III} unitaries makes the induced non-positivity of the circuit implicitly dependent on the reality of the input distribution. We find that classical simulation via KD distributions is a more intricate problem than in the Wigner case, where `conditional' covariance of gates does not occur.

Finally, we note that the above amendments can also be easily incorporated into the sampling algorithm by Mari and Eisert \cite{Mari_2012}, which leverages covariance to efficiently sample from the output distribution of a quantum circuit with positive Wigner function inputs and Clifford gates. Hence, for circuits composed only of \textbf{I, II, III} unitaries, if the input state is KD-positive, then each deterministic modification can be applied and the output distribution can be efficiently sampled. In circuits containing other gates, the type \textbf{III} unitary rule is only justified when the KD distribution of the state preceding the unitary is known to be real.

\textit{Consequences for Resource Theories}. We conclude by combining the results of Sections \ref{section:kdstochasticunitaries} and \ref{section:kdsimulation} into a no-go result for resource theories of KD non-positivity on $V = \mathrm{DFT}^{\otimes n}_{pq}$.

\begin{theorem} \label{thm:resourcetheory}
For KD distributions defined on $V = \mathrm{DFT}^{\otimes n}_{pq}$, any resource theory that takes KD non-positivity as a monotone cannot also declare all positivity-preserving, efficiently simulable unitaries to be free operations. In particular, it must either exclude the unitary $U^{\otimes n}_\star$ from its free operations, or else abandon monotonicity of KD non-positivity under its free operations.
\end{theorem}
\begin{proof}
    As proven in Lemma \ref{cor:compositefourier}, for $d = pq$ the type \textbf{III} unitary $U_\star$ always exists. In Theorem \ref{thm:ustar} and Appendix \ref{app:type3proofs}, we show that $U_\star$ preserves $\mathcal{N}(Q)$ for all KD-positive states. However, it does not generally preserve $\mathcal{N}(Q)$ for non-real states. From $U_\star = P^\dagger (\mathbb{1}_p \otimes \mathrm{DFT}_q) P$ (Equation \eqref{eq:ustargeneral}) we also have that $U_\star^2$ is a permutation and $U_\star^{4} = \mathbb{1}$. Suppose that $U_\star$ is included in the free operations of a resource theory of KD non-positivity. If there exists a state $\rho$ such that $\mathcal{N}(Q(U_\star \rho U_\star^\dagger)) <  \mathcal{N}(Q( \rho ))$, then (writing $\rho_k = U_\star^k \rho (U_\star^\dagger)^k$) it must hold true that $\mathcal{N}(Q(U_\star \rho_k U_\star^\dagger)) > \mathcal{N}(Q(\rho_k))$ for some $\rho_k$, otherwise it would hold that $\mathcal{N}(Q( \rho )) = \mathcal{N}(Q(U_\star \rho_3 U_\star^\dagger)) < \mathcal{N}(Q( \rho ))$, which is a contradiction. Similarly, if $\mathcal{N}(Q(U_\star \rho U_\star^\dagger)) >  \mathcal{N}(Q( \rho ))$, there must exist a state $\rho_k$ such that the total non-positivity decreases under $U_\star$. Unless $U_\star$ is $\mathcal{N}$-preserving for all states (which we have shown is not the case), there will exist states such that the total non-positivity increases under $U_\star$. Thus, $\mathcal{N}(Q)$ is not a monotone under the free operations. On the other hand, we have shown that for any KD-positive state $\rho$ evolution by type \textbf{I, II, III} is classically efficiently simulable, as such unitaries are covariant with respect to positive KD distributions. Hence, if $U_\star$ is not included in the free operations, then there exist positivity-preserving, efficiently simulable unitaries which are not free, and the resource theory fails to capture the full set of classically simulable operations.
\end{proof}

Theorem \ref{thm:resourcetheory} implies that constructing resource theories of KD non-positivity on the $V = \mathrm{DFT}^{\otimes n}_{pq}$ family of KD distributions is likely to be challenging. We remark that as $U_\star$ is KD reality-preserving, one possible avenue to bypass this no-go result is to consider the real part of the KD distribution as a resource, in effect working with the Margenau--Hill distribution \cite{Francica_2022, Pei_2023}. This forms an intriguing open question for future work. 

\

\textit{Note Added.} The results of Remark \ref{remark:haarrandom} and Lemma \ref{lemma:pospresunitaries} were introduced in a revised preprint of our manuscript (2 April 2026). After its posting, a revised version of \cite{langrenez2024} appeared containing analogous results, obtained independently of ours in the PhD thesis of one of its authors \cite{Langrenez_thesis}.

\acknowledgements
We thank numerous anonymous referees for their insightful feedback, which significantly improved the quality of this work. J.B. thanks David Arvidsson-Shukur and Hakop Pashayan for helpful discussions. S.S. acknowledges support from the Royal Society University Research Fellowship and ``Quantum Simulation Algorithms for Quantum Chromodynamics'' grant (ST/W006251/1). 

\appendix 

\begin{figure*}[t!]
    \subfloat{ \begin{quantikz} 
    \lstick{\ket{0}} & \gate{H} & \ctrl{1} & \gate{S^s} & \gate{H} & \meter{} \\
    \lstick{\ket{a_i}} & & \gate[3]{C_3} & & & \\
    \lstick{$\rho$} &  &  & & & \\
    \lstick{\ket{b_j}} &  &  & & &
\end{quantikz}  } \quad \quad \quad 
    \subfloat{ \subfloat{ \begin{quantikz} 
    & & \ctrl{1} &  \\
    &\qwbundle{n} &  \gate{C_n} & 
\end{quantikz} = \begin{quantikz} 
    & \ctrl{1} & \ctrl{2} & \ldots &  \ctrl{4} &  \\
    & \swap{1} & &\ldots  & & \\
    & \targX{} & \swap{1} & \ldots & & \\
    & & \targX{} &\ldots  & & \\
    & & &\ldots & \swap{1} & \\
    & & & \ldots & \targX{} &
\end{quantikz}  }  }
    \caption{\textit{Left:} The cycle test for estimating $\text{Re}[Q(\rho)_{ij}]$ ($s = 0$), and $\text{Im}[Q(\rho)_{ij}]$ ($s = 1$). $H$ is the Hadamard gate, and $S = \text{diag}(1, i)$. After many measurements, $Q(\rho)_{ij}$ is estimated from the outcome frequencies via $p(0)_{s=0} = (1 + \text{Re}[Q(\rho)_{ij}]) / 2$ and $p(0)_{s=1} = (1 + \text{Im}[Q(\rho)_{ij}]) / 2$. \textit{Right:} The cycle gate may be implemented by a cascade of qudit controlled-SWAP gates.}
    \label{fig:cycletest}
\end{figure*}

\section{Estimating KD Superoperator Elements}
\label{appendix:cycletest}

In this section, we extend the cycle test algorithm of \cite{Wagner_2024, oszmaniec2021measuring} to provide a new method for experimental measurement of KD superoperator elements $(\hat{\mathcal{E}}_U)_{ij, kl}$ induced by an unknown unitary $U$ and any informationally complete choice of bases $\mathcal{A, B}$. 

In general, it is not possible to directly sample from the KD distribution $Q(\rho)_{ij} = \text{Tr}( | b_j \rangle \langle b_j | a_i \rangle \langle a_i | \rho)$, since the outer products $\langle b_j | a_i \rangle \times | b_j \rangle \langle a_i |$ are not positive semidefinite, and thus cannot form POVM elements. However, a procedure was given in \cite{Wagner_2024} for estimating the quasiprobabilities $Q(\rho)_{ij}$ using a generalisation of the SWAP test, known as the cycle test \cite{oszmaniec2021measuring}. The circuit is shown in Figure \ref{fig:cycletest}. Similarly to the case of KD distributions, one can estimate elements $(\hat{\mathcal{E}}_U)_{ij, kl}$ of a KD superoperator induced by a unitary $U$, assuming $\mathcal{A, B}$ are informationally complete. To do this, we can modify the cycle test algorithm. First, we rewrite the superoperator elements of Equation \eqref{eq:unitarysuperopelement} as:

\begin{align}
    (\hat{\mathcal{E}}_U)_{ij, kl} = \frac{1}{|\langle b_l | a_k \rangle|^2} \text{Tr} & ( |a_i \rangle \langle a_i | \times  U | a_k \rangle \langle a_k | U^\dagger \nonumber \\
    & \times U | b_l \rangle \langle b_l | U^\dagger \times  | b_j \rangle \langle b_j | ).
\end{align}

\noindent The quantity $|\langle b_l | a_k \rangle|^2 \times (\hat{\mathcal{E}}_U)_{ij, kl}$ then forms a Bargmann invariant, which (given access to $U$ and the ability to prepare states from $\mathcal{A, B}$) can be estimated using a 4-state cycle gate $C_4$, as shown in Figure \ref{fig:cycletest2}. One may estimate this quantity and divide by $|\langle b_l | a_k \rangle|^2$, which may also be estimated separately via the SWAP test if the states from $\mathcal{A, B}$ are unknown. Unfortunately, as this method relies on an insertion of $U^\dagger U = \mathbb{1}$ it does not allow for the estimation of matrix elements of non-unitary CPTP maps; we leave the extension of our procedure to non-unitary CPTP maps as an open question.

\begin{figure}
    \begin{quantikz} 
    \lstick{\ket{0}} & \gate{H} & \ctrl{1} & \gate{S^s} & \gate{H} & \meter{} \\
    \lstick{\ket{a_i}} & & \gate[4]{C_4} & & &  \\
    \lstick{\ket{a_k}} & \gate{U} & & & & \\
    \lstick{\ket{b_l}} & \gate{U} && & & \\
    \lstick{\ket{b_j}} & &  & & & 
    \end{quantikz}
    \caption{\textit{Cycle Test for Estimating KD Superoperator elements}. The circuit estimates the quantity $|\langle b_l | a_k \rangle|^2 \times (\hat{\mathcal{E}}_U)_{ij, kl}$, where $p(0)_{s=0} = (1 + |\langle b_l | a_k \rangle|^2 \times \mathrm{Re}[(\hat{\mathcal{E}}_U)_{ij, kl}])/2$, and $p(0)_{s=1} = (1 + |\langle b_l | a_k \rangle|^2 \times \mathrm{Im}[(\hat{\mathcal{E}}_U)_{ij, kl}])/2$.}
    \label{fig:cycletest2}
\end{figure}

\section{Self-Similarity of KD Distributions}
\label{section:kdconvolutions}

During our study of Fourier-conjugate KD distributions we identified a self-similarity property, which allows for a more compressed representation of KD distributions. We present it here as a standalone result, which may be of independent interest. 

\begin{lemma} \label{thm:kddft}
    The multidimensional discrete Fourier transform $\widehat{Q}$ of a Kirkwood--Dirac distribution on $n$ qudits with transition matrix $V = \mathrm{DFT}_d^{\otimes n}$ is self-similar, satisfying the constraint:
    \begin{equation}
        \widehat{Q}(\mathbf{d - x, d - y}) = \omega^{\mathbf{x \cdot y}} (\widehat{Q}(\mathbf{x, y}))^*. \label{eq:thmkddft}
    \end{equation}
\end{lemma}

\begin{proof} Substituting Equation \eqref{eq:kdanddensity} into $\rho = \rho^\dagger$ and taking appropriate inner products, we obtain the following set of constraints on $Q_{ij}$:

\begin{equation}
    Q_{uv} = \sum_{ij} \frac{\langle a_i | b_v \rangle \langle b_v | a_u \rangle \langle a_u | b_j \rangle}{\langle a_i | b_j \rangle} Q_{ij}^*. \label{eq:kdconstraints}
\end{equation}

\noindent For the Fourier-conjugate KD distribution, we label the reference bases as follows:

\begin{align}
    \mathcal{A} &= \{ |\mathbf{i} \rangle = |i_1 \rangle \otimes |i_2 \rangle \otimes \cdots \otimes |i_n\rangle  \ | \ \mathbf{i} \in \mathbb{Z}_d^n \}  \\
    \mathcal{B} &= \{ \mathrm{DFT}_d^{\otimes n}|\mathbf{j} \rangle = \mathrm{DFT}_d |j_1 \rangle \otimes \cdots \otimes \mathrm{DFT}_d |j_n\rangle  \ | \ \mathbf{j} \in \mathbb{Z}_d^n \}, \nonumber
\end{align}

\noindent where $\mathbf{i}, \mathbf{j} \in \mathbb{Z}_d^n$ are vectors with inner product:

\begin{equation}
    \mathbf{i} \cdot \mathbf{j} = \sum_{k=1}^n i_k j_k \mod d,
\end{equation}

\noindent and entry-wise addition modulo $d$:

\begin{equation}
    \mathbf{i} + \mathbf{j} = (i_1 + j_1 \mod d, \ \ldots , i_n + j_n \mod d).
\end{equation}

\noindent We also write $\mathbf{d} = (d, d, \ldots, d)$ as a vector with $d$ in each entry. The inner products then take the form:

\begin{equation}
    \langle a_\mathbf{i} | b_\mathbf{j} \rangle = d^{-n/2} \omega^{\mathbf{i} \cdot \mathbf{j}}, \ \ \ \omega = e^{\frac{2 \pi i}{d}}.
\end{equation}

\noindent We write elements of the KD distribution as $Q(\mathbf{i}, \mathbf{j})$, where they take the explicit form:

\begin{equation}
    Q(\mathbf{i}, \mathbf{j}) = d^{-n} \sum_{\mathbf{k} \in \mathbb{Z}_d^n} \omega^{\mathbf{k} \cdot \mathbf{j} - \mathbf{i} \cdot \mathbf{j}} \langle \mathbf{i} | \psi \rangle \langle \psi | \mathbf{k} \rangle. \label{eq:kdfordft}
\end{equation}

\noindent The constraints of Equation \eqref{eq:kdconstraints} simplify into the following:

\begin{align}
    Q(\mathbf{u}, \mathbf{v}) &= d^{-n} \sum_{\mathbf{i, j} \in \mathbb{Z}_d^n} \omega^{\mathbf{i} \cdot \mathbf{v} - \mathbf{u} \cdot \mathbf{v} + \mathbf{u} \cdot \mathbf{j} - \mathbf{i} \cdot \mathbf{j}} Q^*(\mathbf{i, j}) \nonumber \\
    &= d^{-n} \sum_{\mathbf{i, j} \in \mathbb{Z}_d^n} \omega^{-(\mathbf{i} - \mathbf{u}) \cdot (\mathbf{j} - \mathbf{v})}  Q^*(\mathbf{i, j}) \nonumber \\
    & = d^{-n} \sum_{\mathbf{a, b} \in \mathbb{Z}_d^n} \omega^{-\mathbf{a} \cdot \mathbf{b}}  Q^*(\mathbf{u - a, v - b}),
\end{align}

\noindent where we have set $\mathbf{i - u = -a}$ and $\mathbf{j - v = -b}$ in the penultimate sum. This is a convolution of $Q^*(\mathbf{u, v})$ with a matrix $\Omega^*$, where $\Omega(\mathbf{a, b}) = d^{-n} \omega^{\mathbf{a \cdot b}} = d^{-n/2} [\mathrm{DFT}_d^{\otimes n}]_{\mathbf{a, b}}$ (here, the superscript $*$ denotes complex conjugation whereas $\star$ denotes a convolution). Taking complex conjugates of both sides, we may write:

\begin{equation}
    Q^*(\mathbf{u, v}) = \Omega \star Q(\mathbf{u, v}).
\end{equation}

\noindent The convolution theorem then implies that: 

\begin{equation}
    \widehat{Q^*}(\mathbf{x, y}) = \widehat{\Omega}(\mathbf{x, y})  \widehat{Q}(\mathbf{x, y}), \label{eq:convolution}
\end{equation}

\noindent where $\widehat{Q}(\mathbf{x, y})$ is the multi-dimensional, discrete Fourier transform of the KD distribution $Q(\mathbf{x, y})$. To obtain Equation \eqref{eq:thmkddft} we calculate $\widehat{\Omega}(\mathbf{x, y})$ and $\widehat{Q^*}(\mathbf{x, y})$:

\begin{align}
    \widehat{\Omega}(\mathbf{x, y}) &= d^{-n} \sum_{\mathbf{a, b} \in \mathbb{Z}_d^n} \omega^{\mathbf{a \cdot b}} \omega^{-\mathbf{a \cdot x - b \cdot y}} \nonumber \\
    &= d^{-n} \sum_{\mathbf{a} \in \mathbb{Z}_d^n} \omega^{-\mathbf{a \cdot x}} \sum_{\mathbf{b} \in \mathbb{Z}_d^n} \omega^{-\mathbf{b \cdot ( y - a)}} \nonumber \\
    &= d^{-n} \sum_{\mathbf{a} \in \mathbb{Z}_d^n} \omega^{-\mathbf{a} \cdot \mathbf{x}} (d^n \delta_{\mathbf{y, a}}) \nonumber \\
    &= \omega^{-\mathbf{x} \cdot \mathbf{y} } \label{eq:dftomega}.
\end{align}

\noindent Similarly, the DFT of $Q^*(\mathbf{a, b})$ is given by:

\begin{align}
    \widehat{Q^*}(\mathbf{x, y}) &= \sum_{\mathbf{a, b} \in \mathbb{Z}_d^n} \omega^{-\mathbf{a \cdot x - b \cdot y }} Q^*(\mathbf{a, b}) \nonumber \\
    &= \sum_{\mathbf{a, b} \in \mathbb{Z}_d^n} \omega^{\mathbf{a \cdot (d - x) + b \cdot (d - y) }} Q^*(\mathbf{a, b}) \nonumber \\ 
    &= \left( \sum_{\mathbf{a, b} \in \mathbb{Z}_d^n} \omega^{\mathbf{-a \cdot (d - x) - b \cdot (d - y) }} Q(\mathbf{a, b}) \right)^* \nonumber \\
    &= (\widehat{Q}(\mathbf{d - x, d - y}))^*. \label{eq:dftofkd}
\end{align}

\noindent Substituting Equations \eqref{eq:dftomega} and \eqref{eq:dftofkd} into \eqref{eq:convolution} gives us the following:

\begin{equation}
    (\widehat{Q}(\mathbf{d - x, d - y}))^* = \omega^{-\mathbf{x} \cdot \mathbf{y} } \widehat{Q}(\mathbf{x, y}) \label{eq:convolutionsubbedin},
\end{equation}

\noindent and the complex conjugate of Equation \eqref{eq:convolutionsubbedin} gives \eqref{eq:thmkddft}.
\end{proof}

Lemma \ref{thm:kddft} shows that $\widehat{Q}$ is determined by choosing one representative from each orbit of the involution $(x, y) \rightarrow (-x, -y)$. For example, if $n=1$ this simplifies to $\widehat{Q}(d - x, d -y ) = e^{\frac{2 \pi i xy}{d}} (\widehat{Q}(x, y))^*$ and the number of independent entries is approximately half of $d^2$.  For odd $d$, there is one fixed point $(0, 0)$, so one may choose $(d^2 + 1)/2$ representatives to describe $\widehat{Q}$. For even $d$, there are four fixed points $(0, 0), (d/2, 0), (0, d/2), (d/2, d/2)$, so one may choose $(d^2 + 4)/2$ representatives. Then, the remainder of the matrix can be reconstructed through taking complex conjugates of the entries and multiplication by the relevant phase factors. In this sense, the discrete Fourier transform `compresses' the KD distribution.

In addition to showing the self-similarity of $\widehat{Q}$,  we can relate this object to the discrete Wigner distribution. Here, it is convenient to use a slightly different convention to Equations \eqref{eq:thmkddft}, \eqref{eq:dftofkd}. We denote the multi-dimensional Fourier transform by $\tilde{Q}(\mathbf{a, b})$ instead. From Equation \eqref{eq:kdfordft}, this is given by:

\begin{align}
\tilde{Q}(\mathbf{a, b}) &= d^{-n} \sum_{\mathbf{i, j} \in \mathbb{Z}_d^n} \omega^{\mathbf{-i \cdot a + j \cdot b}}Q(\mathbf{i, j}) \nonumber \\
&= d^{-2n} \sum_{\mathbf{i, j, k} \in \mathbb{Z}_d^n} \omega^{\mathbf{-i \cdot a + j \cdot b + k \cdot j - i \cdot j}} \langle \mathbf{i} | \psi \rangle \langle \psi | \mathbf{k} \rangle \nonumber \\
&= d^{-2n} \sum_{\mathbf{i, k}\in \mathbb{Z}_d^n} \langle \mathbf{i} | \psi \rangle \langle \psi | \mathbf{k} \rangle \omega^{\mathbf{-i \cdot a}} \underbrace{\sum_\mathbf{j} \omega^{\mathbf{j \cdot (k + b -i)}}}_{\mathbf{k +b - i} =  \mathbf{0} \mod d} \nonumber \\
&= d^{-n} \sum_{\mathbf{k} \in \mathbb{Z}_d^n} \langle \mathbf{k + b} | \psi \rangle \langle \psi | \mathbf{k} \rangle \omega^{\mathbf{-a \cdot (k + b)}} \nonumber \\
&= d^{-n} \sum_{\mathbf{m} \in \mathbb{Z}_d^n} \langle \mathbf{m} | \psi \rangle \langle \psi | \mathbf{m-b} \rangle \omega^{\mathbf{-a \cdot m}} \nonumber \\
&= d^{-n} \text{Tr}\left( | \psi \rangle \langle \psi | (X^\mathbf{b})^\dagger (Z^{\mathbf{a}})^\dagger \right),
\end{align}

\noindent where all of the bitstring addition and subtraction is performed modulo $d$, and $Z^\mathbf{a}, X^\mathbf{b}$ are the Weyl--Heisenberg unitaries, defined for single qudits as:

\begin{align}
    Z &= \sum_{m=0}^{d-1} \omega^m |m \rangle \langle m |, & X &= \sum_{m=0}^{d-1} |m + 1 \rangle \langle m|,
\end{align}

\noindent and for $n$ qudits as:

\begin{align}
Z^\mathbf{a} & = Z^{a_1} \otimes \cdots \otimes Z^{a_n} = \sum_{\mathbf{m} \in \mathbb{Z}_d^n} \omega^{\mathbf{a \cdot m}} |\mathbf{m} \rangle \langle \mathbf{m} | \nonumber \\
X^\mathbf{b} &= X^{b_1} \otimes \cdots \otimes X^{b_n} = \sum_{\mathbf{m} \in \mathbb{Z}_d^n} |\mathbf{m + b} \rangle \langle \mathbf{m} |.
\end{align}

\noindent Once we take the complex conjugate of the final expression, the DFT of our KD distribution forms a matrix of expectation values of $| \psi \rangle $ with respect to the Weyl--Heisenberg unitaries, $ (\tilde{Q}(\mathbf{a, b}))^* = d^{-n} \text{Tr}\left( | \psi \rangle \langle \psi | Z^{\mathbf{a}} X^\mathbf{b} \right)$. We then multiply each entry of the resulting matrix by a phase $\omega^{-(\mathbf{a \cdot b}) \times 2^{-1}}$, where $2^{-1}$ is the multiplicative inverse modulo $d$ (for odd $d$). The resulting matrix $T(\mathbf{a, b})$ contains elements of `phased' Weyl--Heisenberg unitaries:

\begin{align}
    T(\mathbf{a, b}) &= \omega^{-(\mathbf{a \cdot b}) \times 2^{-1}}(\tilde{Q}(\mathbf{a, b}))^* \nonumber \\
    &= d^{-n} \text{Tr}\left( | \psi \rangle \langle \psi | \omega^{-(\mathbf{a \cdot b}) \times 2^{-1}} Z^{\mathbf{a}} X^\mathbf{b} \right).
\end{align}

\noindent Taking the symplectic Fourier transform of this matrix, we obtain a new matrix with elements:

\begin{widetext}
    
\begin{align}
    W(\mathbf{p, q}) &= d^{-n} \sum_{\mathbf{r, s} \in \mathbb{Z}^n_d} \omega^{\mathbf{p \cdot s - q \cdot r}} T(\mathbf{r, s}) \nonumber \\
    &= d^{-2n} \sum_{\mathbf{r, s} \in \mathbb{Z}^n_d}\omega^{\mathbf{p \cdot s - q \cdot r}} \text{Tr}\left( | \psi \rangle \langle \psi | \omega^{-(\mathbf{r \cdot s}) \times 2^{-1}} Z^{\mathbf{r}} X^\mathbf{s} \right)  \nonumber \\
    &= d^{-n} \text{Tr} \left( | \psi \rangle \langle \psi | \times d^{-n} \sum_{\mathbf{r, s} \in \mathbb{Z}^n_d} \omega^{\mathbf{p \cdot s - q \cdot r}} \left( \omega^{-(\mathbf{r \cdot s}) \times 2^{-1}} Z^{\mathbf{r}} X^\mathbf{s}\right) \right) \nonumber \\
    & = d^{-n} \text{Tr} \left( | \psi \rangle \langle \psi | A^{\mathbf{p, q}} \right), \label{eq:kdtowigner}
\end{align}

\end{widetext}

\noindent which for odd $d$ are precisely the elements of the discrete Wigner distribution \cite{Gross_2006, zurel2024}, i.e. expectation values of $| \psi \rangle$ with respect to the phase point operators $A^{\mathbf{p, q}}$ (defined implicitly in the last line of \eqref{eq:kdtowigner}). Performing the inverse of the above similarly allows one to start with a Wigner distribution and obtain the KD distribution for $V = \mathrm{DFT}_d^{\otimes n}$. Thus, the two representations of quantum states are related by an invertible sequence of DFTs and diagonal phase multiplications. This was previously shown for the continuous case in \cite{OCONNELL19859, OConnell1986}.

\section{Bounds on KD Quasiprobabilities}
\label{section:kdbounds}
Here, we offer some general bounds on the magnitude of Kirkwood--Dirac quasiprobabilities, which may be of independent interest. The first result is a general bound on the magnitude of quasiprobabilities in any informationally complete KD distribution. To this end, we define $m_{\mathcal{A,B}}$ and $M_{\mathcal{A,B}}$ as the smallest and largest absolute values of inner products between $\mathcal{A}, \mathcal{B}$:

\begin{align}
    m_\mathcal{A,B} &= \min_{(i, j) \in \llbracket d \rrbracket^2} |\langle a_i | b_j \rangle|, & M_\mathcal{A,B} &= \max_{(i, j) \in \llbracket d \rrbracket^2} |\langle a_i | b_j \rangle|. \label{eq:mmbounds}
\end{align}

\noindent These relate as $0 \leq m_\mathcal{A,B} \leq 1 / \sqrt{d} \leq M_{\mathcal{A,B}} \leq 1$ \cite{De_Bievre_2021}. We will also make use of the support uncertainties $n_\mathcal{A}, n_\mathcal{B}$, defined in Equation \eqref{eq:supunc}. 

\begin{lemma} \label{lemma:bounds}
    The magnitudes of all Kirkwood--Dirac quasiprobabilities $Q(\psi)_{ij}$ for a pure state $| \psi \rangle$ with support uncertainties $(n_\mathcal{A}, n_\mathcal{B})$ are upper bounded as:
    \begin{equation}
        |Q(\psi)_{ij}| \leq M_{\mathcal{A,B}}^3 \sqrt{n_\mathcal{A} n_\mathcal{B}}. \label{eq:kdupperbound}
    \end{equation}
    For KD-positive distributions, we also have the lower bound on the non-zero elements $Q(\psi)_{ij}$:
    \begin{equation}
        m_{\mathcal{A,B}}^3 \leq Q(\psi)_{ij}.
    \end{equation}
\end{lemma}
\begin{proof}

\noindent We will make use of a pair of projectors onto the support of $| \psi \rangle$ in $\mathcal{A}$ and $\mathcal{B}$:

\begin{align}
\Pi_\mathcal{A} &= \sum_{|a_i\rangle \in S_\mathcal{A}} |a_i \rangle \langle a_i |, & \Pi_\mathcal{B} &= \sum_{|b_j\rangle \in S_\mathcal{B}} |b_j \rangle \langle b_j |.
\end{align}

\noindent Clearly, the rank of $\Pi_\mathcal{A}$ is $n_\mathcal{A}$, the rank of $\Pi_\mathcal{B}$ is $n_\mathcal{B}$, and $\Pi_\mathcal{A} | \psi \rangle = \Pi_\mathcal{B} | \psi \rangle = | \psi \rangle$. Inserting these into the expression for $|Q_{ij}|$, we get:

\begin{align}
    |Q&(\psi)_{ij}| = |\langle b_j | a_i \rangle |  | \langle a_i | \Pi_\mathcal{B} | \psi \rangle | | \langle \psi | \Pi_\mathcal{A} | b_j \rangle | \nonumber \\
    &= | \langle b_j | a_i \rangle | \left| \sum_{|b_k\rangle \in S_\mathcal{B}} \langle a_i | b_k \rangle \langle b_k | \psi \rangle \right|  \left| \sum_{|a_l\rangle \in S_\mathcal{A}} \langle \psi | a_l \rangle \langle a_l | b_j \rangle \right| \nonumber \\
    &\leq M_{\mathcal{A,B}}^3 \left(  \sum_{|b_k\rangle \in S_\mathcal{B}} | \langle b_k | \psi \rangle | \right) \left( \sum_{|a_l\rangle \in S_\mathcal{A}} | \langle \psi | a_l \rangle | \right)\nonumber \\
    & \leq M_{\mathcal{A,B}}^3 \sqrt{n_\mathcal{A} n_\mathcal{B}}, 
\end{align}

\noindent where we have used the triangle inequality and definition of $M_\mathcal{A,B}$ in the third line, and the Cauchy–Schwarz inequality in the fourth. 

For the lower bound, we use an argument similar to \cite{De_Bievre_2021}. Using the global phase invariance of the KD distribution, we can adjust the phases and indices of the basis vectors in $\mathcal{A, B}$ so that $\langle a_i | \psi \rangle$ and $\langle b_j | \psi \rangle$ are real and positive for $i,j \in \llbracket n_\mathcal{A} \rrbracket \times \llbracket n_\mathcal{B} \rrbracket$ and $0$ otherwise. It follows that if the KD distribution is positive, the inner products $\langle a_i | b_j \rangle$ are real and non-negative for $|a_i \rangle \in S_\mathcal{A}$, $|b_j \rangle \in S_\mathcal{B}$. We can thus write:

\begin{align}
Q(\psi)_{ij} &=  \langle b_j | a_i \rangle  \sum_{|b_k\rangle \in S_\mathcal{B}} \langle a_i | b_k \rangle \langle b_k | \psi \rangle  \sum_{|a_l\rangle \in S_\mathcal{A}} \langle \psi | a_l \rangle \langle a_l | b_j \rangle  \nonumber \\
& \geq m_{\mathcal{A, B}}^3 \left( \sum_{|b_k\rangle \in S_\mathcal{B}} \langle b_k | \psi \rangle \right) \left( \sum_{|a_l\rangle \in S_\mathcal{A}} \langle \psi | a_l \rangle  \right) \nonumber \\
& \geq m_{\mathcal{A, B}}^3,
\end{align}

\noindent where the last line follows from the $\ell_1$-$\ell_2$ norm inequality $\| |\psi \rangle \|_1 \geq \| |\psi \rangle \|_2$. 
\end{proof}

The upper bound given in Lemma \ref{lemma:bounds} generically constrains the maximum attainable magnitudes of pure state quasiprobabilities in any (including informationally incomplete) KD distribution in terms of the value of $M_{\mathcal{A, B}}$ of the distribution, as well as the support size $n_\mathcal{A} n_\mathcal{B}$ of the state. Recent work characterising Bargmann invariants \cite{PhysRevA.111.022409, PhysRevA.111.042417, hsnv-wpt3, Xu_2026} (which KD quasiprobabilities form an instance of) implies that when more fine-grained information about $Q(\rho)_{ij}$ is known -- namely, the phase $\theta = \arg(Q(\rho)_{ij}) \in [0, 2\pi)$ -- a more precise bound may also be constructed: 

\begin{equation}
|Q(\psi)_{ij}| \leq \cos^3 \left( \frac{\pi}{3} \right) \sec^3 \left( \frac{\theta - \pi}{3} \right). \label{eq:bargmannupperbound}
\end{equation}

\noindent Thus, the bounds \eqref{eq:kdupperbound} and \eqref{eq:bargmannupperbound} are complementary, and may be used at will depending on information available about $Q(\rho)$. An example of their consequence is given below:

\begin{remark}
Let $Q(\psi)$ be a Kirkwood--Dirac distribution of a pure state defined on a pair of MUBs. If $\mathcal{N}(Q(\psi)) = 1$, then all its non-zero quasiprobabilities are uniformly $Q_{ij} = 1/d$. \label{thm:mubuniform}
\end{remark}
\begin{proof}
Here $M_{\mathcal{A, B}} = 1/\sqrt{d}$, and for MUBs all pure KD-positive states satisfy $n_\mathcal{A} n_\mathcal{B} = d$. Substituting these into Equation \eqref{eq:kdupperbound} we get $|Q(\psi)_{ij}| = Q(\psi)_{ij} \leq 1/d$. The number of non-zero $Q(\psi)_{ij}$ is also given by $n_\mathcal{A} n_\mathcal{B} = d$. Because the quasiprobabilities must also sum to $1$, this can all be satisfied only if \eqref{eq:kdupperbound} is saturated, so $Q(\psi)_{ij} = 1/d$ or $0$.
\end{proof}

Remark \ref{thm:mubuniform} further implies that all non-zero amplitudes of KD-positive states on MUBs satisfy $|\langle \psi | a_i \rangle| = \sqrt{1/n_{\mathcal{A}}}$ and $|\langle \psi | b_j \rangle| = \sqrt{1/n_{\mathcal{B}}}$. Because we can write the inner product between any two pure states $| \psi \rangle, | \phi \rangle$ as:

\begin{equation}
    |\langle  \psi | \phi \rangle|^2 = \sum_{ij} \frac{Q^*_{ij}(\psi) Q_{ij}(\phi)}{|\langle a_i | b_j \rangle|^2},
\end{equation}

\noindent the inner product of any two KD-positive states on MUBs satisfies:

\begin{equation}
    |\langle \psi | \phi \rangle| = \sqrt{\frac{k}{d}}, \label{eq:kdmub_innerproduct}
\end{equation}

\noindent where the integer $k$ is the overlap of the support of their distributions. Note that the above holds true for all MUBs, not just the case of $V = \mathrm{DFT}_d$.

Taking a MUB system of $n$ qudits, Remark \ref{thm:mubuniform} tells us that every pure KD-positive state is a uniform mixture on its support with the non-zero quasiprobabilities reading $1/d^n$. In particular, this implies that the distribution becomes exponentially large, with exponentially small probabilities. In the discrete Wigner distribution setting, the exact same property is satisfied by the stabilizer states, i.e. all pure Wigner-positive states \cite{Gross_2013, Gross_2021}. In a recent result Gross, Nezami \& Walter \cite{Gross_2021} have used this to devise an efficient property test for whether an unknown state is a stabilizer state. Whether such tests exist for the $V = \mathrm{DFT}_d^{\otimes n}$ KD distribution (or in general) remains an open question. Finally, we note that Equation \eqref{eq:kdmub_innerproduct} closely resembles Theorem 11 of \cite{garcia2017geometrystabilizerstates} satisfied by stabilizer states.

\section{Proofs for Type III Unitaries} \label{app:type3proofs}
\subsection{Extended Proof of Theorem \ref{thm:ustar}} \label{section:ustarproof}

Here, we prove Theorem \ref{thm:ustar} for the general case of $V = \mathrm{DFT}_d$ where $d = pq$ for any pair of distinct primes $p, q$. 

\begin{proof} The strategy is the same as in the $d=6$ case. We write $|a_i \rangle \in \mathcal{A}$, $|b_j \rangle \in \mathcal{B}$, $|x, y \rangle_{p, q} \in \mathcal{C}_{p,q}$ and $|u, v \rangle_{q, p} \in \mathcal{C}_{q,p}$. The linear decomposition of any $\rho$ such that $Q(\rho)_{ij} \in \mathbb{R}$ (established in \cite{Xu_2025}) then becomes:

\begin{align} 
    \rho &= \sum_{i}  \alpha_i |a_i \rangle \langle a_i| + \sum_{j}  \beta_j |b_j \rangle \langle b_j|  \\ 
    & + \ \sum_{x, y}  \gamma_{x, y} |{x, y} \rangle \langle {x, y}|_{p, q} +  \sum_{u, v}  \varepsilon_{u, v} |{u, v} \rangle \langle {u, v}|_{q, p} .
\end{align}

\noindent Similarly, by linearity of \eqref{eq:kddef} we may write $Q(\rho)$ as a linear combination of the pure state KD distributions:

\begin{align} 
    Q(\rho) &= \sum_{i}  \alpha_i Q(a_i) + \sum_{j}  \beta_j Q(b_j)  \\
    &  + \ \sum_{x, y}  \gamma_{x, y} Q((x, y)_{p, q}) +  \sum_{u, v}  \varepsilon_{u, v} Q((u, v)_{q, p}).
\end{align}

\noindent From Equation \eqref{eq:kddef} and Equation \eqref{eq:dft_inner_product} the pure state distributions can be shown to satisfy:

\begin{align}
Q(a_i)_{kl} &= d^{-1} \delta_{k i}, \\
Q(b_j)_{kl} &= d^{-1} \delta_{l j}, \\
Q((x, y)_{p, q})_{kl} &= d^{-1} \delta_{x, k \bmod p} \ \delta_{y, l \bmod q}, \\
Q((u, v)_{q, p})_{kl} &= d^{-1} \delta_{u, k \bmod q} \ \delta_{v, l \bmod p}.
\end{align}

\noindent Therefore, assuming $\rho \in \mathcal{E}_{\mathrm{KD}+}$, we get $d^2$ linear equations of the form:

\begin{equation}
Q(\rho)_{kl} = \frac{1}{d}\left( \alpha_k+\beta_l+\gamma_{ k\bmod p, l\bmod q}
+\varepsilon_{k\bmod q, l\bmod p} \right).
\end{equation}

\noindent Recall that $U_\star = P^\dagger (\mathbb{1}_p \otimes \mathrm{DFT}_q) P$. We know from Lemma \ref{cor:compositefourier} that $U_\star | a_i \rangle \propto |i \bmod p, p \, i \bmod q \rangle_{p, q}$ and $U_\star | b_j \rangle \propto |- \bar{p} \, j \bmod q,  j \bmod p \rangle_{q, p}$, where $\bar{p} = p^{-1} \bmod q$ and $\bar{q} = q^{-1} \bmod p$. We can explicitly calculate the action of $U_\star$ on $|x, y \rangle_{p, q}$:

\begin{align}
P|x, y \rangle_{p, q} &= \frac{1}{\sqrt{q}} \sum_{k=0}^{q-1} e^{\frac{2 \pi i}{q} k y} P|a_{kp + x} \rangle \\
& = \frac{1}{\sqrt{q}} \sum_{k=0}^{q-1} e^{\frac{2 \pi i}{q} k y} |kp + x \bmod p \rangle \otimes |kp + x \bmod q \rangle  \\
& = |x \rangle \otimes  \frac{1}{\sqrt{q}} \sum_{k=0}^{q-1} e^{\frac{2 \pi i}{q} k y} |k p + x \bmod q \rangle.
\end{align}

\noindent Now set $t = kp + x \bmod q$, so that $k = \bar{p} \, (t - x) \bmod q$. Reindexing the sum, we get:

\begin{align}
P|x, y \rangle_{p, q} &= |x \rangle \otimes  \frac{1}{\sqrt{q}} \sum_{t=0}^{q-1} e^{\frac{2 \pi i}{q} y \bar{p}(t - x)} |t \rangle \\
& \propto |x \rangle \otimes \mathrm{DFT}_q |\bar{p} \, y \rangle.
\end{align}

\noindent Hence, 

\begin{align}
U_\star |x, y \rangle_{p, q} & = P^\dagger (\mathbb{1}_p \otimes \mathrm{DFT}_q) P |x, y \rangle_{p, q}  \\
& \propto P^\dagger (|x \rangle \otimes \mathrm{DFT}^2_q |\bar{p} \, y \rangle) \\
& \propto P^\dagger (|x \rangle \otimes |- \bar{p} \, y \rangle) \\
& \propto |a_{x \, q \bar{q} - \bar{p} \, y \, p \bar{p}} \rangle,
\end{align}

\noindent where $x \, q \bar{q} + (- \bar{p} \,y) \, p \bar{p} \mod d$ is of the form in Equation \eqref{eq:CRT}. Similarly for states in $\mathcal{C}_{q, p}$:

\begin{align}
P|u, v \rangle_{q, p} &= \frac{1}{\sqrt{p}} \sum_{k=0}^{p-1} e^{\frac{2 \pi i}{p} k v} P|a_{kq + u} \rangle \\
& = \frac{1}{\sqrt{p}} \sum_{k=0}^{p-1} e^{\frac{2 \pi i}{p} k v} |kq + u \bmod p \rangle \otimes |kq + u \bmod q \rangle  \\
& =  \left( \frac{1}{\sqrt{p}} \sum_{k=0}^{p-1} e^{\frac{2 \pi i}{p} k v} |kq + u \bmod p \rangle  \right)\otimes |u \rangle \\
& \propto \mathrm{DFT}_p |\bar{q} \, v \rangle \otimes |u \rangle
\end{align}

\noindent So that by Equation \eqref{eq:Ustarb}:

\begin{align}
    U_\star |u, v \rangle_{q, p} &\propto  P^\dagger (\mathrm{DFT}_p |\bar{q} \, v \rangle \otimes \mathrm{DFT}_q |u \rangle) \\
    &\propto |b_{v \, q \bar{q} + p \, u \, p \bar{p}} \rangle .
\end{align}

In summary, we have:

\begin{align}
&U_\star |a_i \rangle \propto |i \bmod p, p \, i \bmod q \rangle_{p, q}, \\
&U_\star |b_j \rangle \propto |- \bar{p} \, j \bmod q,  j \bmod p \rangle_{q, p}, \\
&U_\star |x, y \rangle_{p, q} \propto |a_{x \, q \bar{q} - \bar{p} \, y \, p \bar{p}} \rangle, \\
&U_\star |u, v \rangle_{q, p} \propto |b_{v \, q \bar{q} + p \, u \, p \bar{p}} \rangle.
\end{align}

\noindent Hence, 

\begin{align}
Q(U_\star \rho U_\star^\dagger) &= \sum_{i}  \alpha_i Q((i \bmod p, p \, i \bmod q )_{p, q})  \\
& \ +\sum_{j}  \beta_j Q((- \bar{p} \, j \bmod q,  j \bmod p )_{q, p})  \\
&  + \sum_{x,y} \gamma_{x, y} Q(a_{x \, q \bar{q} - \bar{p} \, y \, p \bar{p}})  \\
& + \sum_{u,v} \varepsilon_{u, v} Q(b_{v \, q \bar{q} + p \, u \, p \bar{p}}).
\end{align}

\noindent Therefore, the entries of $Q(U_\star \rho U_\star^\dagger)$ are given by:

\begin{align}
Q(U_\star \rho &U_\star^\dagger)_{kl} \\
&= \frac{1}{d} \sum_i \alpha_i \delta_{ i\bmod p, k\bmod p} \; \delta_{p \, i \bmod q, l\bmod q} \\
& + \frac{1}{d}\sum_j \beta_j  \delta_{-\bar p \, j\bmod q, k\bmod q} \; \delta_{j\bmod p, l\bmod p} \\
& + \frac{1}{d}\sum_{x,y}\gamma_{x,y} 
\delta_{k,x \,q \bar q- \bar p \, y \, p\bar p} \\
& + \frac{1}{d} \sum_{u,v}\varepsilon_{u,v} 
\delta_{l, v\, q\bar q +p\, u \, p\bar p}.
\end{align}

\noindent We may simplify the sum into the following:

\begin{align}
Q(U_\star \rho U_\star^\dagger)_{kl} &= \frac{1}{d} ( \alpha_{k'} + \beta_{l'} + \gamma_{k \bmod p, -p \, k \bmod q} \\
&+ \ \varepsilon_{\bar{p} \, l \bmod q, l \bmod p} ),
\end{align}

\noindent where we have set:

\begin{align}
k' &= (k \bmod p) q \bar{q} + ( \bar{p} \, l \bmod q) p \bar{p} \label{eq:klmapping}\\
l' &= (l \bmod p) q \bar{q} + ( - p \, k \bmod q) p \bar{p}.
\end{align}

\noindent Our choice of indices $(k', l')$ satisfies:

\begin{align}
k' \bmod p &= k \bmod p, & k' \bmod q &= \bar{p} \, l \bmod q, \\
l' \bmod p &= l \bmod p, & l' \bmod q &= - p \, k \bmod q.
\end{align}

\noindent Giving $Q(U_\star \rho U_\star^\dagger)_{kl} = Q(\rho)_{k' l'}$. Therefore, conjugating $\rho$ by $U_\star$ corresponds to a permutation of the entries of $Q(\rho)$ as $(k, l) \rightarrow (k', l')$ via the bijective mapping defined in Equations \eqref{eq:klmapping}. Assuming $\rho \in \mathcal{E}_{\mathrm{KD}+}$, we have $Q(\rho)_{kl} \geq 0$ for all $k, l$, even if some of the coefficients $(\alpha, \beta, \gamma, \varepsilon)$ are negative. Therefore, $Q(U_\star \rho U_\star^\dagger)_{kl} \geq 0$ for all $k, l$. Hence, $U_\star \rho U_\star^\dagger \in \mathcal{E}_{\mathrm{KD}+}$, and the KD positivity of $\rho$ is preserved under conjugation by $U_\star$. Finally, note that for our proof it does not matter whether $\mathcal{E}_{\mathrm{KD}+} = \mathrm{conv}(\mathcal{E}^{\mathrm{pure}}_{\mathrm{KD}+})$ holds, or not. For $d = pq \neq 6$, to the best of our knowledge, this remains an open question. Similarly to Lemma \ref{cor:compositefourier}, by symmetry the above also applies to $U'_\star = P^\dagger (\mathrm{DFT}_p \otimes \mathbb{1}_q) P$, which is another type \textbf{III} unitary.
\end{proof}

\subsection{Multi-qudit Systems} \label{section:multiqudit}
We now prove that Theorem \ref{thm:ustar} generalises to the multi-qudit case of $V = \mathrm{DFT}_{pq}^{\otimes n}$. Although for such distributions there may exist positivity-preserving unitaries beyond types \textbf{I}, \textbf{II} and \textbf{III}, here we show that the unitary $U_\star^{\otimes{n}}$ acts covariantly and preserves the total non-positivity for any KD-real distribution $Q(\rho)$, and so Theorem \ref{thm:resourcetheory} remains valid for $n > 1$. 

\begin{theorem}
For $V = \mathrm{DFT}^{\otimes n}_{pq}$, the unitary $U^{\otimes n}_\star$ acts covariantly on any KD-real distribution $Q(\rho)$. As a consequence, it is positivity-preserving. 
\end{theorem}

\begin{proof}

We begin similarly to Lemma \ref{cor:compositefourier}, and re-write the action of $U_\star^{\otimes{n}}$ on the state $\rho$ in the $V = \mathrm{DFT}_{pq}^{\otimes n}$ distribution as the action by $\tilde{U}_\star = \mathbb{1}^{\otimes n}_p \otimes \mathrm{DFT}_q^{\otimes n}$ on an identical distribution obtained by a global change of basis by $P^{\otimes n}$. In the new distribution, we write the basis states as $\mathcal{A} = \{ |\mathbf{i}_p \rangle \otimes |\mathbf{i}_q \rangle \}$ and $\mathcal{B} = \{ \mathrm{DFT}^{\otimes n}_p |\bar{q} \, \mathbf{j}_p\rangle \otimes \mathrm{DFT}^{\otimes n}_q |\bar{p} \, \mathbf{j}_q\rangle \}$, where:

\begin{align}
|\mathbf{i}_p \rangle &= |(i_1)_p \rangle \otimes \cdots \otimes |(i_n)_p \rangle \in (\mathbb{C}^p)^{\otimes n}, \\
|\bar{p} \mathbf{j}_q \rangle &= |\bar{p} (j_1)_q \rangle \otimes \cdots \otimes |\bar{p} (j_n)_q \rangle \in (\mathbb{C}^q)^{\otimes n},
\end{align} 

\noindent and similarly for $|\mathbf{i}_q \rangle$, $|\bar{q} \mathbf{j}_p \rangle$. The transformation under $\tilde{U}_\star$ straightforwardly generalises from Equation \eqref{eq:ustartransform}:

\begin{equation}
    Q_{\mathbf{i}_p, \mathbf{i}_q ; \mathbf{j}_p, \mathbf{j}_q}  \rightarrow \frac{1}{q^n}\sum_{\mathbf{s}_q, \mathbf{t}_q \in \mathbb{Z}_q^n} e^{\frac{2 \pi i}{q} (\mathbf{i}_q + \bar{p} \mathbf{t}_q) \cdot (\mathbf{s}_q - \bar{p} \mathbf{j}_q)} Q_{\mathbf{i}_p, \mathbf{s}_q; \mathbf{j}_p, \mathbf{t}_q}. \label{eq:ustartransform_multiqudit}
\end{equation}

Next, we make use of general constraints on quasiprobability distributions from Appendix \ref{section:kdconvolutions}. We assume that the quasiprobability distribution is KD-real, so that $Q_{\mathbf{i}_p, \mathbf{i}_q ; \mathbf{j}_p, \mathbf{j}_q}$ is equal to its complex conjugate. Equation \eqref{eq:kdconstraints} then takes on the form:

\begin{equation}
    Q_{\mathbf{u}_p, \mathbf{u}_q ; \mathbf{v}_p, \mathbf{v}_q} = \sum_{\mathbf{i}_p, \mathbf{j}_p \in \mathbb{Z}_p^n} \sum_{\mathbf{i}_q, \mathbf{j}_q \in \mathbb{Z}_q^n} (\hat{\mathcal{K}})_{\mathbf{u}_p, \mathbf{u}_q ; \mathbf{v}_p, \mathbf{v}_q}^{\mathbf{i}_p, \mathbf{i}_q ; \mathbf{j}_p, \mathbf{j}_q} Q_{\mathbf{i}_p, \mathbf{i}_q ; \mathbf{j}_p, \mathbf{j}_q},
\end{equation}

\noindent where $\hat{\mathcal{K}}$ is a superoperator matrix on $\mathbb{C}^{d^{2n} \times d^{2n}}$ with entries:
\begin{align}
(\hat{\mathcal{K}})_{\mathbf{u}_p, \mathbf{u}_q ; \mathbf{v}_p, \mathbf{v}_q}^{\mathbf{i}_p, \mathbf{i}_q ; \mathbf{j}_p, \mathbf{j}_q} &= \frac{1}{d^n} e^{\frac{2 \pi i}{p} (\mathbf{v}_p - \mathbf{j}_p) \cdot (\mathbf{i}_p - \mathbf{u}_p)} e^{\frac{2 \pi i}{q} (\mathbf{v}_q - \mathbf{j}_q) \cdot (\mathbf{i}_q - \mathbf{u}_q)} \\
& =  \frac{1}{p^n} e^{\frac{2 \pi i}{p} (\mathbf{v}_p - \mathbf{j}_p) \cdot (\mathbf{i}_p - \mathbf{u}_p)} \frac{1}{q^n}  e^{\frac{2 \pi i}{q} (\mathbf{v}_q - \mathbf{j}_q) \cdot (\mathbf{i}_q - \mathbf{u}_q)} \\
&= (\hat{\mathcal{K}}^{(p)})_{\mathbf{u}_p, \mathbf{v}_p}^{\mathbf{i}_p, \mathbf{j}_p} \otimes (\hat{\mathcal{K}}^{(q)})_{\mathbf{u}_q, \mathbf{v}_q}^{\mathbf{i}_q, \mathbf{j}_q}.
\end{align}

\noindent Taking $Q_{\mathbf{i}_p, \mathbf{i}_q ; \mathbf{j}_p, \mathbf{j}_q}$ as a real-valued vector in $\mathbb{C}^{d^{2n}}$, we can write the above in the vectorised picture:

\begin{equation}
\hat{\mathcal{K}} \boldsymbol{|} Q \boldsymbol{\rangle \rangle} = \hat{\mathcal{K}}^{(p)} \otimes \hat{\mathcal{K}}^{(q)} \boldsymbol{|} Q \boldsymbol{\rangle \rangle} = \boldsymbol{|} Q \boldsymbol{\rangle \rangle}. \label{eq:kd_eigenvector}
\end{equation}

\noindent Therefore, by our assumption that $Q(\rho)$ is KD-real we have obtained a linear eigenvector equation \eqref{eq:kd_eigenvector} for the quasiprobability distribution vector. This implies that any KD-real $\boldsymbol{|} Q \boldsymbol{\rangle \rangle}$ resides in the $+1$ eigenspace of $\hat{\mathcal{K}}$, and we may isolate its eigenvectors to obtain the general form of KD-real distributions. Furthermore, since $\hat{\mathcal{K}}$ factorises into a tensor product of two superoperators, the $+1$ eigenspace of $\hat{\mathcal{K}}$ is spanned by the tensor products of eigenspaces of $\hat{\mathcal{K}}^{(p)}$ and $\hat{\mathcal{K}}^{(q)}$ with reciprocal eigenvalues. 

We next find the eigenbasis for $\hat{\mathcal{K}}^{(p)}$. Due to the tensor product structure, its eigenvectors reside in $\mathbb{C}^{p^{2n}}$. Consider the vectors $\boldsymbol{|} \phi_{\mathbf{l}_p, \mathbf{m}_p} \boldsymbol{\rangle \rangle}$ with entries: 

\begin{equation}
\boldsymbol{\langle \langle } \mathbf{i}_p, \mathbf{j}_p \boldsymbol{|} \phi_{\mathbf{l}_p, \mathbf{m}_p} \boldsymbol{\rangle \rangle} = \frac{1}{p^n} e^{\frac{2 \pi i}{p} (\mathbf{i}_p \cdot \mathbf{l}_p + \mathbf{j}_p \cdot \mathbf{m}_p)}.
\end{equation}

\noindent There are $p^n \times p^n$ such vectors; as they are orthonormal, they form a basis for $\mathbb{C}^{p^{2n}}$:

\begin{align}
\boldsymbol{\langle \langle } \phi_{\mathbf{l}_p, \mathbf{m}_p} \boldsymbol{|} \phi_{\mathbf{l}'_p, \mathbf{m}'_p} \boldsymbol{\rangle \rangle} &= \frac{1}{p^{2n}} \sum_{\mathbf{i}_p, \mathbf{j}_p} e^{\frac{2 \pi i}{p} (\mathbf{i}_p \cdot (\mathbf{l}'_p - \mathbf{l}_p) + \mathbf{j}_p \cdot (\mathbf{m}'_p - \mathbf{m}_p))}  \\
&= \frac{1}{p^{2n}} \sum_{\mathbf{i}_p} e^{\frac{2 \pi i}{p} \mathbf{i}_p \cdot (\mathbf{l}'_p - \mathbf{l}_p)} \sum_{\mathbf{j}_p} e^{\frac{2 \pi i}{p} \mathbf{j}_p \cdot (\mathbf{m}'_p - \mathbf{m}_p)} \\
&= \delta_{\mathbf{l}_p, \mathbf{l}'_p} \delta_{\mathbf{m}_p, \mathbf{m}'_p}.
\end{align}

\noindent Each vector $\boldsymbol{|} \phi_{\mathbf{l}_p, \mathbf{m}_p} \boldsymbol{\rangle \rangle}$ forms an eigenvector of $\hat{\mathcal{K}}^{(p)}$:

\begin{align}
& \hat{\mathcal{K}}^{(p)} \boldsymbol{|} \phi_{\mathbf{l}_p, \mathbf{m}_p} \boldsymbol{\rangle \rangle}  = \sum_{\mathbf{i}_p, \mathbf{j}_p} \boldsymbol{\langle \langle } \mathbf{i}_p, \mathbf{j}_p \boldsymbol{|} \phi_{\mathbf{l}_p, \mathbf{m}_p} \boldsymbol{\rangle \rangle} \hat{\mathcal{K}}^{(p)} \boldsymbol{|} \mathbf{i}_p, \mathbf{j}_p \boldsymbol{\rangle \rangle}  \\
& = \sum_{\mathbf{u}_p, \mathbf{v}_p}  \sum_{\mathbf{i}_p, \mathbf{j}_p} \boldsymbol{\langle \langle } \mathbf{i}_p, \mathbf{j}_p \boldsymbol{|} \phi_{\mathbf{l}_p, \mathbf{m}_p} \boldsymbol{\rangle \rangle} (\hat{\mathcal{K}}^{(p)})_{\mathbf{u}_p, \mathbf{v}_p}^{\mathbf{i}_p, \mathbf{j}_p} \boldsymbol{|} \mathbf{u}_p, \mathbf{v}_p \boldsymbol{\rangle \rangle}  \\
&= \frac{1}{p^{2n}} \sum_{\mathbf{u}_p, \mathbf{v}_p} \sum_{\mathbf{i}_p, \mathbf{j}_p} e^{\frac{2 \pi i}{p} ( (\mathbf{v}_p - \mathbf{j}_p) \cdot (\mathbf{i}_p - \mathbf{u}_p) + (\mathbf{i}_p \cdot \mathbf{l}_p + \mathbf{j}_p \cdot \mathbf{m}_p))} \boldsymbol{|} \mathbf{u}_p, \mathbf{v}_p \boldsymbol{\rangle \rangle} \\
&= \frac{1}{p^{2n}} \sum_{\mathbf{u}_p, \mathbf{v}_p} \sum_{\mathbf{i}_p, \mathbf{j}_p} e^{\frac{2 \pi i}{p} (\mathbf{i}_p \cdot (\mathbf{l}_p  + \mathbf{v}_p - \mathbf{j}_p) - \mathbf{u}_p \cdot (\mathbf{v}_p - \mathbf{j}_p) + \mathbf{j}_p \cdot \mathbf{m}_p )} \boldsymbol{|} \mathbf{u}_p, \mathbf{v}_p \boldsymbol{\rangle \rangle} \\
&= \frac{1}{p^n} \sum_{\mathbf{u}_p, \mathbf{v}_p} \sum_{\mathbf{j}_p} \delta_{\mathbf{j}_p, \mathbf{l}_p + \mathbf{v}_p} e^{\frac{2 \pi i}{p} (\mathbf{j}_p \cdot (\mathbf{m}_p + \mathbf{u}_p) - \mathbf{u}_p \cdot \mathbf{v}_p)} \boldsymbol{|} \mathbf{u}_p, \mathbf{v}_p \boldsymbol{\rangle \rangle} \\
&= \frac{1}{p^n} \sum_{\mathbf{u}_p, \mathbf{v}_p} e^{\frac{2 \pi i}{p} (\mathbf{l}_p \cdot \mathbf{m}_p + \mathbf{u}_p \cdot \mathbf{l}_p + \mathbf{v}_p \cdot \mathbf{m}_p)} \boldsymbol{|} \mathbf{u}_p, \mathbf{v}_p \boldsymbol{\rangle \rangle} \\
&= e^{\frac{2 \pi i}{p} \mathbf{l}_p \cdot \mathbf{m}_p} \boldsymbol{|} \phi_{\mathbf{l}_p, \mathbf{m}_p} \boldsymbol{\rangle \rangle}.  
\end{align}

\noindent Similarly, for $\hat{\mathcal{K}}^{(q)}$ each vector $\boldsymbol{|} \psi_{\mathbf{l}_q, \mathbf{m}_q} \boldsymbol{\rangle \rangle}$ with entries $\boldsymbol{\langle \langle } \mathbf{i}_q, \mathbf{j}_q \boldsymbol{|} \psi_{\mathbf{l}_q, \mathbf{m}_q} \boldsymbol{\rangle \rangle} = q^{-n} e^{\frac{2 \pi i}{q} (\mathbf{i}_q \cdot \mathbf{l}_q + \mathbf{j}_q \cdot \mathbf{m}_q)}$ is an eigenvector with eigenvalue $e^{\frac{2 \pi i}{q} \mathbf{l}_q \cdot \mathbf{m}_q}$. Therefore, the $+1$ eigenspace of $\hat{\mathcal{K}}$ is spanned by the tensor products $\boldsymbol{|} \phi_{\mathbf{l}_p, \mathbf{m}_p} \boldsymbol{\rangle \rangle} \otimes \boldsymbol{|} \psi_{\mathbf{l}_q, \mathbf{m}_q} \boldsymbol{\rangle \rangle}$, such that:

\begin{equation}
e^{\frac{2 \pi i}{p} \mathbf{l}_p \cdot \mathbf{m}_p} e^{\frac{2 \pi i}{q} \mathbf{l}_q \cdot \mathbf{m}_q} = 1.
\end{equation}

Since $p$ and $q$ are coprime, this condition is equivalent to requiring that $\mathbf{l}_p \cdot \mathbf{m}_p = 0 \bmod p$ and $\mathbf{l}_q \cdot \mathbf{m}_q = 0 \bmod q$. Therefore, a $+1$ eigenvector of $\hat{\mathcal{K}}$ is a tensor product of $+1$ eigenvectors of $\hat{\mathcal{K}}^{(p)}$ and $\hat{\mathcal{K}}^{(q)}$. 

Next, we study the action of the superoperator $\hat{\mathcal{E}}_{\tilde{U}_\star}$ on the $+1$ eigenspace of $\hat{\mathcal{K}}$. From Equation \eqref{eq:ustartransform_multiqudit} the superoperator elements are given by:

\begin{align}
(\hat{\mathcal{E}}_{\tilde{U}_\star})_{\mathbf{u}_p, \mathbf{u}_q ; \mathbf{v}_p, \mathbf{v}_q}^{\mathbf{i}_p, \mathbf{i}_q ; \mathbf{j}_p, \mathbf{j}_q} &= (\hat{\mathbb{1}}^{(p)})^{\mathbf{i}_p, \mathbf{j}_p}_{\mathbf{u}_p, \mathbf{v}_p} \otimes (\hat{\mathcal{T}}^{(q)})^{\mathbf{i}_q, \mathbf{j}_q}_{\mathbf{u}_q, \mathbf{v}_q} \\
&= (\hat{\mathbb{1}}^{(p)})^{\mathbf{i}_p, \mathbf{j}_p}_{\mathbf{u}_p, \mathbf{v}_p} \otimes \frac{1}{q^n} e^{\frac{2 \pi i}{q} (\mathbf{u}_q + \bar{p} \mathbf{j}_q) \cdot (\mathbf{i}_q - \bar{p} \mathbf{v}_q)}.
\end{align}

\noindent We only need to consider the eigenstates $\boldsymbol{|} \psi_{\mathbf{l}_q, \mathbf{m}_q} \boldsymbol{\rangle \rangle} $:

\begin{align}
&\hat{\mathcal{T}}^{(q)} \boldsymbol{|} \psi_{\mathbf{l}_q, \mathbf{m}_q} \boldsymbol{\rangle \rangle} = \sum_{\mathbf{i}_q, \mathbf{j}_q} \boldsymbol{\langle \langle } \mathbf{i}_q, \mathbf{j}_q \boldsymbol{|} \psi_{\mathbf{l}_q, \mathbf{m}_q} \boldsymbol{\rangle \rangle} \hat{\mathcal{T}}^{(q)} \boldsymbol{|} \mathbf{i}_q, \mathbf{j}_q \boldsymbol{\rangle \rangle} \\
&= \sum_{\mathbf{u}_q, \mathbf{v}_q} \sum_{\mathbf{i}_q, \mathbf{j}_q} \boldsymbol{\langle \langle } \mathbf{i}_q, \mathbf{j}_q \boldsymbol{|} \psi_{\mathbf{l}_q, \mathbf{m}_q} \boldsymbol{\rangle \rangle} (\hat{\mathcal{T}}^{(q)})^{\mathbf{i}_q, \mathbf{j}_q}_{\mathbf{u}_q, \mathbf{v}_q} \boldsymbol{|} \mathbf{u}_q, \mathbf{v}_q \boldsymbol{\rangle \rangle} \\
&= \frac{1}{q^{2n}} \sum_{\mathbf{u}_q, \mathbf{v}_q} \sum_{\mathbf{i}_q, \mathbf{j}_q} e^{\frac{2 \pi i}{q} (\mathbf{i}_q \cdot \mathbf{l}_q + \mathbf{j}_q \cdot \mathbf{m}_q + (\mathbf{u}_q + \bar{p} \mathbf{j}_q) \cdot (\mathbf{i}_q - \bar{p} \mathbf{v}_q))} \boldsymbol{|} \mathbf{u}_q, \mathbf{v}_q \boldsymbol{\rangle \rangle}  \\
&= \frac{1}{q^{2n}} \sum_{\mathbf{u}_q, \mathbf{v}_q} \sum_{\mathbf{i}_q, \mathbf{j}_q} e^{\frac{2 \pi i}{q} ( \mathbf{i}_q \cdot (\mathbf{l}_q + \mathbf{u}_q + \bar{p} \mathbf{j}_q) + \mathbf{j}_q \cdot \mathbf{m}_q - \bar{p} \mathbf{v}_q \cdot (\mathbf{u}_q + \bar{p} \mathbf{j}_q) )} \boldsymbol{|} \mathbf{u}_q, \mathbf{v}_q \boldsymbol{\rangle \rangle}  \\
&= \frac{1}{q^{n}} \sum_{\mathbf{u}_q, \mathbf{v}_q} \sum_{\mathbf{j}_q} \delta_{\mathbf{j}_q, - p (\mathbf{l}_q + \mathbf{u}_q)} e^{\frac{2 \pi i}{q} ( \mathbf{j}_q \cdot \mathbf{m}_q - \bar{p} \mathbf{v}_q \cdot (\mathbf{u}_q + \bar{p} \mathbf{j}_q) )} \boldsymbol{|} \mathbf{u}_q, \mathbf{v}_q \boldsymbol{\rangle \rangle} \\
&= \frac{1}{q^{n}} \sum_{\mathbf{u}_q, \mathbf{v}_q} e^{\frac{2 \pi i}{q} ( - p (\mathbf{l}_q + \mathbf{u}_q) \cdot \mathbf{m}_q + \bar{p} \mathbf{v}_q \cdot \mathbf{l}_q )} \boldsymbol{|} \mathbf{u}_q, \mathbf{v}_q \boldsymbol{\rangle \rangle} \\
&= \frac{1}{q^{n}} e^{\frac{-2 \pi i}{q} p \mathbf{l}_q \cdot \mathbf{m}_q} \sum_{\mathbf{u}_q, \mathbf{v}_q} e^{\frac{2 \pi i}{q} ( \mathbf{u}_q \cdot (- p \mathbf{m}_q)  + \mathbf{v}_q \cdot ( \bar{p} \mathbf{l}_q) )} \boldsymbol{|} \mathbf{u}_q, \mathbf{v}_q \boldsymbol{\rangle \rangle} \\
& = e^{\frac{-2 \pi i}{q} p \mathbf{l}_q \cdot \mathbf{m}_q} \boldsymbol{|} \psi_{- p \mathbf{m}_q, \bar{p} \mathbf{l}_q} \boldsymbol{\rangle \rangle}.
\end{align}

\noindent In general, this is a phased permutation of the eigenbasis of $\hat{\mathcal{K}}^{(q)}$. However, if $\boldsymbol{|} \psi_{\mathbf{l}_q, \mathbf{m}_q} \boldsymbol{\rangle \rangle} $ is a $+1$ eigenvector then $\mathbf{l}_q \cdot \mathbf{m}_q = 0 \bmod q$, and the phase factor is unity. The resulting vector $\boldsymbol{|} \psi_{- p \mathbf{m}_q, \bar{p} \mathbf{l}_q} \boldsymbol{\rangle \rangle}$ is also a $+1$ eigenvector of $\hat{\mathcal{K}}^{(q)}$, since $(- p \mathbf{m}_q) \cdot (\bar{p} \mathbf{l}_q) = - p \bar{p} (\mathbf{m}_q \cdot \mathbf{l}_q) = 0 \bmod q$. Therefore, the superoperator $\hat{\mathcal{E}}_{\tilde{U}_\star}$ permutes the $+1$ eigenspace of $\hat{\mathcal{K}}$ and preserves KD-reality.

Finally, we explicitly show that the action of $\hat{\mathcal{E}}_{\tilde{U}_\star}$ permutes the entries of any KD-real distribution. The entries of a KD distribution are given by:

\begin{equation}
Q(\tilde{\rho})_{\mathbf{i}_p, \mathbf{i}_q ; \mathbf{j}_p, \mathbf{j}_q} = \boldsymbol{ \langle \langle } \mathbf{i}_p, \mathbf{i}_q ; \mathbf{j}_p, \mathbf{j}_q \boldsymbol{|} Q(\tilde{\rho}) \boldsymbol{\rangle \rangle} 
\end{equation}

\noindent Then, 

\begin{align}
Q(\tilde{U}_\star \tilde{\rho} \tilde{U}_\star^\dagger)_{\mathbf{i}_p, \mathbf{i}_q ; \mathbf{j}_p, \mathbf{j}_q} &= \boldsymbol{ \langle \langle } \mathbf{i}_p, \mathbf{i}_q ; \mathbf{j}_p, \mathbf{j}_q \boldsymbol{|} Q(\tilde{U}_\star \tilde{\rho} \tilde{U}_\star^\dagger) \boldsymbol{\rangle \rangle} \\
& = \boldsymbol{ \langle \langle } \mathbf{i}_p, \mathbf{i}_q ; \mathbf{j}_p, \mathbf{j}_q \boldsymbol{|} \hat{\mathcal{E}}_{\tilde{U}_\star} \boldsymbol{|} Q(\tilde{\rho}) \boldsymbol{\rangle \rangle}.
\end{align}

\noindent Since $Q(\tilde{\rho})$ is KD-real, it expands over the $+1$ eigenvectors of $\hat{\mathcal{K}}$ as:

\begin{equation}
\boldsymbol{|} Q(\tilde{\rho}) \boldsymbol{\rangle \rangle} = \sum_{\substack{\mathbf{l}_p \cdot \mathbf{m}_p = 0 \bmod p, \\ \mathbf{l}_q \cdot \mathbf{m}_q = 0 \bmod q}} c_{\mathbf{l}_p, \mathbf{m}_p, \mathbf{l}_q, \mathbf{m}_q} \boldsymbol{|} \phi_{\mathbf{l}_p, \mathbf{m}_p} \boldsymbol{\rangle \rangle} \otimes \boldsymbol{|} \psi_{\mathbf{l}_q, \mathbf{m}_q} \boldsymbol{\rangle \rangle}.
\end{equation}

\noindent It follows that:

\begin{align}
\hat{\mathcal{E}}_{\tilde{U}_\star} \boldsymbol{|} Q(\tilde{\rho}) \boldsymbol{\rangle \rangle} &= \hat{\mathbb{1}}^{(p)} \otimes \hat{\mathcal{T}}^{(q)} \boldsymbol{|} Q(\tilde{\rho}) \boldsymbol{\rangle \rangle} \\
&= \sum_{\substack{\mathbf{l}_p \cdot \mathbf{m}_p = 0, \\ \mathbf{l}_q \cdot \mathbf{m}_q = 0}} c_{\mathbf{l}_p, \mathbf{m}_p, \mathbf{l}_q, \mathbf{m}_q} \boldsymbol{|} \phi_{\mathbf{l}_p, \mathbf{m}_p} \boldsymbol{\rangle \rangle} \otimes \boldsymbol{|} \psi_{- p \mathbf{m}_q, \bar{p} \mathbf{l}_q} \boldsymbol{\rangle \rangle}.
\end{align}

\noindent Furthermore, we have:

\begin{align}
\boldsymbol{\langle \langle } \mathbf{i}_q, \mathbf{j}_q \boldsymbol{|} \psi_{- p \mathbf{m}_q, \bar{p} \mathbf{l}_q}  \boldsymbol{\rangle \rangle} &= q^{-n} e^{\frac{2 \pi i}{q} (-p \mathbf{i}_q \cdot \mathbf{m}_q + \bar{p} \mathbf{j}_q \cdot \mathbf{l}_q)}\\
&= \boldsymbol{\langle \langle } \bar{p} \mathbf{j}_q, - p \mathbf{i}_q \boldsymbol{|} \psi_{\mathbf{l}_q, \mathbf{m}_q}  \boldsymbol{\rangle \rangle}.
\end{align}

\noindent Putting everything together, we obtain:

\begin{align}
 \quad & Q(\tilde{U}_\star \tilde{\rho} \tilde{U}_\star^\dagger)_{\mathbf{i}_p, \mathbf{i}_q ; \mathbf{j}_p, \mathbf{j}_q} \\
&= \boldsymbol{ \langle \langle } \mathbf{i}_p, \mathbf{i}_q ; \mathbf{j}_p, \mathbf{j}_q \boldsymbol{|} \hat{\mathcal{E}}_{\tilde{U}_\star} \boldsymbol{|} Q(\tilde{\rho}) \boldsymbol{\rangle \rangle} \\
&= \sum_{\substack{\mathbf{l}_p \cdot \mathbf{m}_p = 0, \\ \mathbf{l}_q \cdot \mathbf{m}_q = 0}} c_{\mathbf{l}_p, \mathbf{m}_p, \mathbf{l}_q, \mathbf{m}_q} \boldsymbol{\langle \langle } \mathbf{i}_p, \mathbf{j}_p \boldsymbol{|}  \phi_{\mathbf{l}_p, \mathbf{m}_p}  \boldsymbol{\rangle \rangle} \cdot  \boldsymbol{\langle \langle } \mathbf{i}_q, \mathbf{j}_q \boldsymbol{|}  \psi_{- p \mathbf{m}_q, \bar{p} \mathbf{l}_q}  \boldsymbol{\rangle \rangle}  \\
&= \sum_{\substack{\mathbf{l}_p \cdot \mathbf{m}_p = 0, \\ \mathbf{l}_q \cdot \mathbf{m}_q = 0}} c_{\mathbf{l}_p, \mathbf{m}_p, \mathbf{l}_q, \mathbf{m}_q} \boldsymbol{\langle \langle } \mathbf{i}_p, \mathbf{j}_p \boldsymbol{|}  \phi_{\mathbf{l}_p, \mathbf{m}_p}  \boldsymbol{\rangle \rangle} \cdot  \boldsymbol{\langle \langle } \bar{p} \mathbf{j}_q, - p \mathbf{i}_q \boldsymbol{|} \psi_{\mathbf{l}_q, \mathbf{m}_q}  \boldsymbol{\rangle \rangle}  \\
&= Q(\tilde{\rho})_{\mathbf{i}_p, \bar{p} \mathbf{j}_q ; \mathbf{j}_p, - p \mathbf{i}_q}.
\end{align}

\noindent This is a permutation of the index set. A permutation of entries preserves the total non-positivity as well as KD-reality, hence the total non-positivity of a KD-real $Q(\tilde{\rho})$ -- and equivalently of a KD-real $Q(\rho)$ -- is preserved under the action of $\tilde{U}_\star$. This proves that $U_\star^{\otimes n}$ is positivity-preserving for the $V = \mathrm{DFT}_{pq}^{\otimes n}$ distribution. The induced bijection $\Gamma^{-1}$ between indices takes on the same form entrywise as in the $n=1$ case (Equation \eqref{eq:klmapping}):

\begin{equation}
\Gamma^{-1}(\mathbf{i}, \mathbf{j}) = (\mathbf{i}_p  q \bar{q} + \bar{p} \mathbf{j}_q p \bar{p}, \ \mathbf{j}_p q \bar{q} - p \mathbf{i}_q p \bar{p}).
\end{equation}

\end{proof}


\begin{thebibliography}{64}%
\makeatletter
\providecommand \@ifxundefined [1]{%
 \@ifx{#1\undefined}
}%
\providecommand \@ifnum [1]{%
 \ifnum #1\expandafter \@firstoftwo
 \else \expandafter \@secondoftwo
 \fi
}%
\providecommand \@ifx [1]{%
 \ifx #1\expandafter \@firstoftwo
 \else \expandafter \@secondoftwo
 \fi
}%
\providecommand \natexlab [1]{#1}%
\providecommand \enquote  [1]{``#1''}%
\providecommand \bibnamefont  [1]{#1}%
\providecommand \bibfnamefont [1]{#1}%
\providecommand \citenamefont [1]{#1}%
\providecommand \href@noop [0]{\@secondoftwo}%
\providecommand \href [0]{\begingroup \@sanitize@url \@href}%
\providecommand \@href[1]{\@@startlink{#1}\@@href}%
\providecommand \@@href[1]{\endgroup#1\@@endlink}%
\providecommand \@sanitize@url [0]{\catcode `\\12\catcode `\$12\catcode `\&12\catcode `\#12\catcode `\^12\catcode `\_12\catcode `\%12\relax}%
\providecommand \@@startlink[1]{}%
\providecommand \@@endlink[0]{}%
\providecommand \url  [0]{\begingroup\@sanitize@url \@url }%
\providecommand \@url [1]{\endgroup\@href {#1}{\urlprefix }}%
\providecommand \urlprefix  [0]{URL }%
\providecommand \Eprint [0]{\href }%
\providecommand \doibase [0]{https://doi.org/}%
\providecommand \selectlanguage [0]{\@gobble}%
\providecommand \bibinfo  [0]{\@secondoftwo}%
\providecommand \bibfield  [0]{\@secondoftwo}%
\providecommand \translation [1]{[#1]}%
\providecommand \BibitemOpen [0]{}%
\providecommand \bibitemStop [0]{}%
\providecommand \bibitemNoStop [0]{.\EOS\space}%
\providecommand \EOS [0]{\spacefactor3000\relax}%
\providecommand \BibitemShut  [1]{\csname bibitem#1\endcsname}%
\let\auto@bib@innerbib\@empty
\bibitem [{\citenamefont {Gottesman}(1998)}]{gottesman1998}%
  \BibitemOpen
  \bibfield  {author} {\bibinfo {author} {\bibfnamefont {D.}~\bibnamefont {Gottesman}},\ }\href {https://arxiv.org/abs/quant-ph/9807006} {\bibinfo {title} {The heisenberg representation of quantum computers}} (\bibinfo {year} {1998}),\ \Eprint {https://arxiv.org/abs/quant-ph/9807006} {arXiv:quant-ph/9807006 [quant-ph]} \BibitemShut {NoStop}%
\bibitem [{\citenamefont {Spekkens}(2008)}]{Spekkens_2008}%
  \BibitemOpen
  \bibfield  {author} {\bibinfo {author} {\bibfnamefont {R.~W.}\ \bibnamefont {Spekkens}},\ }\bibfield  {title} {\bibinfo {title} {Negativity and contextuality are equivalent notions of nonclassicality},\ }\bibfield  {journal} {\bibinfo  {journal} {Physical Review Letters}\ }\textbf {\bibinfo {volume} {101}},\ \href {https://doi.org/10.1103/physrevlett.101.020401} {10.1103/physrevlett.101.020401} (\bibinfo {year} {2008})\BibitemShut {NoStop}%
\bibitem [{\citenamefont {Ferraro}\ and\ \citenamefont {Paris}(2012)}]{Ferraro_2012}%
  \BibitemOpen
  \bibfield  {author} {\bibinfo {author} {\bibfnamefont {A.}~\bibnamefont {Ferraro}}\ and\ \bibinfo {author} {\bibfnamefont {M.~G.~A.}\ \bibnamefont {Paris}},\ }\bibfield  {title} {\bibinfo {title} {Nonclassicality criteria from phase-space representations and information-theoretical constraints are maximally inequivalent},\ }\bibfield  {journal} {\bibinfo  {journal} {Physical Review Letters}\ }\textbf {\bibinfo {volume} {108}},\ \href {https://doi.org/10.1103/physrevlett.108.260403} {10.1103/physrevlett.108.260403} (\bibinfo {year} {2012})\BibitemShut {NoStop}%
\bibitem [{\citenamefont {Mari}\ and\ \citenamefont {Eisert}(2012)}]{Mari_2012}%
  \BibitemOpen
  \bibfield  {author} {\bibinfo {author} {\bibfnamefont {A.}~\bibnamefont {Mari}}\ and\ \bibinfo {author} {\bibfnamefont {J.}~\bibnamefont {Eisert}},\ }\bibfield  {title} {\bibinfo {title} {Positive wigner functions render classical simulation of quantum computation efficient},\ }\bibfield  {journal} {\bibinfo  {journal} {Physical Review Letters}\ }\textbf {\bibinfo {volume} {109}},\ \href {https://doi.org/10.1103/physrevlett.109.230503} {10.1103/physrevlett.109.230503} (\bibinfo {year} {2012})\BibitemShut {NoStop}%
\bibitem [{\citenamefont {Gross}(2006)}]{Gross_2006}%
  \BibitemOpen
  \bibfield  {author} {\bibinfo {author} {\bibfnamefont {D.}~\bibnamefont {Gross}},\ }\bibfield  {title} {\bibinfo {title} {Hudson's theorem for finite-dimensional quantum systems},\ }\bibfield  {journal} {\bibinfo  {journal} {Journal of Mathematical Physics}\ }\textbf {\bibinfo {volume} {47}},\ \href {https://doi.org/10.1063/1.2393152} {10.1063/1.2393152} (\bibinfo {year} {2006})\BibitemShut {NoStop}%
\bibitem [{\citenamefont {Ferrie}\ and\ \citenamefont {Emerson}(2008)}]{Ferrie_2008}%
  \BibitemOpen
  \bibfield  {author} {\bibinfo {author} {\bibfnamefont {C.}~\bibnamefont {Ferrie}}\ and\ \bibinfo {author} {\bibfnamefont {J.}~\bibnamefont {Emerson}},\ }\bibfield  {title} {\bibinfo {title} {Frame representations of quantum mechanics and the necessity of negativity in quasi-probability representations},\ }\href {https://doi.org/10.1088/1751-8113/41/35/352001} {\bibfield  {journal} {\bibinfo  {journal} {Journal of Physics A: Mathematical and Theoretical}\ }\textbf {\bibinfo {volume} {41}},\ \bibinfo {pages} {352001} (\bibinfo {year} {2008})}\BibitemShut {NoStop}%
\bibitem [{\citenamefont {Veitch}\ \emph {et~al.}(2012)\citenamefont {Veitch}, \citenamefont {Ferrie}, \citenamefont {Gross},\ and\ \citenamefont {Emerson}}]{Veitch_2012}%
  \BibitemOpen
  \bibfield  {author} {\bibinfo {author} {\bibfnamefont {V.}~\bibnamefont {Veitch}}, \bibinfo {author} {\bibfnamefont {C.}~\bibnamefont {Ferrie}}, \bibinfo {author} {\bibfnamefont {D.}~\bibnamefont {Gross}},\ and\ \bibinfo {author} {\bibfnamefont {J.}~\bibnamefont {Emerson}},\ }\bibfield  {title} {\bibinfo {title} {Negative quasi-probability as a resource for quantum computation},\ }\href {https://doi.org/10.1088/1367-2630/14/11/113011} {\bibfield  {journal} {\bibinfo  {journal} {New Journal of Physics}\ }\textbf {\bibinfo {volume} {14}},\ \bibinfo {pages} {113011} (\bibinfo {year} {2012})}\BibitemShut {NoStop}%
\bibitem [{\citenamefont {Galv\~ao}(2005)}]{PhysRevA.71.042302}%
  \BibitemOpen
  \bibfield  {author} {\bibinfo {author} {\bibfnamefont {E.~F.}\ \bibnamefont {Galv\~ao}},\ }\bibfield  {title} {\bibinfo {title} {Discrete wigner functions and quantum computational speedup},\ }\href {https://doi.org/10.1103/PhysRevA.71.042302} {\bibfield  {journal} {\bibinfo  {journal} {Phys. Rev. A}\ }\textbf {\bibinfo {volume} {71}},\ \bibinfo {pages} {042302} (\bibinfo {year} {2005})}\BibitemShut {NoStop}%
\bibitem [{\citenamefont {Veitch}\ \emph {et~al.}(2014)\citenamefont {Veitch}, \citenamefont {Hamed~Mousavian}, \citenamefont {Gottesman},\ and\ \citenamefont {Emerson}}]{Veitch_2014}%
  \BibitemOpen
  \bibfield  {author} {\bibinfo {author} {\bibfnamefont {V.}~\bibnamefont {Veitch}}, \bibinfo {author} {\bibfnamefont {S.~A.}\ \bibnamefont {Hamed~Mousavian}}, \bibinfo {author} {\bibfnamefont {D.}~\bibnamefont {Gottesman}},\ and\ \bibinfo {author} {\bibfnamefont {J.}~\bibnamefont {Emerson}},\ }\bibfield  {title} {\bibinfo {title} {The resource theory of stabilizer quantum computation},\ }\href {https://doi.org/10.1088/1367-2630/16/1/013009} {\bibfield  {journal} {\bibinfo  {journal} {New Journal of Physics}\ }\textbf {\bibinfo {volume} {16}},\ \bibinfo {pages} {013009} (\bibinfo {year} {2014})}\BibitemShut {NoStop}%
\bibitem [{\citenamefont {Delfosse}\ \emph {et~al.}(2015)\citenamefont {Delfosse}, \citenamefont {Allard~Guerin}, \citenamefont {Bian},\ and\ \citenamefont {Raussendorf}}]{PhysRevX.5.021003}%
  \BibitemOpen
  \bibfield  {author} {\bibinfo {author} {\bibfnamefont {N.}~\bibnamefont {Delfosse}}, \bibinfo {author} {\bibfnamefont {P.}~\bibnamefont {Allard~Guerin}}, \bibinfo {author} {\bibfnamefont {J.}~\bibnamefont {Bian}},\ and\ \bibinfo {author} {\bibfnamefont {R.}~\bibnamefont {Raussendorf}},\ }\bibfield  {title} {\bibinfo {title} {Wigner function negativity and contextuality in quantum computation on rebits},\ }\href {https://doi.org/10.1103/PhysRevX.5.021003} {\bibfield  {journal} {\bibinfo  {journal} {Phys. Rev. X}\ }\textbf {\bibinfo {volume} {5}},\ \bibinfo {pages} {021003} (\bibinfo {year} {2015})}\BibitemShut {NoStop}%
\bibitem [{\citenamefont {Howard}\ and\ \citenamefont {Campbell}(2017)}]{Howard_2017}%
  \BibitemOpen
  \bibfield  {author} {\bibinfo {author} {\bibfnamefont {M.}~\bibnamefont {Howard}}\ and\ \bibinfo {author} {\bibfnamefont {E.}~\bibnamefont {Campbell}},\ }\bibfield  {title} {\bibinfo {title} {Application of a resource theory for magic states to fault-tolerant quantum computing},\ }\bibfield  {journal} {\bibinfo  {journal} {Physical Review Letters}\ }\textbf {\bibinfo {volume} {118}},\ \href {https://doi.org/10.1103/physrevlett.118.090501} {10.1103/physrevlett.118.090501} (\bibinfo {year} {2017})\BibitemShut {NoStop}%
\bibitem [{\citenamefont {Raussendorf}\ \emph {et~al.}(2020)\citenamefont {Raussendorf}, \citenamefont {Bermejo-Vega}, \citenamefont {Tyhurst}, \citenamefont {Okay},\ and\ \citenamefont {Zurel}}]{PhysRevA.101.012350}%
  \BibitemOpen
  \bibfield  {author} {\bibinfo {author} {\bibfnamefont {R.}~\bibnamefont {Raussendorf}}, \bibinfo {author} {\bibfnamefont {J.}~\bibnamefont {Bermejo-Vega}}, \bibinfo {author} {\bibfnamefont {E.}~\bibnamefont {Tyhurst}}, \bibinfo {author} {\bibfnamefont {C.}~\bibnamefont {Okay}},\ and\ \bibinfo {author} {\bibfnamefont {M.}~\bibnamefont {Zurel}},\ }\bibfield  {title} {\bibinfo {title} {Phase-space-simulation method for quantum computation with magic states on qubits},\ }\href {https://doi.org/10.1103/PhysRevA.101.012350} {\bibfield  {journal} {\bibinfo  {journal} {Phys. Rev. A}\ }\textbf {\bibinfo {volume} {101}},\ \bibinfo {pages} {012350} (\bibinfo {year} {2020})}\BibitemShut {NoStop}%
\bibitem [{\citenamefont {Weedbrook}\ \emph {et~al.}(2012)\citenamefont {Weedbrook}, \citenamefont {Pirandola}, \citenamefont {Garc\'{\i}a-Patr\'on}, \citenamefont {Cerf}, \citenamefont {Ralph}, \citenamefont {Shapiro},\ and\ \citenamefont {Lloyd}}]{RevModPhys.84.621}%
  \BibitemOpen
  \bibfield  {author} {\bibinfo {author} {\bibfnamefont {C.}~\bibnamefont {Weedbrook}}, \bibinfo {author} {\bibfnamefont {S.}~\bibnamefont {Pirandola}}, \bibinfo {author} {\bibfnamefont {R.}~\bibnamefont {Garc\'{\i}a-Patr\'on}}, \bibinfo {author} {\bibfnamefont {N.~J.}\ \bibnamefont {Cerf}}, \bibinfo {author} {\bibfnamefont {T.~C.}\ \bibnamefont {Ralph}}, \bibinfo {author} {\bibfnamefont {J.~H.}\ \bibnamefont {Shapiro}},\ and\ \bibinfo {author} {\bibfnamefont {S.}~\bibnamefont {Lloyd}},\ }\bibfield  {title} {\bibinfo {title} {Gaussian quantum information},\ }\href {https://doi.org/10.1103/RevModPhys.84.621} {\bibfield  {journal} {\bibinfo  {journal} {Rev. Mod. Phys.}\ }\textbf {\bibinfo {volume} {84}},\ \bibinfo {pages} {621} (\bibinfo {year} {2012})}\BibitemShut {NoStop}%
\bibitem [{\citenamefont {Pashayan}\ \emph {et~al.}(2015)\citenamefont {Pashayan}, \citenamefont {Wallman},\ and\ \citenamefont {Bartlett}}]{Pashayan_2015}%
  \BibitemOpen
  \bibfield  {author} {\bibinfo {author} {\bibfnamefont {H.}~\bibnamefont {Pashayan}}, \bibinfo {author} {\bibfnamefont {J.~J.}\ \bibnamefont {Wallman}},\ and\ \bibinfo {author} {\bibfnamefont {S.~D.}\ \bibnamefont {Bartlett}},\ }\bibfield  {title} {\bibinfo {title} {Estimating outcome probabilities of quantum circuits using quasiprobabilities},\ }\bibfield  {journal} {\bibinfo  {journal} {Physical Review Letters}\ }\textbf {\bibinfo {volume} {115}},\ \href {https://doi.org/10.1103/physrevlett.115.070501} {10.1103/physrevlett.115.070501} (\bibinfo {year} {2015})\BibitemShut {NoStop}%
\bibitem [{\citenamefont {Jozsa}\ and\ \citenamefont {Miyake}(2008)}]{Jozsa_2008}%
  \BibitemOpen
  \bibfield  {author} {\bibinfo {author} {\bibfnamefont {R.}~\bibnamefont {Jozsa}}\ and\ \bibinfo {author} {\bibfnamefont {A.}~\bibnamefont {Miyake}},\ }\bibfield  {title} {\bibinfo {title} {Matchgates and classical simulation of quantum circuits},\ }\href {https://doi.org/10.1098/rspa.2008.0189} {\bibfield  {journal} {\bibinfo  {journal} {Proceedings of the Royal Society A: Mathematical, Physical and Engineering Sciences}\ }\textbf {\bibinfo {volume} {464}},\ \bibinfo {pages} {3089–3106} (\bibinfo {year} {2008})}\BibitemShut {NoStop}%
\bibitem [{\citenamefont {Aaronson}\ and\ \citenamefont {Arkhipov}(2010)}]{aaronson2010}%
  \BibitemOpen
  \bibfield  {author} {\bibinfo {author} {\bibfnamefont {S.}~\bibnamefont {Aaronson}}\ and\ \bibinfo {author} {\bibfnamefont {A.}~\bibnamefont {Arkhipov}},\ }\href {https://arxiv.org/abs/1011.3245} {\bibinfo {title} {The computational complexity of linear optics}} (\bibinfo {year} {2010}),\ \Eprint {https://arxiv.org/abs/1011.3245} {arXiv:1011.3245 [quant-ph]} \BibitemShut {NoStop}%
\bibitem [{\citenamefont {Terhal}\ and\ \citenamefont {DiVincenzo}(2004)}]{terhal2004}%
  \BibitemOpen
  \bibfield  {author} {\bibinfo {author} {\bibfnamefont {B.~M.}\ \bibnamefont {Terhal}}\ and\ \bibinfo {author} {\bibfnamefont {D.~P.}\ \bibnamefont {DiVincenzo}},\ }\href {https://arxiv.org/abs/quant-ph/0205133} {\bibinfo {title} {Adaptive quantum computation, constant depth quantum circuits and arthur-merlin games}} (\bibinfo {year} {2004}),\ \Eprint {https://arxiv.org/abs/quant-ph/0205133} {arXiv:quant-ph/0205133 [quant-ph]} \BibitemShut {NoStop}%
\bibitem [{\citenamefont {Kirkwood}(1933)}]{PhysRev.44.31}%
  \BibitemOpen
  \bibfield  {author} {\bibinfo {author} {\bibfnamefont {J.~G.}\ \bibnamefont {Kirkwood}},\ }\bibfield  {title} {\bibinfo {title} {Quantum statistics of almost classical assemblies},\ }\href {https://doi.org/10.1103/PhysRev.44.31} {\bibfield  {journal} {\bibinfo  {journal} {Phys. Rev.}\ }\textbf {\bibinfo {volume} {44}},\ \bibinfo {pages} {31} (\bibinfo {year} {1933})}\BibitemShut {NoStop}%
\bibitem [{\citenamefont {Dirac}(1945)}]{RevModPhys.17.195}%
  \BibitemOpen
  \bibfield  {author} {\bibinfo {author} {\bibfnamefont {P.~A.~M.}\ \bibnamefont {Dirac}},\ }\bibfield  {title} {\bibinfo {title} {On the analogy between classical and quantum mechanics},\ }\href {https://doi.org/10.1103/RevModPhys.17.195} {\bibfield  {journal} {\bibinfo  {journal} {Rev. Mod. Phys.}\ }\textbf {\bibinfo {volume} {17}},\ \bibinfo {pages} {195} (\bibinfo {year} {1945})}\BibitemShut {NoStop}%
\bibitem [{\citenamefont {Lostaglio}\ \emph {et~al.}(2023)\citenamefont {Lostaglio}, \citenamefont {Belenchia}, \citenamefont {Levy}, \citenamefont {Hernández-Gómez}, \citenamefont {Fabbri},\ and\ \citenamefont {Gherardini}}]{Lostaglio_2023}%
  \BibitemOpen
  \bibfield  {author} {\bibinfo {author} {\bibfnamefont {M.}~\bibnamefont {Lostaglio}}, \bibinfo {author} {\bibfnamefont {A.}~\bibnamefont {Belenchia}}, \bibinfo {author} {\bibfnamefont {A.}~\bibnamefont {Levy}}, \bibinfo {author} {\bibfnamefont {S.}~\bibnamefont {Hernández-Gómez}}, \bibinfo {author} {\bibfnamefont {N.}~\bibnamefont {Fabbri}},\ and\ \bibinfo {author} {\bibfnamefont {S.}~\bibnamefont {Gherardini}},\ }\bibfield  {title} {\bibinfo {title} {Kirkwood-dirac quasiprobability approach to the statistics of incompatible observables},\ }\href {https://doi.org/10.22331/q-2023-10-09-1128} {\bibfield  {journal} {\bibinfo  {journal} {Quantum}\ }\textbf {\bibinfo {volume} {7}},\ \bibinfo {pages} {1128} (\bibinfo {year} {2023})}\BibitemShut {NoStop}%
\bibitem [{\citenamefont {Johansen}(2007)}]{PhysRevA.76.012119}%
  \BibitemOpen
  \bibfield  {author} {\bibinfo {author} {\bibfnamefont {L.~M.}\ \bibnamefont {Johansen}},\ }\bibfield  {title} {\bibinfo {title} {Quantum theory of successive projective measurements},\ }\href {https://doi.org/10.1103/PhysRevA.76.012119} {\bibfield  {journal} {\bibinfo  {journal} {Phys. Rev. A}\ }\textbf {\bibinfo {volume} {76}},\ \bibinfo {pages} {012119} (\bibinfo {year} {2007})}\BibitemShut {NoStop}%
\bibitem [{\citenamefont {Bamber}\ and\ \citenamefont {Lundeen}(2014)}]{PhysRevLett.112.070405}%
  \BibitemOpen
  \bibfield  {author} {\bibinfo {author} {\bibfnamefont {C.}~\bibnamefont {Bamber}}\ and\ \bibinfo {author} {\bibfnamefont {J.~S.}\ \bibnamefont {Lundeen}},\ }\bibfield  {title} {\bibinfo {title} {Observing dirac's classical phase space analog to the quantum state},\ }\href {https://doi.org/10.1103/PhysRevLett.112.070405} {\bibfield  {journal} {\bibinfo  {journal} {Phys. Rev. Lett.}\ }\textbf {\bibinfo {volume} {112}},\ \bibinfo {pages} {070405} (\bibinfo {year} {2014})}\BibitemShut {NoStop}%
\bibitem [{\citenamefont {Arvidsson-Shukur}\ \emph {et~al.}(2020)\citenamefont {Arvidsson-Shukur}, \citenamefont {Yunger~Halpern}, \citenamefont {Lepage}, \citenamefont {Lasek}, \citenamefont {Barnes},\ and\ \citenamefont {Lloyd}}]{Arvidsson2020}%
  \BibitemOpen
  \bibfield  {author} {\bibinfo {author} {\bibfnamefont {D.~R.~M.}\ \bibnamefont {Arvidsson-Shukur}}, \bibinfo {author} {\bibfnamefont {N.}~\bibnamefont {Yunger~Halpern}}, \bibinfo {author} {\bibfnamefont {H.~V.}\ \bibnamefont {Lepage}}, \bibinfo {author} {\bibfnamefont {A.~A.}\ \bibnamefont {Lasek}}, \bibinfo {author} {\bibfnamefont {C.~H.~W.}\ \bibnamefont {Barnes}},\ and\ \bibinfo {author} {\bibfnamefont {S.}~\bibnamefont {Lloyd}},\ }\bibfield  {title} {\bibinfo {title} {Quantum advantage in postselected metrology},\ }\bibfield  {journal} {\bibinfo  {journal} {Nature Communications}\ }\textbf {\bibinfo {volume} {11}},\ \href {https://doi.org/10.1038/s41467-020-17559-w} {10.1038/s41467-020-17559-w} (\bibinfo {year} {2020})\BibitemShut {NoStop}%
\bibitem [{\citenamefont {Das}\ \emph {et~al.}(2023)\citenamefont {Das}, \citenamefont {Modak},\ and\ \citenamefont {Bera}}]{Das_2023}%
  \BibitemOpen
  \bibfield  {author} {\bibinfo {author} {\bibfnamefont {S.}~\bibnamefont {Das}}, \bibinfo {author} {\bibfnamefont {S.}~\bibnamefont {Modak}},\ and\ \bibinfo {author} {\bibfnamefont {M.~N.}\ \bibnamefont {Bera}},\ }\bibfield  {title} {\bibinfo {title} {Saturating quantum advantages in postselected metrology with the positive kirkwood-dirac distribution},\ }\bibfield  {journal} {\bibinfo  {journal} {Physical Review A}\ }\textbf {\bibinfo {volume} {107}},\ \href {https://doi.org/10.1103/physreva.107.042413} {10.1103/physreva.107.042413} (\bibinfo {year} {2023})\BibitemShut {NoStop}%
\bibitem [{\citenamefont {Lupu-Gladstein}\ \emph {et~al.}(2022)\citenamefont {Lupu-Gladstein}, \citenamefont {Yilmaz}, \citenamefont {Arvidsson-Shukur}, \citenamefont {Brodutch}, \citenamefont {Pang}, \citenamefont {Steinberg},\ and\ \citenamefont {Halpern}}]{Lupu_Gladstein_2022}%
  \BibitemOpen
  \bibfield  {author} {\bibinfo {author} {\bibfnamefont {N.}~\bibnamefont {Lupu-Gladstein}}, \bibinfo {author} {\bibfnamefont {Y.~B.}\ \bibnamefont {Yilmaz}}, \bibinfo {author} {\bibfnamefont {D.~R.}\ \bibnamefont {Arvidsson-Shukur}}, \bibinfo {author} {\bibfnamefont {A.}~\bibnamefont {Brodutch}}, \bibinfo {author} {\bibfnamefont {A.~O.}\ \bibnamefont {Pang}}, \bibinfo {author} {\bibfnamefont {A.~M.}\ \bibnamefont {Steinberg}},\ and\ \bibinfo {author} {\bibfnamefont {N.~Y.}\ \bibnamefont {Halpern}},\ }\bibfield  {title} {\bibinfo {title} {Negative quasiprobabilities enhance phase estimation in quantum-optics experiment},\ }\bibfield  {journal} {\bibinfo  {journal} {Physical Review Letters}\ }\textbf {\bibinfo {volume} {128}},\ \href {https://doi.org/10.1103/physrevlett.128.220504} {10.1103/physrevlett.128.220504} (\bibinfo {year} {2022})\BibitemShut {NoStop}%
\bibitem [{\citenamefont {Arvidsson-Shukur}\ \emph {et~al.}(2021)\citenamefont {Arvidsson-Shukur}, \citenamefont {Drori},\ and\ \citenamefont {Halpern}}]{Arvidsson_Shukur_2021}%
  \BibitemOpen
  \bibfield  {author} {\bibinfo {author} {\bibfnamefont {D.~R.~M.}\ \bibnamefont {Arvidsson-Shukur}}, \bibinfo {author} {\bibfnamefont {J.~C.}\ \bibnamefont {Drori}},\ and\ \bibinfo {author} {\bibfnamefont {N.~Y.}\ \bibnamefont {Halpern}},\ }\bibfield  {title} {\bibinfo {title} {Conditions tighter than noncommutation needed for nonclassicality},\ }\href {https://doi.org/10.1088/1751-8121/ac0289} {\bibfield  {journal} {\bibinfo  {journal} {Journal of Physics A: Mathematical and Theoretical}\ }\textbf {\bibinfo {volume} {54}},\ \bibinfo {pages} {284001} (\bibinfo {year} {2021})}\BibitemShut {NoStop}%
\bibitem [{\citenamefont {Yunger~Halpern}(2017)}]{PhysRevA.95.012120}%
  \BibitemOpen
  \bibfield  {author} {\bibinfo {author} {\bibfnamefont {N.}~\bibnamefont {Yunger~Halpern}},\ }\bibfield  {title} {\bibinfo {title} {Jarzynski-like equality for the out-of-time-ordered correlator},\ }\href {https://doi.org/10.1103/PhysRevA.95.012120} {\bibfield  {journal} {\bibinfo  {journal} {Phys. Rev. A}\ }\textbf {\bibinfo {volume} {95}},\ \bibinfo {pages} {012120} (\bibinfo {year} {2017})}\BibitemShut {NoStop}%
\bibitem [{\citenamefont {Allahverdyan}(2014)}]{PhysRevE.90.032137}%
  \BibitemOpen
  \bibfield  {author} {\bibinfo {author} {\bibfnamefont {A.~E.}\ \bibnamefont {Allahverdyan}},\ }\bibfield  {title} {\bibinfo {title} {Nonequilibrium quantum fluctuations of work},\ }\href {https://doi.org/10.1103/PhysRevE.90.032137} {\bibfield  {journal} {\bibinfo  {journal} {Phys. Rev. E}\ }\textbf {\bibinfo {volume} {90}},\ \bibinfo {pages} {032137} (\bibinfo {year} {2014})}\BibitemShut {NoStop}%
\bibitem [{\citenamefont {Miller}\ and\ \citenamefont {Anders}(2017)}]{Miller_2017}%
  \BibitemOpen
  \bibfield  {author} {\bibinfo {author} {\bibfnamefont {H.~J.~D.}\ \bibnamefont {Miller}}\ and\ \bibinfo {author} {\bibfnamefont {J.}~\bibnamefont {Anders}},\ }\bibfield  {title} {\bibinfo {title} {Time-reversal symmetric work distributions for closed quantum dynamics in the histories framework},\ }\href {https://doi.org/10.1088/1367-2630/aa703f} {\bibfield  {journal} {\bibinfo  {journal} {New Journal of Physics}\ }\textbf {\bibinfo {volume} {19}},\ \bibinfo {pages} {062001} (\bibinfo {year} {2017})}\BibitemShut {NoStop}%
\bibitem [{\citenamefont {Levy}\ and\ \citenamefont {Lostaglio}(2020)}]{Levy_2020}%
  \BibitemOpen
  \bibfield  {author} {\bibinfo {author} {\bibfnamefont {A.}~\bibnamefont {Levy}}\ and\ \bibinfo {author} {\bibfnamefont {M.}~\bibnamefont {Lostaglio}},\ }\bibfield  {title} {\bibinfo {title} {Quasiprobability distribution for heat fluctuations in the quantum regime},\ }\bibfield  {journal} {\bibinfo  {journal} {PRX Quantum}\ }\textbf {\bibinfo {volume} {1}},\ \href {https://doi.org/10.1103/prxquantum.1.010309} {10.1103/prxquantum.1.010309} (\bibinfo {year} {2020})\BibitemShut {NoStop}%
\bibitem [{\citenamefont {Lostaglio}(2020)}]{PhysRevLett.125.230603}%
  \BibitemOpen
  \bibfield  {author} {\bibinfo {author} {\bibfnamefont {M.}~\bibnamefont {Lostaglio}},\ }\bibfield  {title} {\bibinfo {title} {Certifying quantum signatures in thermodynamics and metrology via contextuality of quantum linear response},\ }\href {https://doi.org/10.1103/PhysRevLett.125.230603} {\bibfield  {journal} {\bibinfo  {journal} {Phys. Rev. Lett.}\ }\textbf {\bibinfo {volume} {125}},\ \bibinfo {pages} {230603} (\bibinfo {year} {2020})}\BibitemShut {NoStop}%
\bibitem [{\citenamefont {Gonz\'alez~Alonso}\ \emph {et~al.}(2019)\citenamefont {Gonz\'alez~Alonso}, \citenamefont {Yunger~Halpern},\ and\ \citenamefont {Dressel}}]{PhysRevLett.122.040404}%
  \BibitemOpen
  \bibfield  {author} {\bibinfo {author} {\bibfnamefont {J.~R.}\ \bibnamefont {Gonz\'alez~Alonso}}, \bibinfo {author} {\bibfnamefont {N.}~\bibnamefont {Yunger~Halpern}},\ and\ \bibinfo {author} {\bibfnamefont {J.}~\bibnamefont {Dressel}},\ }\bibfield  {title} {\bibinfo {title} {Out-of-time-ordered-correlator quasiprobabilities robustly witness scrambling},\ }\href {https://doi.org/10.1103/PhysRevLett.122.040404} {\bibfield  {journal} {\bibinfo  {journal} {Phys. Rev. Lett.}\ }\textbf {\bibinfo {volume} {122}},\ \bibinfo {pages} {040404} (\bibinfo {year} {2019})}\BibitemShut {NoStop}%
\bibitem [{\citenamefont {Wagner}\ \emph {et~al.}(2024)\citenamefont {Wagner}, \citenamefont {Schwartzman-Nowik}, \citenamefont {Paiva}, \citenamefont {Te'eni}, \citenamefont {Ruiz-Molero}, \citenamefont {Barbosa}, \citenamefont {Cohen},\ and\ \citenamefont {Galvão}}]{Wagner_2024}%
  \BibitemOpen
  \bibfield  {author} {\bibinfo {author} {\bibfnamefont {R.}~\bibnamefont {Wagner}}, \bibinfo {author} {\bibfnamefont {Z.}~\bibnamefont {Schwartzman-Nowik}}, \bibinfo {author} {\bibfnamefont {I.~L.}\ \bibnamefont {Paiva}}, \bibinfo {author} {\bibfnamefont {A.}~\bibnamefont {Te'eni}}, \bibinfo {author} {\bibfnamefont {A.}~\bibnamefont {Ruiz-Molero}}, \bibinfo {author} {\bibfnamefont {R.~S.}\ \bibnamefont {Barbosa}}, \bibinfo {author} {\bibfnamefont {E.}~\bibnamefont {Cohen}},\ and\ \bibinfo {author} {\bibfnamefont {E.~F.}\ \bibnamefont {Galvão}},\ }\bibfield  {title} {\bibinfo {title} {Quantum circuits for measuring weak values, kirkwood-dirac quasiprobability distributions, and state spectra},\ }\href {https://doi.org/10.1088/2058-9565/ad124c} {\bibfield  {journal} {\bibinfo  {journal} {Quantum Science and Technology}\ }\textbf {\bibinfo {volume} {9}},\ \bibinfo {pages} {015030} (\bibinfo {year} {2024})}\BibitemShut {NoStop}%
\bibitem [{\citenamefont {Schmid}\ \emph {et~al.}(2024)\citenamefont {Schmid}, \citenamefont {Baldijão}, \citenamefont {Yīng}, \citenamefont {Wagner},\ and\ \citenamefont {Selby}}]{schmid2024kirkwooddirac}%
  \BibitemOpen
  \bibfield  {author} {\bibinfo {author} {\bibfnamefont {D.}~\bibnamefont {Schmid}}, \bibinfo {author} {\bibfnamefont {R.~D.}\ \bibnamefont {Baldijão}}, \bibinfo {author} {\bibfnamefont {Y.}~\bibnamefont {Yīng}}, \bibinfo {author} {\bibfnamefont {R.}~\bibnamefont {Wagner}},\ and\ \bibinfo {author} {\bibfnamefont {J.~H.}\ \bibnamefont {Selby}},\ }\href@noop {} {\bibinfo {title} {Kirkwood-dirac representations beyond quantum states (and their relation to noncontextuality)}} (\bibinfo {year} {2024}),\ \Eprint {https://arxiv.org/abs/2405.04573} {arXiv:2405.04573} \BibitemShut {NoStop}%
\bibitem [{\citenamefont {Pusey}(2014)}]{PhysRevLett.113.200401}%
  \BibitemOpen
  \bibfield  {author} {\bibinfo {author} {\bibfnamefont {M.~F.}\ \bibnamefont {Pusey}},\ }\bibfield  {title} {\bibinfo {title} {Anomalous weak values are proofs of contextuality},\ }\href {https://doi.org/10.1103/PhysRevLett.113.200401} {\bibfield  {journal} {\bibinfo  {journal} {Phys. Rev. Lett.}\ }\textbf {\bibinfo {volume} {113}},\ \bibinfo {pages} {200401} (\bibinfo {year} {2014})}\BibitemShut {NoStop}%
\bibitem [{\citenamefont {Williams}\ and\ \citenamefont {Jordan}(2008)}]{Williams_2008}%
  \BibitemOpen
  \bibfield  {author} {\bibinfo {author} {\bibfnamefont {N.~S.}\ \bibnamefont {Williams}}\ and\ \bibinfo {author} {\bibfnamefont {A.~N.}\ \bibnamefont {Jordan}},\ }\bibfield  {title} {\bibinfo {title} {Weak values and the leggett-garg inequality in solid-state qubits},\ }\bibfield  {journal} {\bibinfo  {journal} {Physical Review Letters}\ }\textbf {\bibinfo {volume} {100}},\ \href {https://doi.org/10.1103/physrevlett.100.026804} {10.1103/physrevlett.100.026804} (\bibinfo {year} {2008})\BibitemShut {NoStop}%
\bibitem [{\citenamefont {Hofmann}(2024)}]{Hofmann2024}%
  \BibitemOpen
  \bibfield  {author} {\bibinfo {author} {\bibfnamefont {H.~F.}\ \bibnamefont {Hofmann}},\ }\bibfield  {title} {\bibinfo {title} {Statistical signatures of quantum contextuality},\ }\href {https://doi.org/10.3390/e26090725} {\bibfield  {journal} {\bibinfo  {journal} {Entropy}\ }\textbf {\bibinfo {volume} {26}},\ \bibinfo {pages} {725} (\bibinfo {year} {2024})}\BibitemShut {NoStop}%
\bibitem [{\citenamefont {Arvidsson-Shukur}\ \emph {et~al.}(2024)\citenamefont {Arvidsson-Shukur}, \citenamefont {Jr.}, \citenamefont {Bievre}, \citenamefont {Dressel}, \citenamefont {Jordan}, \citenamefont {Langrenez}, \citenamefont {Lostaglio}, \citenamefont {Lundeen},\ and\ \citenamefont {Halpern}}]{arvidssonshukur2024properties}%
  \BibitemOpen
  \bibfield  {author} {\bibinfo {author} {\bibfnamefont {D.~R.~M.}\ \bibnamefont {Arvidsson-Shukur}}, \bibinfo {author} {\bibfnamefont {W.~F.~B.}\ \bibnamefont {Jr.}}, \bibinfo {author} {\bibfnamefont {S.~D.}\ \bibnamefont {Bievre}}, \bibinfo {author} {\bibfnamefont {J.}~\bibnamefont {Dressel}}, \bibinfo {author} {\bibfnamefont {A.~N.}\ \bibnamefont {Jordan}}, \bibinfo {author} {\bibfnamefont {C.}~\bibnamefont {Langrenez}}, \bibinfo {author} {\bibfnamefont {M.}~\bibnamefont {Lostaglio}}, \bibinfo {author} {\bibfnamefont {J.~S.}\ \bibnamefont {Lundeen}},\ and\ \bibinfo {author} {\bibfnamefont {N.~Y.}\ \bibnamefont {Halpern}},\ }\href@noop {} {\bibinfo {title} {Properties and applications of the kirkwood-dirac distribution}} (\bibinfo {year} {2024}),\ \Eprint {https://arxiv.org/abs/2403.18899} {arXiv:2403.18899} \BibitemShut {NoStop}%
\bibitem [{\citenamefont {Raussendorf}\ \emph {et~al.}(2017)\citenamefont {Raussendorf}, \citenamefont {Browne}, \citenamefont {Delfosse}, \citenamefont {Okay},\ and\ \citenamefont {Bermejo-Vega}}]{Raussendorf_2017}%
  \BibitemOpen
  \bibfield  {author} {\bibinfo {author} {\bibfnamefont {R.}~\bibnamefont {Raussendorf}}, \bibinfo {author} {\bibfnamefont {D.~E.}\ \bibnamefont {Browne}}, \bibinfo {author} {\bibfnamefont {N.}~\bibnamefont {Delfosse}}, \bibinfo {author} {\bibfnamefont {C.}~\bibnamefont {Okay}},\ and\ \bibinfo {author} {\bibfnamefont {J.}~\bibnamefont {Bermejo-Vega}},\ }\bibfield  {title} {\bibinfo {title} {Contextuality and wigner-function negativity in qubit quantum computation},\ }\bibfield  {journal} {\bibinfo  {journal} {Physical Review A}\ }\textbf {\bibinfo {volume} {95}},\ \href {https://doi.org/10.1103/physreva.95.052334} {10.1103/physreva.95.052334} (\bibinfo {year} {2017})\BibitemShut {NoStop}%
\bibitem [{\citenamefont {Bièvre}\ \emph {et~al.}(2025)\citenamefont {Bièvre}, \citenamefont {Langrenez},\ and\ \citenamefont {Radchenko}}]{debievre2025}%
  \BibitemOpen
  \bibfield  {author} {\bibinfo {author} {\bibfnamefont {S.~D.}\ \bibnamefont {Bièvre}}, \bibinfo {author} {\bibfnamefont {C.}~\bibnamefont {Langrenez}},\ and\ \bibinfo {author} {\bibfnamefont {D.}~\bibnamefont {Radchenko}},\ }\href {https://arxiv.org/abs/2501.12252} {\bibinfo {title} {The kirkwood-dirac representation associated to the fourier transform for finite abelian groups: positivity}} (\bibinfo {year} {2025}),\ \Eprint {https://arxiv.org/abs/2501.12252} {arXiv:2501.12252 [quant-ph]} \BibitemShut {NoStop}%
\bibitem [{\citenamefont {Spriet}(2026)}]{Spriet2026}%
  \BibitemOpen
  \bibfield  {author} {\bibinfo {author} {\bibfnamefont {M.}~\bibnamefont {Spriet}},\ }\href {https://arxiv.org/abs/2507.23628} {\bibinfo {title} {Characterizing the kirkwood-dirac positivity on second countable lca groups}} (\bibinfo {year} {2026}),\ \Eprint {https://arxiv.org/abs/2507.23628} {arXiv:2507.23628 [quant-ph]} \BibitemShut {NoStop}%
\bibitem [{\citenamefont {Langrenez}\ \emph {et~al.}(2026)\citenamefont {Langrenez}, \citenamefont {Salmon}, \citenamefont {De~Bièvre}, \citenamefont {Thio}, \citenamefont {Long},\ and\ \citenamefont {Arvidsson-Shukur}}]{langrenez2024}%
  \BibitemOpen
  \bibfield  {author} {\bibinfo {author} {\bibfnamefont {C.}~\bibnamefont {Langrenez}}, \bibinfo {author} {\bibfnamefont {W.}~\bibnamefont {Salmon}}, \bibinfo {author} {\bibfnamefont {S.}~\bibnamefont {De~Bièvre}}, \bibinfo {author} {\bibfnamefont {J.~J.}\ \bibnamefont {Thio}}, \bibinfo {author} {\bibfnamefont {C.~K.}\ \bibnamefont {Long}},\ and\ \bibinfo {author} {\bibfnamefont {D.~R.~M.}\ \bibnamefont {Arvidsson-Shukur}},\ }\bibfield  {title} {\bibinfo {title} {Almost sure minimality of the set of experiments with classical kirkwood-dirac representations},\ }\bibfield  {journal} {\bibinfo  {journal} {Physical Review A}\ }\textbf {\bibinfo {volume} {113}},\ \href {https://doi.org/10.1103/v7z4-qsz8} {10.1103/v7z4-qsz8} (\bibinfo {year} {2026})\BibitemShut {NoStop}%
\bibitem [{\citenamefont {Langrenez}\ \emph {et~al.}(2024)\citenamefont {Langrenez}, \citenamefont {Arvidsson-Shukur},\ and\ \citenamefont {De~Bièvre}}]{Langrenez_2024_characterising}%
  \BibitemOpen
  \bibfield  {author} {\bibinfo {author} {\bibfnamefont {C.}~\bibnamefont {Langrenez}}, \bibinfo {author} {\bibfnamefont {D.~R.~M.}\ \bibnamefont {Arvidsson-Shukur}},\ and\ \bibinfo {author} {\bibfnamefont {S.}~\bibnamefont {De~Bièvre}},\ }\bibfield  {title} {\bibinfo {title} {Characterizing the geometry of the kirkwood–dirac-positive states},\ }\bibfield  {journal} {\bibinfo  {journal} {Journal of Mathematical Physics}\ }\textbf {\bibinfo {volume} {65}},\ \href {https://doi.org/10.1063/5.0164672} {10.1063/5.0164672} (\bibinfo {year} {2024})\BibitemShut {NoStop}%
\bibitem [{\citenamefont {De~Bièvre}(2021)}]{De_Bievre_2021}%
  \BibitemOpen
  \bibfield  {author} {\bibinfo {author} {\bibfnamefont {S.}~\bibnamefont {De~Bièvre}},\ }\bibfield  {title} {\bibinfo {title} {Complete incompatibility, support uncertainty, and kirkwood-dirac nonclassicality},\ }\bibfield  {journal} {\bibinfo  {journal} {Physical Review Letters}\ }\textbf {\bibinfo {volume} {127}},\ \href {https://doi.org/10.1103/physrevlett.127.190404} {10.1103/physrevlett.127.190404} (\bibinfo {year} {2021})\BibitemShut {NoStop}%
\bibitem [{\citenamefont {Xu}(2024)}]{Xu_2024}%
  \BibitemOpen
  \bibfield  {author} {\bibinfo {author} {\bibfnamefont {J.}~\bibnamefont {Xu}},\ }\bibfield  {title} {\bibinfo {title} {Kirkwood-dirac classical pure states},\ }\href {https://doi.org/10.1016/j.physleta.2024.129529} {\bibfield  {journal} {\bibinfo  {journal} {Physics Letters A}\ }\textbf {\bibinfo {volume} {510}},\ \bibinfo {pages} {129529} (\bibinfo {year} {2024})}\BibitemShut {NoStop}%
\bibitem [{\citenamefont {Yang}\ \emph {et~al.}(2024)\citenamefont {Yang}, \citenamefont {Yao}, \citenamefont {Geng}, \citenamefont {Wang},\ and\ \citenamefont {Chen}}]{Yang_2024}%
  \BibitemOpen
  \bibfield  {author} {\bibinfo {author} {\bibfnamefont {Y.-H.}\ \bibnamefont {Yang}}, \bibinfo {author} {\bibfnamefont {S.}~\bibnamefont {Yao}}, \bibinfo {author} {\bibfnamefont {S.-J.}\ \bibnamefont {Geng}}, \bibinfo {author} {\bibfnamefont {X.-L.}\ \bibnamefont {Wang}},\ and\ \bibinfo {author} {\bibfnamefont {P.-Y.}\ \bibnamefont {Chen}},\ }\bibfield  {title} {\bibinfo {title} {Geometry of kirkwood–dirac classical states: a case study based on discrete fourier transform},\ }\href {https://doi.org/10.1088/1751-8121/ad819a} {\bibfield  {journal} {\bibinfo  {journal} {Journal of Physics A: Mathematical and Theoretical}\ }\textbf {\bibinfo {volume} {57}},\ \bibinfo {pages} {435303} (\bibinfo {year} {2024})}\BibitemShut {NoStop}%
\bibitem [{\citenamefont {Oszmaniec}\ \emph {et~al.}(2021)\citenamefont {Oszmaniec}, \citenamefont {Brod},\ and\ \citenamefont {Galvão}}]{oszmaniec2021measuring}%
  \BibitemOpen
  \bibfield  {author} {\bibinfo {author} {\bibfnamefont {M.}~\bibnamefont {Oszmaniec}}, \bibinfo {author} {\bibfnamefont {D.~J.}\ \bibnamefont {Brod}},\ and\ \bibinfo {author} {\bibfnamefont {E.~F.}\ \bibnamefont {Galvão}},\ }\href@noop {} {\bibinfo {title} {Measuring relational information between quantum states, and applications}} (\bibinfo {year} {2021}),\ \Eprint {https://arxiv.org/abs/2109.10006} {arXiv:2109.10006} \BibitemShut {NoStop}%
\bibitem [{\citenamefont {Xu}(2025)}]{Xu_2025}%
  \BibitemOpen
  \bibfield  {author} {\bibinfo {author} {\bibfnamefont {J.}~\bibnamefont {Xu}},\ }\bibfield  {title} {\bibinfo {title} {Hermitian kirkwood–dirac-real operators for discrete fourier transformations},\ }\bibfield  {journal} {\bibinfo  {journal} {Journal of Mathematical Physics}\ }\textbf {\bibinfo {volume} {66}},\ \href {https://doi.org/10.1063/5.0256339} {10.1063/5.0256339} (\bibinfo {year} {2025})\BibitemShut {NoStop}%
\bibitem [{Sup()}]{SupplementalMaterial}%
  \BibitemOpen
  \href@noop {} {}\bibinfo {note} {See Supplemental Material at http://link.aps.org/supplemental/10.1103/rsy9-12fx for a Mathematica notebook containing calculations supporting Theorem 2.}\BibitemShut {Stop}%
\bibitem [{\citenamefont {Ding}\ and\ \citenamefont {Rhee}(2013)}]{Ding01032013}%
  \BibitemOpen
  \bibfield  {author} {\bibinfo {author} {\bibfnamefont {J.}~\bibnamefont {Ding}}\ and\ \bibinfo {author} {\bibfnamefont {N.~H.}\ \bibnamefont {Rhee}},\ }\bibfield  {title} {\bibinfo {title} {Teaching tip: When a matrix and its inverse are stochastic},\ }\href {https://doi.org/10.4169/college.math.j.44.2.108} {\bibfield  {journal} {\bibinfo  {journal} {The College Mathematics Journal}\ }\textbf {\bibinfo {volume} {44}},\ \bibinfo {pages} {108} (\bibinfo {year} {2013})},\ \Eprint {https://arxiv.org/abs/https://doi.org/10.4169/college.math.j.44.2.108} {https://doi.org/10.4169/college.math.j.44.2.108} \BibitemShut {NoStop}%
\bibitem [{\citenamefont {Thio}\ \emph {et~al.}(2026)\citenamefont {Thio}, \citenamefont {Yang}, \citenamefont {Halpern}, \citenamefont {Bièvre}, \citenamefont {Barnes},\ and\ \citenamefont {Arvidsson-Shukur}}]{Thio_2026}%
  \BibitemOpen
  \bibfield  {author} {\bibinfo {author} {\bibfnamefont {J.~J.}\ \bibnamefont {Thio}}, \bibinfo {author} {\bibfnamefont {S.}~\bibnamefont {Yang}}, \bibinfo {author} {\bibfnamefont {N.~Y.}\ \bibnamefont {Halpern}}, \bibinfo {author} {\bibfnamefont {S.~D.}\ \bibnamefont {Bièvre}}, \bibinfo {author} {\bibfnamefont {C.~H.~W.}\ \bibnamefont {Barnes}},\ and\ \bibinfo {author} {\bibfnamefont {D.~R.~M.}\ \bibnamefont {Arvidsson-Shukur}},\ }\href {https://arxiv.org/abs/2506.08092} {\bibinfo {title} {Kirkwood-dirac nonpositivity is a necessary resource for quantum computing}} (\bibinfo {year} {2026}),\ \Eprint {https://arxiv.org/abs/2506.08092} {arXiv:2506.08092 [quant-ph]} \BibitemShut {NoStop}%
\bibitem [{\citenamefont {Francica}(2022)}]{Francica_2022}%
  \BibitemOpen
  \bibfield  {author} {\bibinfo {author} {\bibfnamefont {G.}~\bibnamefont {Francica}},\ }\bibfield  {title} {\bibinfo {title} {Most general class of quasiprobability distributions of work},\ }\bibfield  {journal} {\bibinfo  {journal} {Physical Review E}\ }\textbf {\bibinfo {volume} {106}},\ \href {https://doi.org/10.1103/physreve.106.054129} {10.1103/physreve.106.054129} (\bibinfo {year} {2022})\BibitemShut {NoStop}%
\bibitem [{\citenamefont {Pei}\ \emph {et~al.}(2023)\citenamefont {Pei}, \citenamefont {Chen},\ and\ \citenamefont {Quan}}]{Pei_2023}%
  \BibitemOpen
  \bibfield  {author} {\bibinfo {author} {\bibfnamefont {J.-H.}\ \bibnamefont {Pei}}, \bibinfo {author} {\bibfnamefont {J.-F.}\ \bibnamefont {Chen}},\ and\ \bibinfo {author} {\bibfnamefont {H.~T.}\ \bibnamefont {Quan}},\ }\bibfield  {title} {\bibinfo {title} {Exploring quasiprobability approaches to quantum work in the presence of initial coherence: Advantages of the margenau-hill distribution},\ }\bibfield  {journal} {\bibinfo  {journal} {Physical Review E}\ }\textbf {\bibinfo {volume} {108}},\ \href {https://doi.org/10.1103/physreve.108.054109} {10.1103/physreve.108.054109} (\bibinfo {year} {2023})\BibitemShut {NoStop}%
\bibitem [{\citenamefont {Langrenez}(2025)}]{Langrenez_thesis}%
  \BibitemOpen
  \bibfield  {author} {\bibinfo {author} {\bibfnamefont {C.}~\bibnamefont {Langrenez}},\ }\emph {\bibinfo {title} {{La distribution de Kirkwood-Dirac comme outil de l'information quantique}}},\ \href {https://hal.science/tel-05395404} {\bibinfo {type} {Theses}},\ \bibinfo  {school} {{Universit{\'e} de Lille}} (\bibinfo {year} {2025})\BibitemShut {NoStop}%
\bibitem [{\citenamefont {Zurel}\ and\ \citenamefont {Heimendahl}(2024)}]{zurel2024}%
  \BibitemOpen
  \bibfield  {author} {\bibinfo {author} {\bibfnamefont {M.}~\bibnamefont {Zurel}}\ and\ \bibinfo {author} {\bibfnamefont {A.}~\bibnamefont {Heimendahl}},\ }\href {https://arxiv.org/abs/2407.10349} {\bibinfo {title} {Efficient classical simulation of quantum computation beyond wigner positivity}} (\bibinfo {year} {2024}),\ \Eprint {https://arxiv.org/abs/2407.10349} {arXiv:2407.10349 [quant-ph]} \BibitemShut {NoStop}%
\bibitem [{\citenamefont {O'Connell}\ and\ \citenamefont {Wang}(1985)}]{OCONNELL19859}%
  \BibitemOpen
  \bibfield  {author} {\bibinfo {author} {\bibfnamefont {R.}~\bibnamefont {O'Connell}}\ and\ \bibinfo {author} {\bibfnamefont {L.}~\bibnamefont {Wang}},\ }\bibfield  {title} {\bibinfo {title} {A new parametrized quantum distribution function and its time development},\ }\href {https://doi.org/https://doi.org/10.1016/0375-9601(85)90235-X} {\bibfield  {journal} {\bibinfo  {journal} {Physics Letters A}\ }\textbf {\bibinfo {volume} {107}},\ \bibinfo {pages} {9} (\bibinfo {year} {1985})}\BibitemShut {NoStop}%
\bibitem [{\citenamefont {O'Connell}(1986)}]{OConnell1986}%
  \BibitemOpen
  \bibfield  {author} {\bibinfo {author} {\bibfnamefont {R.~F.}\ \bibnamefont {O'Connell}},\ }\bibinfo {title} {Quantum distribution functions in non-equilibrium statistical mechanics},\ in\ \href {https://doi.org/10.1007/978-1-4613-2181-1_5} {\emph {\bibinfo {booktitle} {Frontiers of Nonequilibrium Statistical Physics}}},\ \bibinfo {editor} {edited by\ \bibinfo {editor} {\bibfnamefont {G.~T.}\ \bibnamefont {Moore}}\ and\ \bibinfo {editor} {\bibfnamefont {M.~O.}\ \bibnamefont {Scully}}}\ (\bibinfo  {publisher} {Springer US},\ \bibinfo {address} {Boston, MA},\ \bibinfo {year} {1986})\ pp.\ \bibinfo {pages} {83--95}\BibitemShut {NoStop}%
\bibitem [{\citenamefont {Li}\ and\ \citenamefont {Tan}(2025)}]{PhysRevA.111.022409}%
  \BibitemOpen
  \bibfield  {author} {\bibinfo {author} {\bibfnamefont {M.-S.}\ \bibnamefont {Li}}\ and\ \bibinfo {author} {\bibfnamefont {Y.-X.}\ \bibnamefont {Tan}},\ }\bibfield  {title} {\bibinfo {title} {Bargmann invariants for quantum imaginarity},\ }\href {https://doi.org/10.1103/PhysRevA.111.022409} {\bibfield  {journal} {\bibinfo  {journal} {Phys. Rev. A}\ }\textbf {\bibinfo {volume} {111}},\ \bibinfo {pages} {022409} (\bibinfo {year} {2025})}\BibitemShut {NoStop}%
\bibitem [{\citenamefont {Zhang}\ \emph {et~al.}(2025)\citenamefont {Zhang}, \citenamefont {Xie},\ and\ \citenamefont {Li}}]{PhysRevA.111.042417}%
  \BibitemOpen
  \bibfield  {author} {\bibinfo {author} {\bibfnamefont {L.}~\bibnamefont {Zhang}}, \bibinfo {author} {\bibfnamefont {B.}~\bibnamefont {Xie}},\ and\ \bibinfo {author} {\bibfnamefont {B.}~\bibnamefont {Li}},\ }\bibfield  {title} {\bibinfo {title} {Geometry of sets of bargmann invariants},\ }\href {https://doi.org/10.1103/PhysRevA.111.042417} {\bibfield  {journal} {\bibinfo  {journal} {Phys. Rev. A}\ }\textbf {\bibinfo {volume} {111}},\ \bibinfo {pages} {042417} (\bibinfo {year} {2025})}\BibitemShut {NoStop}%
\bibitem [{\citenamefont {Pratapsi}\ \emph {et~al.}(2025)\citenamefont {Pratapsi}, \citenamefont {Gouveia}, \citenamefont {Novo},\ and\ \citenamefont {Galv\~ao}}]{hsnv-wpt3}%
  \BibitemOpen
  \bibfield  {author} {\bibinfo {author} {\bibfnamefont {S.~S.}\ \bibnamefont {Pratapsi}}, \bibinfo {author} {\bibfnamefont {J.~a.}\ \bibnamefont {Gouveia}}, \bibinfo {author} {\bibfnamefont {L.}~\bibnamefont {Novo}},\ and\ \bibinfo {author} {\bibfnamefont {E.~F.}\ \bibnamefont {Galv\~ao}},\ }\bibfield  {title} {\bibinfo {title} {Elementary characterization of bargmann invariants},\ }\href {https://doi.org/10.1103/hsnv-wpt3} {\bibfield  {journal} {\bibinfo  {journal} {Phys. Rev. A}\ }\textbf {\bibinfo {volume} {112}},\ \bibinfo {pages} {042421} (\bibinfo {year} {2025})}\BibitemShut {NoStop}%
\bibitem [{\citenamefont {Xu}(2026)}]{Xu_2026}%
  \BibitemOpen
  \bibfield  {author} {\bibinfo {author} {\bibfnamefont {J.}~\bibnamefont {Xu}},\ }\bibfield  {title} {\bibinfo {title} {Numerical ranges of bargmann invariants},\ }\href {https://doi.org/10.1016/j.physleta.2025.131091} {\bibfield  {journal} {\bibinfo  {journal} {Physics Letters A}\ }\textbf {\bibinfo {volume} {565}},\ \bibinfo {pages} {131091} (\bibinfo {year} {2026})}\BibitemShut {NoStop}%
\bibitem [{\citenamefont {Gross}\ and\ \citenamefont {Walter}(2013)}]{Gross_2013}%
  \BibitemOpen
  \bibfield  {author} {\bibinfo {author} {\bibfnamefont {D.}~\bibnamefont {Gross}}\ and\ \bibinfo {author} {\bibfnamefont {M.}~\bibnamefont {Walter}},\ }\bibfield  {title} {\bibinfo {title} {Stabilizer information inequalities from phase space distributions},\ }\bibfield  {journal} {\bibinfo  {journal} {Journal of Mathematical Physics}\ }\textbf {\bibinfo {volume} {54}},\ \href {https://doi.org/10.1063/1.4818950} {10.1063/1.4818950} (\bibinfo {year} {2013})\BibitemShut {NoStop}%
\bibitem [{\citenamefont {Gross}\ \emph {et~al.}(2021)\citenamefont {Gross}, \citenamefont {Nezami},\ and\ \citenamefont {Walter}}]{Gross_2021}%
  \BibitemOpen
  \bibfield  {author} {\bibinfo {author} {\bibfnamefont {D.}~\bibnamefont {Gross}}, \bibinfo {author} {\bibfnamefont {S.}~\bibnamefont {Nezami}},\ and\ \bibinfo {author} {\bibfnamefont {M.}~\bibnamefont {Walter}},\ }\bibfield  {title} {\bibinfo {title} {Schur–weyl duality for the clifford group with applications: Property testing, a robust hudson theorem, and de finetti representations},\ }\href {https://doi.org/10.1007/s00220-021-04118-7} {\bibfield  {journal} {\bibinfo  {journal} {Communications in Mathematical Physics}\ }\textbf {\bibinfo {volume} {385}},\ \bibinfo {pages} {1325–1393} (\bibinfo {year} {2021})}\BibitemShut {NoStop}%
\bibitem [{\citenamefont {García}\ \emph {et~al.}(2017)\citenamefont {García}, \citenamefont {Markov},\ and\ \citenamefont {Cross}}]{garcia2017geometrystabilizerstates}%
  \BibitemOpen
  \bibfield  {author} {\bibinfo {author} {\bibfnamefont {H.~J.}\ \bibnamefont {García}}, \bibinfo {author} {\bibfnamefont {I.~L.}\ \bibnamefont {Markov}},\ and\ \bibinfo {author} {\bibfnamefont {A.~W.}\ \bibnamefont {Cross}},\ }\href {https://arxiv.org/abs/1711.07848} {\bibinfo {title} {On the geometry of stabilizer states}} (\bibinfo {year} {2017}),\ \Eprint {https://arxiv.org/abs/1711.07848} {arXiv:1711.07848 [quant-ph]} \BibitemShut {NoStop}%
\end{thebibliography}
\end{document}